\RequirePackage{fix-cm}
\documentclass[smallextended]{svjour3}       % onecolumn (second format)
\smartqed  % flush right qed marks, e.g. at end of proof
\usepackage{graphicx}
\usepackage{natbib}
\usepackage{latexsym}
\usepackage{amssymb}
\usepackage{amsmath}
\usepackage{gensymb}
\usepackage{bm}
\usepackage[normalem]{ulem}
 \newcommand{\aap}{A\&A}
 \newcommand{\aapr}{A\&A Review}
 \newcommand{\mnras}{MNRAS}
 \newcommand{\apjs}{ApJS}
 
 \newcommand{\apj}{ApJ}
 \newcommand{\araa}{ARA\&A}
 
 \newcommand{\aj}{AJ}
 \newcommand{\pre}{Phys. Rev. E}
 
 \newcommand{\apjl}{ApJ}
 \newcommand{\nat}{Nature}
 
 \newcommand{\pasp}{PASP}
 
 \newcommand{\pasj}{PASJ}
 \newcommand{\ssr}{Space Science Reviews}
 \newcommand{\zap}{Zeitschrift f\"{u}r Astrophysik}

\def\aaps{A\&A Suppl.}
\def\pasa{PASA}
\def\bain{BAN}

\def\araa{ARA\&A}

\def\mnras{MNRAS}

\def\prl{PhysRevLett}

\def\doi{doi}
\def\urlprefix{}
\def\url{}
\def\eprint{arXiv}
\def\textbackslashSigma{Sigma}

\def\msun{\ifmmode M_\odot \else M$_\odot$ \fi}
\def\rsun{\ifmmode R_\odot \else R$_\odot$ \fi}
\def\lsun{\ifmmode L_\odot \else L$_\odot$ \fi}
\def\e{\ifmmode ^{-1} \else $^{-1}$ \fi}
\newcommand{\PD}[2]{\frac{\partial #1}{\partial #2}}
\newcommand{\vektor}[1]{{\bm #1}}

\usepackage[usenames,dvipsnames,svgnames,table]{xcolor}

\begin{document}

\title{Physical Processes in Star Formation}

\author{Philipp Girichidis \and
        Stella~S.~R.~Offner \and
        Alexei G. Kritsuk \and
        Ralf S.\ Klessen \and 
        Patrick Hennebelle \and
        J.~M.~Diederik~Kruijssen \and
        Martin G. H. Krause \and
        Simon C. O. Glover \and
        Marco Padovani
}

\authorrunning{Girichidis et al.}

\institute{
Philipp~Girichidis \at
              Leibniz-Institut f\"{u}r Astrophysik (AIP) - Potsdam, Germany
              \email{philipp@girichidis.com}
            \and
Stella~S.~R.~Offner \at
              The University of Texas at Austin - Austin TX, USA
            \and
Alexei G. Kritsuk \at
              University of California, San Diego; La Jolla, California, 92093-0424, USA
              \and
Ralf S.\ Klessen \at
            Universit\"{a}t Heidelberg, Zentrum f\"{u}r Astronomie, Institut f\"{u}r Theoretische Astrophysik,  Albert-Ueberle-Str. 2, 69120 Heidelberg, Germany
            \and
Patrick Hennebelle \at
            Laboratoire AIM, Paris-Saclay, CEA/IRFU/SAp – CNRS – Universit\'{e} Paris Diderot,91191 Gif-sur-Yvette Cedex, France LERMA (UMR CNRS 8112), Ecole NormaleSup\'{e}rieure, 75231 Paris Cedex, France
            \and
J.~M.~Diederik~Kruijssen \at
              Astronomisches Rechen-Institut, Zentrum f\"{u}r Astronomie der Universit\"{a}t Heidelberg - Heidelberg, Germany
            \and
Martin G. H. Krause \at
              Centre for Astrophysics Research, 
              School of Physics, Astronomy and Mathematics, 
              University of Hertfordshire, College Lane,
              Hatfield, Hertfordshire AL10 9AB, UK
            \and
Simon C. O. Glover \at
            Universit\"{a}t Heidelberg, Zentrum f\"{u}r Astronomie, Institut f\"{u}r Theoretische Astrophysik,  Albert-Ueberle-Str. 2, 69120 Heidelberg, Germany
            \and
Marco Padovani \at
              INAF-Osservatorio Astrofisico di Arcetri - Largo E. Fermi, 5 - 50125 Firenze, Italy
}

\date{Received: date / Accepted: date}

\maketitle

\begin{abstract}
Star formation is a complex multi-scale phenomenon that is of significant importance for astrophysics in general. Stars and star formation are key pillars in observational astronomy from local star forming regions in the Milky Way up to high-redshift galaxies. From a theoretical perspective, star formation and feedback processes (radiation, winds, and supernovae) play a pivotal role in advancing our understanding of the physical
processes at work, both individually and of their interactions. In this review we will give an overview of the main processes that are important for the understanding of star formation. We start with an observationally motivated view on star formation from a global perspective and outline the general paradigm of the life-cycle of molecular clouds, in which star formation is the key process to close the cycle. After that we focus on the thermal and chemical aspects in star forming regions, discuss turbulence and magnetic fields as well as gravitational forces. Finally, we review the most important stellar feedback mechanisms. 
\end{abstract}

\section{Introduction}

In this section we provide a general overview of the star formation process on global scales as well as the current paradigm of the life cycle of molecular clouds, in which the formation of stars marks the turning point between the cooling of gas, the condensation and finally the collapse on the one hand and the feedback processes on the other hand, in which the stars influence their environment and reheat the gas to complete the cycle. There are large variety of physical processes involved star formation, which interact in a complicated and highly non-linear manner and on very different spatial and dynamical scales. Nonetheless, the global process of star formation in galaxies seems to follow rather simple relations, which points towards a self-regulated rather than an unstable or chaotic process.

\subsection{Star formation in galaxies}
Star formation takes place in molecular clouds \citep{KennicuttEvans2012}. As a result, galaxies exhibit a tight relation between the star formation rate (SFR) surface density, $\Sigma_{\rm SFR}=\equiv\int_{-\infty}^{\infty}\rho_\mathrm{SFR}(z)\mathrm{d}z$ (typically expressed in units of $\mathrm{M}_\odot\,\mathrm{yr}^{-1}\,\mathrm{kpc}^{-2}$), and the (molecular) gas surface density, $\Sigma\equiv\int_{-\infty}^{\infty}\rho_\mathrm{gas}(z)\mathrm{d}z$ (typically expressed in units of $\mathrm{M}_\odot\,\mathrm{pc}^{-2}$) \citep{Schmidt1959,kennicutt89,KennicuttSchmidt1998}, where $z$ is the coordinate perpendicular to the galactic disc. In a spatially-resolved sense, this `star formation' relation persists down to scales of $\sim500$~pc \citep{BigielEtAl2008,kennicutt07,LeroyEtAl2013}, below which the stochasticity introduced by the time evolution of individual molecular clouds and star-forming regions causes the relation to break down \citep{SchrubaEtAl2010,LiuEtAl2011,feldmann11,KruijssenLongmore2014}. The star formation relation follows a power law $\Sigma_{\rm SFR}\propto\Sigma^N$ with $N=1.0{-}1.5$ and a normalisation implying a molecular `gas depletion time' (i.e.\ the time required to turn the entire gas reservoir into stars at the current SFR) of $t_{\rm dep}\equiv\Sigma/\Sigma_{\rm SFR}\sim2$~Gyr. Efforts to physically interpret this relation have focused on two main questions.
\begin{enumerate}
    \item
Which physical processes set the slope of the star formation relation? A power close to $N=1.5$ suggests that the free-fall time (related to the gas volume density as $t_{\rm ff}\propto\rho^{-1/2}$) is important, assuming a constant scale height gas disc so that $\Sigma\propto\rho$. In turn, this leaves the question whether the slope of the relation is exclusively set by free-fall collapse or whether it is affected by the balance between gas heating and cooling or the balance between energy dissipation and injection by feedback.
    \item 
Why is the molecular gas depletion time a factor of $\sim100$ longer than the dynamical time of molecular clouds (i.e.\ the turbulent crossing time or gravitational free-fall time). Does this mean that star formation in clouds takes place over many dynamical times at a high integrated star formation efficiency (i.e.\ the integrated fraction of gas that turns into stars) or that it takes place over a single dynamical time and achieves only a low star formation efficiency? Because the normalisation of the star formation relation reflects the gas depletion time, the answer to this question will also explain the proportionality constant of the star formation relation. 
\end{enumerate}
While both questions are related, as are the physical mechanisms setting the star formation relation slope and normalisation, there are different ways of addressing them.

The idea that the slope of the star formation relation is set by free-fall collapse goes back decades \citep[e.g.][]{madore77,Krumholz2005}. This interpretation has been complicated by the suggestion that the slope may vary as a function of spatial scale, going from clouds \citep[$N\gtrsim2$, e.g.][]{gutermuth11,lada13} to galaxies \citep[$N=1{-}1.5$, e.g.][]{KennicuttEvans2012}. The cloud-scale and galactic-scale star formation relations are not necessarily expected to be compatible, because cloud-scale relations select a single snapshot in the evolutionary timeline of molecular cloud evolution and star formation \citep{kruijssen18}. An additional complication is that the slope varies across the range of surface densities probed, with a steeper slope ($N\sim2$) at low ($\Sigma\lesssim10~{\rm M}_\odot~{\rm pc}^{-2}$) and high ($\Sigma\gtrsim200~{\rm M}_\odot~{\rm pc}^{-2}$) gas surface densities \citep[e.g.][]{BigielEtAl2008,daddi10,genzel10,KennicuttEvans2012}, It was suggested by \citet[also see \citealt{BacchiniEtAl2019}]{krumholz12} that the different incarnations of the star formation relation may be united by normalising the gas surface density to the appropriate version of the free-fall time. Specifically, if molecular clouds exhibit a large density contrast relative to the galactic midplane and thus exist in relative isolation, which happens mostly at low gas pressures, then their evolution takes place in a local free-fall time, which is much shorter than the average midplane gas free-fall time. If molecular clouds have densities similar to that of the average midplane gas, their evolution is affected by galactic dynamics and the relevant free-fall time is that of the midplane \citep[also see][]{jeffreson18}.

While these considerations address the scale dependence of the star formation relation, there is an increasing body of literature that suggests that the changing slope with surface density also has an important physical meaning. Most prominently, these studies suggest that the balance between self-gravity and stellar feedback sets the slope and normalisation of the star formation relation \citep[e.g.][]{OstrikerMcKeeLeroy2010,OstrikerShetty2011,HopkinsEtAl2014,hayward17,krumholz18,orr18} and predict multiple physical regimes. In understanding the different regimes predicted by these models, it is important to realise that star formation is broadly speaking an accelerating process, in which the collapse time decreases as the gas contracts. This means that the slowest, rate-limiting step is presented by the first bottleneck in the evolution of the interstellar medium towards star formation. At low gas surface densities ($\Sigma\lesssim10~{\rm M}_\odot~{\rm pc}^{-2}$) in the atomic gas-dominated regime the rate-limiting step is the condensation of molecular gas out of the atomic medium. Cooling competes with heating from stellar feedback, such that equating the heating and cooling rates gives:
\begin{equation}
    \Sigma_{\rm SFR}\propto \Sigma^2,
\end{equation}
where the first term is the heating rate (driven by massive stars and therefore proportional to $\Sigma_{\rm SFR}$), the second term is the cooling rate (driven by collisions and therefore proportional to $\Sigma^2$), and we have omitted an additional dependence on metallicity \citep[also see e.g.][]{schaye04,krumholz09b,krumholz09c,OstrikerMcKeeLeroy2010,hayward17}. At higher gas surface densities, the gas becomes predominantly molecular. Because the molecular gas in galaxies is supersonically turbulent, the rate-limiting step towards star formation becomes the turbulent energy dissipation rate. Equating this to the momentum injection rate by stellar feedback, we obtain:
\begin{align}
    \Sigma_{\rm SFR}&\propto \Sigma(\Sigma+\Sigma_\star) &  \\
    &\propto \Sigma & {\rm for}~ \Sigma\ll\Sigma_\star \nonumber\\
    &\propto \Sigma^2 & {\rm for}~ \Sigma\gg\Sigma_\star, \nonumber
\end{align}
where $\Sigma_\star$ is the stellar surface density, the first term is the momentum injection rate (driven by massive stars and therefore proportional to $\Sigma_{\rm SFR}$), the second term is the turbulent dissipation rate [in hydrostatic equilibrium, this is set by the motion of the gas under the influence of the total potential and is therefore proportional to $\Sigma(\Sigma+\Sigma_\star)$], and we have omitted an additional dependence on the \citet{Toomre1964} $Q$ stability parameter \citep[also see e.g.][]{OstrikerShetty2011,fauchergiguere13,KimOstriker2015,orr18}. The addition inside the parentheses implies two different regimes. At intermediate gas surface densities (lower than the stellar surface density, which in practice means $10\lesssim\Sigma/{\rm M}_\odot~{\rm pc}^{-2}\lesssim100$ for star-forming main sequence galaxies in the local Universe), the star formation relation is predicted to be linear, i.e.\ $\Sigma_{\rm SFR}\propto\Sigma$. At high gas surface densities ($\Sigma\gtrsim100~{\rm M}_\odot~{\rm pc}^{-2}$), the star formation relation is predicted to become super-linear, i.e.\ $\Sigma_{\rm SFR}\propto\Sigma^2$. These differences in slope between roughly-defined regimes of low, intermediate, and high gas surface density are in reasonable qualitative agreement with the observed star formation relation \citep{KennicuttEvans2012}.

In the equilibrium-based models discussed above, the normalisation of the star formation relation is set by the balance between the heating or momentum injection rate and the cooling or turbulence dissipation rate. In the molecular regime, this implies that the normalisation of the star formation relation is set by the momentum input rate per unit stellar mass, which is a single number that depends on the details of stellar evolution and the porosity of the interstellar medium \citep{OstrikerShetty2011,krumholz18}. More vigorous feedback drives the star formation relation to longer gas depletion times, whereas weaker feedback allows a higher SFR per unit gas mass.

While the concept of energy and momentum balance can explain the overall normalisation of the star formation relation, it does not necessarily explain whether star formation in molecular clouds takes place over many free-fall times (achieving high star formation efficiencies) or a single one (achieving a low star formation efficiency). Whichever the answer, feedback plays a key role. In the case of slow and efficient star formation, feedback would act as a homogeneous pressure term that adiabatically resists gravitational collapse, even within individual molecular clouds. In the case of rapid and inefficient star formation, feedback would act as an impulsive disruptor that disperses molecular clouds and halts the star formation process. Historically, it has been extremely challenging to measure the timescales governing the molecular cloud lifecycle, but the combination of new methodology \citep{kruijssen18} and high-resolution imaging of molecular gas in nearby galaxies \citep[e.g.][]{sun18} now enables this question to be resolved. By analysing the spatial offset between tracers of molecular gas and massive star formation \citep[e.g.][]{SchrubaEtAl2010,kreckel18,schinnerer19} it is possible to quantify the underlying cloud lifecycle. Across the local galaxy population, it is found that molecular clouds live for about a dynamical time (10--30~Myr) and achieve low star formation efficiencies \citep[1--10\%,][also see below]{KruijssenEtAl2019,ChevanceEtAl2019}.

The above result implies that the star formation relation is an ensemble average of the population of molecular clouds and star-forming regions, where each is individually subject to highly dynamical processes driving rapid evolutionary cycling \citep{KruijssenEtAl2019}. Star formation in galaxies thus represents a truly multi-scale system -- the cloud-scale evolution is sensitive to the large-scale energy and momentum balance, which sets e.g.\ the rate of cloud formation, yet the large-scale balance is also influenced by the rate and efficiency of star formation and feedback on the cloud scale, which sets the energy and momentum input rate.

\subsection{The lifecycle of molecular clouds}
Star formation in galaxies is a continuous process which is tightly coupled to the life cycle of molecular clouds. Most of the volume in galaxies is filled with hot, low density gas \citep{HaffnerEtAl2009}. The mass fraction of the low-density ($\sim10^{-3}-10^{-2}\,\mathrm{cm}^{-3}$) gas is small. However, due to the large temperatures of $\sim10^6\,\mathrm{K}$ this phase contains a significant fraction of the thermal energy \citep{ferriere01}. Collisional excitation and radiative cooling allow the hot phase to cool to $10^4\,\mathrm{K}$ with typical densities of approximately $1\,\mathrm{cm}^{-3}$, which is the warm diffuse phase. As the cooling times in the hot phase are long compared to the dynamical time \citep[e.g.][]{McKeeOstriker1977,GnatFerland2012}, cooling is assisted by turbulent compression. Neither the hot nor the warm gas are dominated by self-gravity and are stabilized against gravitational collapse by thermal pressure, magnetic fields and turbulent motions. We note that turbulence on the one hand creates over-densities and assist cooling and on the other hand can support the gas by an effective turbulent pressure \citep[e.g.][]{MacLowKlessen2004}. The diffuse atomic gas can further cool down to form a colder phase with temperatures of a few $10-100\,\mathrm{K}$ with densities of $\gtrsim10^3\,\mathrm{cm}^{-3}$, which consists mainly of molecular rather than atomic hydrogen \citep{HennebelleFalgarone2012,klessen16}. Molecular clouds are turbulent structures with typical spatial extents of $10-50\,\mathrm{pc}$ and masses ranging from $10^2-10^6\,M_\odot$ \citep{MivilleDeschenesMurrayLee2017}. We highlight that the cold phase and, in particular, molecular clouds and clumps are not isolated entities. Instead, they condense out of a complicated filamentary network \citep[e.g.][]{ArzoumanianEtAl2011}. These elongated and complex filaments are in good agreement with structures that form because of the cascade of turbulent motions \citep[e.g.][]{MacLowKlessen2004,HennebelleFalgarone2012}. We note that molecular clouds may have formed due to gravitational attraction in the galaxy, but are globally not bound by self-gravity \citep{HeyerEtAl2009, HeyerDame2015}. The virial parameter
\begin{equation}
\alpha_{\mathrm{vir}}=\frac{2 T}{|W|} \approx \frac{5 \sigma_{v}^{2} R}{G M}
\end{equation}
relates the kinetic ($T$) to the gravitational energy ($W$) of a cloud. Assuming a homogeneous sphere this ratio can be approximated as the last term where $\sigma_v$ is the one-dimensional velocity dispersion, $R$ is the radius, $M$ is the mass, and $G$ the gravitational constant. Observed values of $\alpha_{\mathrm{vir}}$ for clouds in the Milky Way span two orders of magnitude from $1-100$ \citep{MivilleDeschenesMurrayLee2017}.  However, we note that there is no consensus on how important gravity is for cloud formation and dynamical driving. As the warm and the cold phase can coexist in pressure equilibrium \citep{Field1965,McKeeOstriker1977}, turbulent motions and gravity play an important role in shaping molecular clouds.

The next step towards star formation is the gravitational instability that -- together with turbulence, rotation and magnetic fields -- determines the fragmentation of gas and the resulting spatial and mass distribution of collapsing clumps that are the sites of star and star cluster formation. Once fragmented regions start to locally collapse due to self-gravity, the opposing forces like thermal pressure and magnetic fields become less significant and the star-forming process can be described as a gravo-turbulent process. During the fragmentation and contraction of the molecular cloud it is also very likely that many individual regions collapse simultaneously. As a result, cores and stars typically form in groups, i.e. clusters and associations, \citep{LadaLada2003, BressertEtAl2010, Kruijssen2012}. However, the fraction of how many stars form in clusters and details of the clustering are still  debated \citep[e.g.][]{longmore2014,krumholz2019,ward2019}. The first gravitational instability only sets the seeds of star formation. The actual masses of stars are further influenced by accretion from the ambient reservoir. In the case of high-mass star formation two extreme models have been proposed and found in numerical simulations. In the \emph{competitive accretion} model \citep{BonnellEtAl2001} gas is funneled into the centre of the gravitational potential. Accretion onto centrally located stars is then favoured, which allows them to grow into the most massive stars. The opposite effect occurs in the case of \emph{fragmentation induced starvation}, where gas on its way to the centre of the cluster fragments \citep{PetersEtAl2010c, GirichidisEtAl2012a, KruijssenEtAl2012}. The new stars accrete in-flowing material, thereby depriving the central stars of material and halting their runaway growth. Which model is favoured depends on the details of the accretion flow, the nature of the turbulence, the available gas in the vicinity of the local gravitational centre, the position of the collapsing regions with respect to neighbouring condensations or the position relative to the centre of the molecular cloud core \citep{GirichidisEtAl2011}, and, finally, the onset of stellar activity and early (proto) stellar feedback, which can halt accretion \citep{PetersEtAl2010b, GeenEtAl2015}. The multitude of all of these processes eventually produces a remarkably universal distribution of stellar masses \citep{Kroupa01, Chabrier2003}. The \emph{stellar initial mass function} can be described by a lognormal distribution around a peak at $\sim0.2\,\mathrm{M}_\odot$ and a high-mass power law with a slope of $dN/dM \propto M^{-2.3}$ \citep{Salpeter1955}. However, simulations predict the mass function of the first stars in the early universe is a notable exception, since the lack of metal coolants changes the fragmentation behaviour and is likely to result in a more top-heavy IMF, i.e. an overabundance in massive stars compared to the present-day IFM \citep{GreifEtAl2011, ClarkEtAl2011b, StacyBromm2013, Susa2013, HiranoEtAl2014, StacyBrommLee2016}.

There are several forms of stellar feedback: protostellar outflows, radiation, stellar winds, and supernovae, where the latter two processes are mainly relevant for massive stars. The combined interaction of stellar feedback keeps the integrated star-formation efficiency low, at a few percent, and eventually leads to the dissolution of molecular clouds after about a dynamical time \citep[10--30~Myr, see e.g.][]{WalchEtAl2012,DaleErcolanoBonnell2013,KruijssenEtAl2019,ChevanceEtAl2019}. A fraction of the gas can be pushed out of the galactic disc to form a fountain flow or even escape from the galaxy as a wind \citep[e.g.][]{HillEtAl2012, WalchEtAl2015, GirichidisEtAl2016b}. With mass outflow rates that can be comparable to the star formation rate these outflows have a strong dynamical impact on the evolution of galaxies and the redistribution of gas in the ISM \citep[see e.g. reviews by][]{VeilleuxCecilBlandHawthorn2005,SomervilleDave2015,NaabOstriker2017}. The turbulent hot gas, which is enriched with metals produced in the stars then forms the reservoir from which the next cycle of molecular clouds forms. It is important to note that the cycle of gas in the interstellar medium is not a truly periodic cycle but rather a continuous process, in which different regions of the interstellar medium pass through the hot, warm and cold phases at different rates.

One important aspect of stellar feedback is that it is not {\it only} disruptive. Instead, pressure waves, ejected material and turbulence created by feedback can also locally trigger the onset of gravitational collapse and thereby cause star formation by increasing the gas density and accelerating cooling, a result known as triggered star formation.

\section{Composition and thermodynamic behavior of the ISM}
\label{sec:ISM}

The interstellar medium is composed almost entirely of hydrogen and helium, with the former accounting for around 70\% of the total mass and the latter for 28\%. All other elements contribute the remaining 2\% \citep{draine11}. While most of the light elements are found in the gas phase (Section \ref{sec:gas}), a large fraction of the heavier elements can be locked up in dust grains, corresponding on average to about $1$\% of the total mass (see \S\ref{sec:dust}). We provide an overview of the most important chemical reactions in \S\ref{sec:chem} and of the relevant cooling and heating processes in \S\ref{sec:cooling} and \S\ref{sec:heating}.

The total gas mass in the Milky Way is difficult to estimate but is probably close to $10^{10} \: {\rm M_{\odot}}$ \citep{kk09}. The majority of the volume of the ISM is occupied by ionized gas, but the total mass associated with this component is not more than around 25\% of the total gas mass. The majority of the mass is located in regions dominated by neutral atomic gas (H, He) or molecular gas (H$_{2}$). Much of the atomic gas and all of the molecular gas is found in the form of dense clouds that occupy only 1--2\% of the total ISM volume \cite[see  e.g.][]{ferriere01}.

\subsection{Different gas phases}
\label{sec:gas}

The changes in the chemical make-up of the ISM also go hand in hand with different thermal phases as discuss in more detail in Section \ref{sec:thermal-structure}. Starting with atomic gas, often called warm neutral medium (in short WNM), the main transitions are collisional ionization, leading to the warm ionized medium (WIM) when the temperature exceeds 10$^4\,$K, or ionization due to ultraviolet photons in the vicinity of high-mass stars, creating classical HII regions. Similarly, the formation of the hydrogen molecule marks the build-up of molecular clouds with densities above $\sim 100\,$cm$^{-3}$ and temperatures below $\sim 100\,$K. 

The simplest model of the ISM phase structure was suggested by \citet{fgh69}. If one assumes that the atomic gas in the ISM is in thermal equilibrium and in pressure balance, then there are two thermally stable solutions for a wide range of pressures. The cold dense phase corresponds to the cold neutral medium (CNM) introduced earlier, and the warm, diffuse phase is the warm neutral medium (WNM) discussed above. In the \citet{fgh69} model, gas at intermediate temperatures is thermally unstable: depending on its density it will either cool and get denser until it joins the CNM, or heat and become more tenuous until it joins the WNM. At high densities, there is considerable overlap between the physical parameters of the CNM and of molecular clouds, which are often thought to a have an envelope of dense neutral atomic gas. This two-phase model was extended by \citet{McKeeOstriker1977}, who pointed out that supernovae exploding in the ISM would create large, ionized bubbles filled with very hot gas ($T \sim 10^{6}\,$K). Although this gas would eventually cool, the temperature dependence of the atomic cooling curve at high temperatures is such that the cooling time around $T \sim 10^{6}$~K is considerably longer than the cooling time in the temperature range $10^{4}\,\mathrm{K} < T < 10^{6}\,$K (see \S~\ref{sec:chem} below). It is also longer than the time needed for supernovae to produce appreciable amounts of ionized material. And so, while gas at the low end of the temperature range quickly cools to join the WIM, gas close to $\sim 10^{6}\,$K  remains at these temperatures, effectively forming a third phase known as the hot ionized medium (HIM).

The Galactic distribution of molecular gas can be estimated by combining data from CO observations, which trace clouds with high concentrations of both H$_{2}$ and CO, with measurements of C$^{+}$, which trace so-called ``dark molecular gas'', i.e.\ clouds with high H$_{2}$ fractions but little CO \citep[see e.g.][]{pineda13}. The molecular gas surface density shows a  peak within the central few hundred parsec of the Galaxy, a region known as the Central Molecular Zone (CMZ). It then falls off sharply between 0.5 and 3~kpc, possibly owing to the influence of the Milky Way's central stellar bar \citep{ms96}, before peaking again at a Galactocentric radius of around 4--6~kpc in a structure known as the Molecular Ring. Outside the Molecular Ring, the surface density of molecular gas declines exponentially, but it can still be traced out to distances of at least 12--13~kpc \citep{heyer98}. 

Observational evidence for the presence of the WIM comes from free-free absorption of the Galactic synchrotron background \citep{he63}, the dispersion of radio signals from pulsars \citep{reynolds89,gae08}, and collisionally excited faint optical emission lines produced by ionized species such as O$^{+}$ and N$^{+}$ \citep{reynolds73,mie06}. Its density is comparable to the WNM with a scale-height of the order of 1~kpc \citep[see e.g.][]{reynolds89}. It is thought that over 90\% of the total ionized gas within the ISM is located in the WIM \citep{HaffnerEtAl2009}. We note that the ionized gas in classical H{\textsc{ii}} regions surrounding O stars is generally not considered to be part of the WIM. As mentioned above, the material in the WIM is collisionally ionized, whereas the high ionization degree in classical HII regions is due to ultraviolet photons from massive stars. We summarize the main physical properties of the different ISM phases  in Table~\ref{ism-phases}. We note, however, that the picture of a simple three-phase medium is a severe over-simplification for many real applications. In many regions of the galaxy, these phases appear strongly intermixed. This is due to the fact that the ISM is a highly turbulent medium. Turbulence in the ISM is driven by a number of different physical processes, including thermal instability (Section \ref{sec:TI}), supernova feedback \citep[see e.g.][]{MacLowKlessen2004,klessen16}, and the inflow of gas onto the disk \citep{klessen10,eb10}. It acts to mix what would otherwise be  distinct ISM phases \citep[see e.g.][]{joung09,ssn11}. We discuss the role that turbulence plays in structuring the ISM in \S\ref{sec:turb}.

\begin{table}[th]
\begin{center}
\caption{{Phases of the ISM} \label{ism-phases}}
\vspace{.1in}
\begin{tabular}{r @{\hspace{0.5cm}}  r @{\hspace{0.5cm}} r @{\hspace{0.5cm}} r}
\hline
Component  & Temperature & Density  & Fractional  \\
  &  (K) &  (${\rm cm^{-3}}$) &  ionization \\
\hline
Molecular gas & 10--20 & $> 10^{2}$ & $< 10^{-6}$ \\
Cold neutral medium (CNM) & 50--100 & 20--50 & $\sim 10^{-4}$ \\
Warm neutral medium (WNM) & 6000--10000 & 0.2--0.5 & $\sim$0.1 \\
Warm ionized medium (WIM) & $\sim 8000$ & 0.2--0.5 & 1.0 \\
Hot ionized medium (HIM) & $\sim 10^{6}$ & $\sim 10^{-2}$ & 1.0 \\
\hline\\[-0.2cm]
\end{tabular}
\\
{\footnotesize Adapted from \citet{ferriere01}, \citet{cas98},
\citet{wolfire2003}, and \citet{jenk13}.}
\end{center}
\end{table}

\subsection{Interstellar dust}
\label{sec:dust}
The reddening of starlight in the ISM points towards  an additional component, responsible for absorbing light over a wide range of frequencies. There are distinct features in the extinction curve, for example the 217.5\ {nm} bump that tend to be extremely broad, very different from the narrow lines and structured bands that we expect from atoms or small molecules. Furthermore, abundance measurements show that a number of elements, notably silicon and iron, are strongly depleted in the gas-phase when compared to the solar value. These are typically also the most refractory elements \citet{draine11}. Finally, mid-infrared and far-infrared observations show widespread continuum emission, with a spectrum close to that of a blackbody, and an intensity that correlates well with the hydrogen column density. We conclude that besides the ionized, atomic and molecular constituents of the ISM there is an addition component, which we usually call dust.

We gain further insight into the nature of dust by looking at the spectral shape of the extinction curve that it produces. To first approximation, individual dust grains absorb only those photons with wavelengths smaller than the physical size of the grain. From the fact that we see a large amount of absorption in the ultraviolet, somewhat less in the optical and even less at infrared wavelengths, we can directly infer that there are many more small dust grains than there are large ones. In addition, we often associate particular spectral features in the extinction curve with particular types of dust grain, e.g. graphite in the case of the 217.5\ {nm}~bump \citep{mrn77} and amorphous silicates in the case of the infrared bands at $9.7\ \mu$m and $18\ \mu$m \citep[e.g.][]{dl84,draine07}. 

This can be used to derive more quantitative constraints on the size distribution of interstellar dust grains. For example, \citet{mrn77} reproduced the ISM extinction curve between 0.1--1$\,\mu$m with a mixture of spherical graphite and silicate grains with a size distribution
\begin{equation}
N(a) da \propto a^{-3.5} da\,,
\label{eqn:MRN}
\end{equation}
where $a$ is the grain radius and the distribution extends over a range of radii from $a_{\rm min} \sim 50\;$nm to $a_{\rm max} \sim 0.25\;\mu$m. We note however that actual grains are not spherical, as evidenced by polarization of starlight in extinction and of thermal emission in the submillimetre. Subsequent studies have improved on this simple description \citep[see e.g.][]{dl84,wd01a}, but it remains a useful first approximation to the properties of interstellar dust. Due to the steep distribution of grain sizes, see (\ref{eqn:MRN}), all models predict that the total mass is dominated by the contribution from large grains, while the total surface area is dominated by the contribution made by small grains \cite[for a more detailed discussion, see][]{draine11}.

The total mass in dust grains is difficult to constrain, but if we combine absorption measurements with the observed elemental depletion patterns in the cold ISM, we find that the total mass of metals locked up in grains is roughly the same as the total mass of metals in the gas phase. The dust therefore accounts for around 1\% of the total ISM mass. Therefore, when we attempt to model the thermal and chemical behavior of the ISM dust can play a role that is as important or more important than the gas-phase metals.

\subsection{Chemistry of molecular cloud formation}
\label{sec:chem}

As star formation in the local universe takes place in molecular clouds, we focus our attention now on the chemical processes that lead to the build-up of these clouds. We note that there are two main chemical transitions, occurring at different points in their assembly history, which we can use to identify molecular clouds. The first and most fundamental of these is the transition from atomic to molecular hydrogen. We define a molecular cloud as a region where most of the hydrogen content is in the form of H$_{2}$ rather than H. However, due to the symmetric structure of the H$_{2}$ molecule has no electric dipole transitions but only (forbidden) higher-order ones, and so it does not emit at the  temperatures of a few 10~K to 100~K typical of Galactic clouds. Therefore, it is common to use a different, observationally-motivated definition, which refers to the moment when the cloud becomes visible in CO emission. This requires understanding the chemical transition from C$^{+}$ to C to CO within the assembling cloud, which we discuss in \S\ref{sec:ctoco}). 

\subsubsection{Transition from H to H$_{2}$}
\label{hh2}
The simplest way to form H$_{2}$ in the ISM is via the radiative association of two hydrogen atoms, i.e.\
\begin{equation}
{\rm H + H} \rightarrow {\rm H_{2}} + \gamma.
\end{equation}
However, the rate coefficient for this reaction is extremely small \citep{latter91}, so that very little H$_2$ forms in this way. Ion-neutral reaction pathways
\begin{eqnarray}
{\rm H + e^{-}} & \rightarrow & {\rm H^{-} + \gamma}, \\
{\rm H^{-} + H} & \rightarrow & {\rm H_{2} + e^{-}},
\end{eqnarray}
and
\begin{eqnarray}
{\rm H + H^{+}} & \rightarrow & {\rm H_{2}^{+} + \gamma}, \\
{\rm H_{2}^{+} + H} & \rightarrow & {\rm H_{2} + H^{+}}\;,
\end{eqnarray}
are also highly inefficient. And so it is difficult to produce a H$_{2}$ fractional abundances larger than $f_{\rm H_{2}} \sim 10^{-2}$ with these reactions, even in the most optimal conditions \citep[see e.g.][]{teg97}. Moreover, photodetachment of H$^{-}$ and photodissociation of H$_{2}^{+}$ by photons from the interstellar radiation field render these pathways considerably less effective \citep{glover03}.  Consequently, the  formation of H$_{2}$ in the gas phase is extremely inefficient under typical ISM conditions. 

Formation of H$_2$ in the Milky Way and in essentially all galaxies in the low-redshift Universe therefore follows a very different pathway: molecular hydrogen assembles on the surface of dust grains \citep{gs63}. This is an exothermic reaction with the energy released being absorbed in part by the dust grain and in part being converted to internal energy and kinetic energy of the H$_2$ molecule \cite[see  e.g.][]{bron14}.  Association reactions between adsorbed hydrogen atoms occur readily on grain surfaces; the rate at which H$_2$ forms is then limited primarily by the rate at which H atoms are adsorbed onto the surface. For typical Milky Way conditions, the resulting H$_{2}$ formation rate is approximately \citep{jura75}
\begin{equation}
R_{\rm H_{2}} \sim 3 \times 10^{-17} n n_{\rm H} \: {\rm s^{-1} cm^{-3} }.
\label{eqn:H2-form-rate}
\end{equation}
Here, $n$ is the total number density of  gas particles, while $n_{\rm H}$ is the number density of atomic hydrogen, all given in units of particles per cm$^{3}$. For atomic hydrogen gas, both quantities are identical if we neglect contributions from helium and possibly metals. Note that $n_{\rm H}$ declines as the molecular fraction increases, while $n$ remains the same in the absence of compression or expansion.  The H$_{2}$ formation timescale corresponding to the formation rate (\ref{eqn:H2-form-rate}) is
approximately
\begin{equation}
t_{\rm form} = \frac{n_{\rm H}}{R_{\rm H_{2}}} \sim 10^{9} n^{-1} \: {\rm yr}.
\end{equation}
We point out that at low density this timescale is considerably longer than the dynamical timescale of the system, such as the gravitational free-fall time or the  turbulent crossing time. In order to form molecular clouds within a timescale of several million years as inferred by observations \citep{fukui02, clark14}, we must again invoke interstellar turbulence. The intermittent compression of gas due to supersonic turbulent gas motions can shorten $t_{\rm form}$ considerably \cite[see  e.g.][]{gm07b,micic12}.

Complementary to the processes that lead to H$_2$ formation, we must also consider H$_2$ destruction. Destruction can happen by collision with another atom or molecule in the cloud or by ultraviolet photodissociation. Focusing on the main cloud constituents, collisional dissociation can be formulated as 
\begin{eqnarray}
{\rm H_{2} + H} & \rightarrow & {\rm H + H + H}, \\
{\rm H_{2} + H_{2}} & \rightarrow & {\rm H + H + H_{2}}.
\end{eqnarray}
We note that these reactions do not play an important role in regulating the molecular content of the ISM. They are only effective at destroying H$_{2}$ in warm, dense gas, such as observed in molecular outflows \citep[see e.g.][]{flower03}. Consequently, the key reaction for our purpose here is photodissociation of H$_{2}$, which occurs via a process known as spontaneous radiative dissociation \citep{sw67,vd87}. The H$_{2}$ molecule first absorbs a UV photon with energy $E > 11.2$~eV, placing it in an excited electronic state. The excited H$_{2}$ molecule then undergoes a radiative transition back to the electronic ground state. This transition can occur either into a  bound ro-vibrational level in the ground state, in which case the molecule survives, or into the vibrational continuum, in which case it dissociates. Altogether the  dissociation probability is around 15\% per UV photon absorption \citep{db96}. The decay back to the rovibrational ground state produces a discrete set of UV absorption lines which are known as the Lyman-Werner bands.

Because H$_2$ photodissociation is line-based, rather than continuum-based, the H$_2$ photodissociation rate in the ISM is highly sensitive to self-shielding. Lyman-Werner photons of the ambient radiation field with energies corresponding to the main absorption lines are mostly absorbed by H$_{2}$ on the surface of the molecular cloud, with only a few photons remaining in the regions further in. Consequently, the H$_{2}$ photodissociation rate drops by a large factor compared to the rate in the unshielded, optically thin gas. This becomes important once the H$_{2}$ column density exceeds $N_{\rm H_{2}} \sim 10^{14} \: {\rm cm^{-2}}$ \citep{db96}. The total column density of hydrogen, $N$, depends on the strength of the interstellar radiation field and on the density $n$ of the gas. When we express the radiation field in Habing units $G_{0}$ \citep{Habing1968} we find that self-shielding becomes important for column densities exceeding a value of 
\begin{equation}
N = 10^{20} G_{0} n^{-1} \: {\rm cm^{-2}}\;.
\end{equation}
We note that the visual extinction required to reduce the H$_{2}$ photodissociation rate by a factor of ten is  approximately $A_{\rm V} \approx 0.65$. In the diffuse ISM  this corresponds to a total hydrogen column density $N \sim 10^{21} \: {\rm cm^{-2}}$. Consequently, H$_2$ self-shielding becomes important at lower total column densities than dust shielding in conditions when $G_{0} / n$ is small, such as in the CNM far away from regions of massive star formation. On the other hand, if $G_{0} / n$ is large, e.g. in the photodissociation regions surrounding massive stars  dust extinction typically dominates.

\subsubsection{Transition from C$^{+}$ to C to CO}\label{sec:ctoco}
\label{cco}
The chemistry involved in the transition from C$^{+}$ to C is very simple. Atomic carbon
forms via the radiative recombination of C$^{+}$,
\begin{equation}
{\rm C^{+} + e^{-}} \rightarrow {\rm C + \gamma},
\end{equation}
and is destroyed by photoionization,
\begin{equation}
{\rm C + \gamma} \rightarrow {\rm C^{+} + e^{-}}.
\end{equation}

The formation of CO is considerably more complicated, because there is no single dominant process but instead a number of different pathways compete to build up CO. The two main routes to CO formation involve either hydroxyl (OH) or its positive ion (OH$^{+}$) as chemical intermediates, or they rely on simple hydrocarbons such as CH or CH$_{2}$ and their positive ions. A brief summary of these reaction rates is provided by \citet{klessen16}. For more details and a more comprehensive discussion we refer the readers to the classic papers by \citet{gl75}, \citet{l76}, \citet{db76}, \citet{th85} and \citet{sd95}.

Both main CO-formation pathways share one key feature. The rate-limiting step is the formation of the initial chemical intermediate. In the first case, it is the build-up of OH and the OH$^{+}$ ion. In the second case, it is the formation of CH, CH$_{2}$, CH$^{+}$ or CH$_{2}^{+}$ by radiative association or the formation of H$_{3}^{+}$ as a consequence of the cosmic ray ionization of H$_2$. Once the initial molecular ion or radical forms, the remainder of the reactions that lead to CO proceed relatively quickly. This behavior forms the basis of several simplified methods for treating CO formation \cite[for a comparison of different approaches, see][]{GloverClark2012b}. In addition, all of the different pathways  to proceed from C$^{+}$ or C to CO  rely on the presence of molecular hydrogen. It implies that substantial quantities of CO will form only in regions that already have high H$_{2}$ fractions. As a consequence, the non-equilibrium, time-dependent behavior of H$_2$ chemistry can be important also for CO, despite the fact that the characteristic timescales of the chemical reactions involved in CO formation are generally shorter than the H$_{2}$ formation time. 

In the surface layers of molecular clouds, i.e.\ regions with a low visual extinction, the destruction of CO is dominated by photodissociation,
\begin{equation}
{\rm CO} + \gamma \rightarrow {\rm C + O}.
\end{equation}
The molecule first absorbs a UV photon with energy $E > 11.09 \: {\rm eV}$, placing it in an excited electronic state \citep{vd87}. From here, it can either return to the ground state via radiative decay, or it can undergo a transition to a repulsive electronic state via a radiationless process. In the latter case, the molecule very rapidly dissociates. In the case of CO, dissociation is typically far more likely than decay back to the ground state \citep{vdb88,visser09}. Consequently, the lifetimes of the excited electronic states are very short. This is important, as Heisenberg's uncertainty principle then implies that their energy is comparatively uncertain. The UV absorption lines associated with the photodissociation of CO are therefore much broader than the lines associated with H$_{2}$ photodissociation. As a result, more photons need to be absorbed to prevent molecules further into the cloud from being photodissociated, and consequently CO self-shielding is less effective than the analogous process for H$_{2}$ discussed above.

Once the visual extinction $A_{\rm V}$ at the cloud surface due to dust absorption and H$_2$ shielding becomes large, very few photons remain in the cloud interior to photodissociate CO, and so two other processes become important in balancing CO formation. First, cosmic ray ionization of hydrogen molecules or hydrogen atoms produces energetic photo-electrons (see Section \ref{sec:CR}). If these collide with other hydrogen molecules before dissipating their energy, they can excite the H$_{2}$ molecules into excited electronic states. The subsequent radiative decay of the molecules back to the ground state produces UV photons that can initiate localized photodissociation of CO and other molecules \citep{pt83,gld87,gredel89}.  Second, CO is also destroyed via dissociative charge transfer with He$^{+}$ ions 
\begin{equation}
{\rm CO + He^{+}} \rightarrow {\rm C^{+} + O + He}.
\end{equation}
The He$^{+}$ ions required by this reaction are again produced by cosmic ray ionization of neutral helium. Altogether, CO destruction in high $A_{\rm V}$ regions is controlled by the cosmic ray ionization rate. This is relatively small \citep{vv00} in typical clouds, and so almost all of the carbon is found in the form of CO. The situation changes in regions of high cosmic ray flux, such as inferred for the Central Molecular Zone of the Galaxy, where the CO fraction can be significantly suppressed even in well-shielded gas \citep[see e.g.][]{clark13}.

\subsection{Cooling processes}
\label{sec:cooling}

In this section we summarize the key heating and cooling processes that determine the thermal evolution of the ISM. Most cooling processes are based on the fact that collisions can excite internal degrees of freedom of the atom, molecule, ion or dust grain under consideration, therefore removing kinetic energy from the system, and that there are de-excitation pathways that involve the emission of photons. If these photons are absorbed nearby i.e. there are optically thick conditions, the energy remains in the system and there is no net cooling effect. If the photons leave the system the medium is optically thin, and their energy is carried away and the system is cooler. We note that also adiabatic expansion can lead to a decrease of the kinetic temperature of the medium.  Most heating processes are based on the inverse of these processes. However, we note that reality can be more complicated due to the fact that some chemical phase changes involve latent heat, which needs to be included in the energy budget. There are other potential heating and cooling mechanisms that are not based on collisional or radiative coupling or on cloud dynamics but instead may involve  interaction with a magnetic field to exchange energy, such as heating by magnetic reconnection. 

For the microphysical treatment of the collisional and radiative coupling of multi-level atoms and molecules, we refer to the relevant textbooks, e.g.\ to  \citet{rybicki86}, \citet{osterbrock89}, \citet{tielens2010}, and \citet{draine11}, or to a recent compilation provided by \citet{klessen16}.

\subsubsection{Cooling by permitted transitions}
At high temperatures in regions dominated by atomic or ionized gas, the cooling of the ISM takes place largely via permitted (i.e.\ dipole-allowed) electronic transitions of various atoms and ions. At temperatures close to $10^{4}$~K, excitation of the Lyman series lines of atomic hydrogen is the dominant process leading to a cooling rate 
per unit volume \citep{black81,cen92} of
\begin{equation}
\Lambda_{\rm H} = 7.5 \times 10^{-19} \mathrm{erg}\;\! \mathrm{s}^{-1} \mathrm{cm}^{-3}\,\frac{1}{1 + (T / 10^{5})^{1/2}} \exp \left(\frac{-118348}{T} \right)
n_{\rm e} n_{\rm H}\;,
\end{equation}
where $n_{\rm e}$ and $n_{\rm H}$ are the number densities of free electrons and atomic hydrogen given in units of cm$^{-3}$, respectively. At temperatures $T \sim 3 \times 10^{4}$~K and above, however, the abundance of atomic hydrogen generally becomes very small, and other elements, particularly C, O, Ne and Fe, start to dominate the cooling \citep[see e.g.][]{GnatFerland2012}. 

We note that there are many cases in which the  assumption of collisional ionization equilibrium does  not apply. For example, consider gas in the H{\textsc{ii}} regions around massive stars, where the ionization state of the gas is determined primarily by photoionization rather than collisional ionization. The  assumption of collisional ionization equlibrium also breaks down whenever the gas cools rapidly, i.e.\ when the cooling time becomes shorter than the recombination time, or if gas is heated more rapidly than it is collisionally ionized, such as in a very strong shock.  There are many efforts to account for non-equilibrium effects, either by explicitly solving for the non-equilibrium ionization state  \citep[see e.g.][]{cf06,db12,os13,rso14} or with an ionization state dominated by photoionization rather than collisional ionization \citep[e.g.][]{wiersma09,gh12}. 

\subsubsection{Cooling by fine structure lines}
\label{subsubsec:fine-structure-lines}
At temperatures far below $10^{4}$~K, it becomes extremely difficult for the gas to cool via radiation from permitted atomic transitions, as the number of electrons available with sufficient energy to excite these transitions declines exponentially with decreasing temperature. In this regime forbidden transitions between different fine structure energy levels become important. 

Fine structure splitting is caused by the coupling between the orbital and spin angular momenta of the electrons in an atom \citep[see e.g.][]{atkins11}. Each electron within an atom has a magnetic moment due to its orbital motion and also an intrinsic magnetic moment due to its spin. States where these magnetic moments are parallel have higher energy than states where they are anti-parallel. In order for an atom or ion to display fine structure splitting in its ground state, the electrons in the outermost shell must have both non-zero total orbital angular momentum (i.e.\ $L > 0$) and non-zero total spin angular momentum (i.e.\ $S > 0$), or else the spin-orbit coupling term in the Hamiltonian, which is proportional to $\vec{L} \cdot \vec{S}$, will vanish. The resulting splitting of the energy levels is small, with energy separations of the order of $10^{-2}$~eV. This corresponds to a temperature of the order of 100~K, meaning that it is possible to excite these transitions even at relatively low gas temperatures. 

The corresponding quantum mechanical transition matrix elements are of the order of $\alpha^{2} \approx 5 \times 10^{-5}$ times smaller than for electric dipole transitions, where $\alpha$ is the fine structure constant. Furthermore, the spontaneous transition rates  scale as  $A_{ij} \propto \nu_{ij}^{3}$. Since associated frequencies are about a thousand times smaller than those of the most important permitted electronic transitions, such as Lyman-$\alpha$, one expects the spontaneous transition rates to be a factor of $10^{9}$ smaller. These two effects combined result in transition rates  that are of order of $10^{14}$ times smaller than those of the permitted atomic transitions. Consequently, the critical densities associated with many of the important fine structure lines are relatively low: $n_{\rm crit} \sim 10^{2}$--$10^{6} \: {\rm cm^{-3}}$ in conditions when collisions with H or H$_{2}$ dominate,  and up to two to three orders of magnitude smaller when collisions with electrons dominate \citep{hm89}. Therefore fine structure emission is effective in the WNM or CNM but not so important at the much higher densities found in gravitationally collapsing regions within molecular clouds.

Since hydrogen and helium have no fine structure in their ground states, fine structure cooling in the diffuse ISM is dominated by the contribution from the next most abundant elements: carbon and oxygen in their atomic and low ionization states \citep{wolf95}. Data on the collisional excitation rates of the fine structure transitions of C$^{+}$, C and O can be found in a number of places in the literature. Compilations of excitation rate data are given in \citet{hm89}, \citet{gj07} and \citet{maio07}, as well as in the LAMDA database \citep{sch05}.

\subsubsection{Carbon monoxide} 
\label{subsubsec:CO}
In order to study cold and dense molecular clouds, we need to resort to low-energy rotational and vibrational transitions of molecular species. Molecular hydrogen is a symmetric molecule and therefore has no permanent dipole moment. The next order transition based on the quadrupole moment requires higher temperatures than are usually found in Galactic molecular clouds and are typically not observed. Luckily, the second most abundant molecular species, CO, has a significant dipole moment. It is readily rotationally excited even at very low gas temperatures, $T < 20$~K, owing to very small energy separations between its excited rotational levels. It therefore plays a key role in the thermal balance of cold, dense clouds.

We note that at low densities, fine structure cooling from neutral atomic carbon is more effective than CO cooling, and that at $T \sim 20$~K and above the contribution from C$^{+}$ also becomes significant. The overall importance of CO therefore depends strongly on the chemical state of the gas. If the gas-phase carbon is primarily in the form of C or C$^{+}$, then fine structure emission from these species will dominate, implying that CO becomes important only once the fraction of carbon in CO becomes large. Consequently, CO cooling only dominates once the gas density exceeds $n \sim 1000 \: {\rm cm^{-3}}$. Furthermore, the relevant density range is quite limited. For $n \gg 1000 \: {\rm cm^{-3}}$ the relative populations of the lowest rotational levels quickly reach their LTE levels. Furthermore, the lines of the most abundant species, $^{12}$CO become optically thick and the molecule tends to freeze out on dust grains and its gas phase abundance drops significantly \citep{Goldsmith01}.

\subsubsection{Gas-grain energy transfer}
\label{subsubsec:gas-grain-transfer}
At high temperatures dust can also play an important role in the cooling of the ISM \citep{gk74,l75}.
Individual dust grains are extremely efficient radiators, and so the mean temperature of a population of dust grains very quickly relaxes to an equilibrium value given by the balance between radiative heating caused by  absorption of  photons from the interstellar radiation field (ISRF) and radiative cooling via thermal emission from the grains. If the resulting dust temperature, $T_{\rm d}$, differs from the gas temperature, $T_{\rm K}$, then collisions between gas particles and dust grains  lead to a net flow of energy from one component to the other, potentially changing both $T_{\rm K}$ and $T_{\rm d}$. 

We can write the cooling rate per unit volume due to energy transfer from the gas to the dust as
\begin{equation}
\Lambda_{\rm gd} = \pi \sigma_{\rm d} \bar{v} \bar{\alpha} (2kT_{\rm K} - 2kT_{\rm d}) n_{\rm tot} n_{\rm d}.
\end{equation}
Here, $\sigma_{\rm d}$ is the mean cross-sectional area of a dust grain, $n_{\rm d}$ is the number density of dust grains, $n_{\rm tot}$ is the number density of particles, and $\bar{v}$ is the mean thermal velocity of the particles in the gas. Although it is common to discuss this in terms of cooling, if $T_{\rm d} > T_{\rm K}$ then energy will flow from the dust to the gas, i.e.\ this will become a heating rate. $\Lambda_{\rm gd}$ is often expressed in the form
\begin{equation}
\Lambda_{\rm gd} = C_{\rm gd} T_{\rm K}^{1/2} \left( T_{\rm K} - T_{\rm d} \right) n^{2}
\: {\rm erg \: s^{-1} \: cm^{-3}},
\end{equation}
where $n$ is the number density of hydrogen nuclei and $C_{\rm gd}$ is a cooling rate coefficient  given by
\begin{equation}
C_{\rm gd} = 2 \pi k \sigma_{\rm d} \left(\frac{\bar{v}}{T_{\rm K}^{1/2}} \right) \bar{\alpha} \frac{n_{\rm tot} n_{\rm d}}{n^{2}}. 
\end{equation}
The value of $C_{\rm gd}$ is largely determined by the assumptions that we make regarding the  chemical state of the gas and the nature of the dust grain population, but in principle it also depends on temperature through the temperature dependence of the mean accommodation coefficient, $\bar{\alpha}$, which quantifies how efficiently energy is shared between dust and gas.  There is some debate in the literature concerning the exact functional form of $C_{\rm gd}$ and its normalization \cite[for example, compare the discussions in][]{hm89,th85,Goldsmith01,klm11}; however, most values for typical Milky Way conditions are in the range of $C_{\rm gd} = 10^{-33} \: {\rm erg \: s^{-1} \: cm^{3} \: K^{-3/2}}$ with deviations of about one order of magnitude. The uncertainty in $C_{\rm gd}$ becomes even greater as we move to lower metallicity, as less is known about the properties of the dust. It is often assumed that  $C_{\rm gd}$ scales linearly with metallicity \citep[e.g.][]{gc12c}, but this is at best a crude approximation, particularly as the dust abundance appears to scale non-linearly with metallicity in metal-poor galaxies \citep{gala11,hc12}.

\subsection{Heating processes}
\label{sec:heating}

\subsubsection{Photoelectric heating}
\label{subsec:photoelectric-heating}
One of the most important forms of radiative heating in the diffuse ISM is photoelectric heating. It is caused by the absorption of UV photons by dust grains which subsequently emit photo-electrons. Their kinetic energy is equal to the difference between the energy of the photon and the energy  barrier for detaching the electron from the grain. This difference can be large (of the order of an eV or more), and the released energy is efficiently redistributed amongst the other gas particles by collisions, causing the gas to heat up. 

For standard interstellar dust, following \citet{bt94} the photoelectric heating rate  per unit volume can be written as
\begin{equation}
\Gamma_{\rm pe} = 1.3 \times 10^{-24} \epsilon G_{0} n  \: {\rm erg} \: {\rm s^{-1}} \: {\rm cm^{-3}},
\end{equation}
where $\epsilon$ is the photoelectric heating efficiency, given by
\begin{equation}
\epsilon = \frac{0.049}{1 + (\psi / 1925)^{0.73}} + \frac{0.037 (T / 10000)^{0.7}}{1 + (\psi / 5000)},
\end{equation}
 $G_{0}$ is the strength of the interstellar radiation field in units of the \citet{Habing1968} field, and $T$ is the gas temperature.  The parameter 
\begin{equation}
\psi \equiv \frac{G_{0} T^{1/2}}{n_{\rm e}}\;
\end{equation}
is related to the charge of the dust grains in the ISM  \citep{ds87, wd01b}. The interpretation is as follows: a  high photon flux or a low numbers of of free electrons will lead to dust grains being more positively charged, while the converse will lead to grains being more negatively charged. In the latter case electrons can more easily detach and the heating rate is high.  The proportionality $T^{1/2}$ simply reflects the temperature dependence of  the rate coefficient for electron recombination with the grains. For  small $\psi$, $\epsilon \approx 0.05$ when the temperature is low, and $\epsilon \approx 0.09$ when the temperature is high. Since the photons required to eject electrons must be energetic, with minimum energies typically around 6~eV, the photoelectric heating rate is highly sensitive to the dust extinction. It becomes ineffective once the visual extinction exceeds values of $A_{V} \sim 1$--2~mag.

\subsubsection{Heating by ultraviolet radiation}
Besides the photoelectric effect, UV photons heat the ISM in two other important ways. First, the photodissociation of H$_{2}$ results in a transfer of energy to the gas. The absorption process produces a molecule in an excited state, which can break apart and convert the excess energy into kinetic energy through non-radiative decay. As the hydrogen atoms produced in this process have average kinetic energies that are larger than that of the gas particles. The energy release varies depending on which rovibrational level of the excited electronic state was involved in the dissociation \citep{sd73,abg00}. The average heating rate is around 0.4~eV per dissociation \citep{bd77}.  Second, UV irradiation of molecular hydrogen can lead to heating via a process known as UV pumping. The absorption of a UV photon by H$_{2}$ leads to photodissociation only around 15\% of the time \citep{db96}. The rest of the time, the H$_{2}$ molecule decays back into a bound rovibrational state. Although the molecule sometimes goes back directly into the $v=0$ vibrational ground state, it is far more likely to end up in a vibrationally excited level. In low density gas, it then radiatively decays back to the rovibrational ground state, producing a number of near infrared photons in the process which do not contribute to the overall heating. In high density gas, collisional de-excitation occurs more rapidly, and so most of the excitation energy is converted into heat. In this case, the resulting heating rate is around 2~eV per pumping event, compared to around 10--11~eV per photodissociation \citep[see e.g.][]{bht90}. This process becomes significant only above a critical density of $n_{\rm crit} \sim 10^{4} \: {\rm cm^{-3}}$. It is thus not a major heat source at typical molecular cloud densities but can become important in dense cores exposed to strong UV radiation fields.

\subsubsection{Cosmic rays}
\label{sec:CR}

In the deep interior of molecular clouds, where the gas is well shielded from the interstellar radiation field, both of the above processes become unimportant and the same holds for photoelectric heating. In this case, cosmic rays provide the main source of heat. They can penetrate deeply into molecular clouds and collisionally ionize hydrogen or helium atoms or H$_{2}$ molecules. The resulting ions and electrons typically have very large velocities and so subsequent collisions can trigger secondary ionization events. Eventually the excess kinetic energy gets converted into heat, with the amount of energy transferred depending on the composition of the gas \citep{dyl99,ggp12}, but it is typically around 10--20~eV. Most models of thermal balance in dark clouds adopt a heating rate that is a simply fixed multiple of the primary hydrogen cosmic ray ionization rate \citep[see e.g.][]{goldsmith78,Goldsmith01,g10,klm11}. A commonly adopted parameterization is
\begin{equation}
\Gamma_{\rm cr} \sim 3.2 \times 10^{-28} (\zeta_{\rm H} / 10^{-17} \: {\rm s^{-1}}) \,n \:\:\: {\rm erg} \: {\rm cm^{-3} s^{-1}}\;,
\label{eqn:CR-rate}
\end{equation}
where the cosmic ray ionization rate of atomic hydrogen $\zeta_{\rm H}$ is scaled by its typical value of $10^{-17}\,$s$^{-1}$ and $n$ is the number density of hydrogen nuclei. Note that the uncertainty introduced by averaging is typically much smaller than the current uncertainty in the actual cosmic ray ionization rate in the considered region. We note that in most applications the cosmic ray ionization and heating rates are assumed to be constant throughout the cloud. A better approach is 
\begin{equation}\label{eqn:CR-heatingrate}
\Gamma_{\rm cr}=nQ(n)\zeta(n)\,,
\end{equation} 
with density dependent heating rate $Q(n)$ and cosmic ray flux $\zeta(n)$. Whereas $Q$ only slightly depends on density $n$ \citep{ggp12}, $\zeta$ decreases by orders of magnitude in the densest parts of a cloud \citep[see e.g.][]{Padovani2009,Padovani2018}. The use of equation (\ref{eqn:CR-heatingrate}) leads to more accurate values of $\Gamma_{\rm cr}$ up to a factor of $\sim7$.

\subsubsection{X-rays}

X-rays can heat the interstellar medium in a very similar fashion to cosmic rays:  X-ray ionization produces an energetic electron that can cause a cascade of secondary ionization events with some fraction of the excess energy going into
heat. Unlike cosmic rays, X-rays are sensitive to the effects of absorption, because their mean free path is much smaller. Hence, they can be an important heat source in the diffuse ISM \citep[see e.g.][]{wolf95}, but they are generally thought to be unimportant in the dense gas inside molecular clouds, unless these clouds are located close to a strong X-ray source such as an AGN \citep[see e.g.][]{hs10}.

\subsubsection{Chemical reactions}

The latent heat associated with chemical reactions can also contribute to the overall thermal energy balance in certain phases of the ISM. For example, the formation of a new chemical bond, such as that between the two hydrogen nuclei in a H$_{2}$ molecule, releases energy. Much of this energy will be channeled into rotational and/or vibrational modes of the newly-formed molecule. In low density environments, it will rapidly be lost by radiation. At high densities, however, collisional de-excitation can convert this energy into heat before it can be lost via radiation. 

Some of the energy released in a reaction may also go into motion of the newly-formed molecule, and this will also be rapidly converted into heat via collisions. Many of the reactions occurring in interstellar gas lead to heating in this way, but for the most part, their effects are unimportant. The one case in which this process can become significant is the formation of H$_{2}$ at high densities. Each event releases a total energy of 4.48~eV. With the typical H$_2$ formation rates given by (\ref{eqn:H2-form-rate}) we obtain 
\begin{equation}
\Gamma_{\rm H_{2} form} \sim 2 \times 10^{-28} \epsilon_{\rm H_{2}}  n n_{\rm H} \: {\rm erg} \: {\rm cm^{-3} s^{-1}}
\end{equation}
for the corresponding heating rate, assuming that the efficiency with which this energy is converted into heat is $\epsilon_{\rm H_{2}}$. Comparing with the cosmic ray heating rate (\ref{eqn:CR-rate}), we see that H$_{2}$ formation heating will dominate whenever $\epsilon_{\rm H_{2}}  n_{\rm H} > (\zeta_{\rm H} / 10^{-17} \: {\rm s^{-1}})$. In principle, H$_{2}$ formation heating can be important provided that the efficiency factor $\epsilon_{\rm H_{2}}$ is not too small. Unfortunately, the value of $\epsilon_{\rm H_{2}}$ is highly uncertain \citep[see e.g.][and references therein]{leb12,roser03,con09}, and more work is required to constrain it.

\begin{figure}
\includegraphics[width=\textwidth]{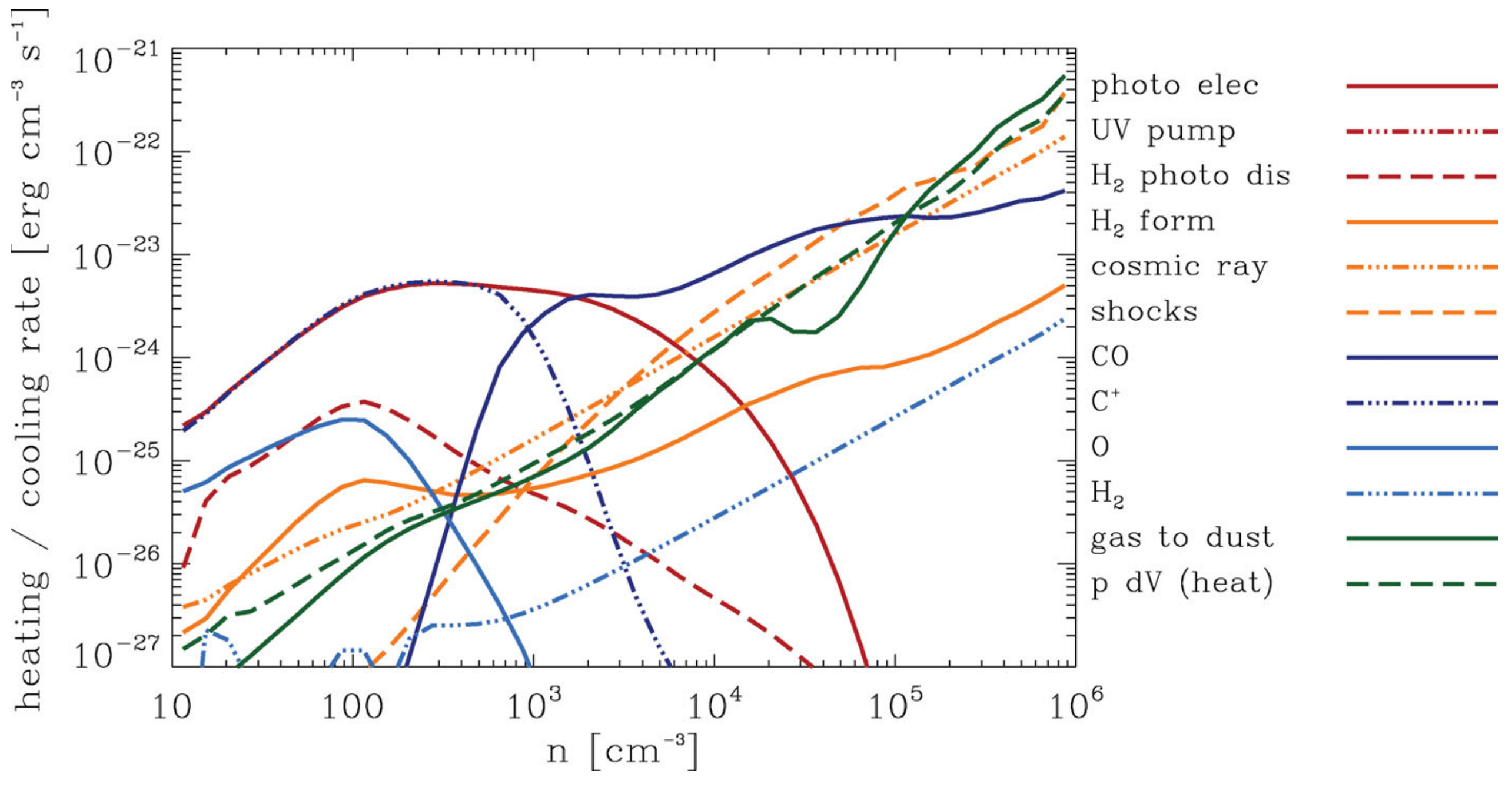}
\caption{Illustration of the relative importance of the various heating and cooling processes in the ISM, plotted as a function of the hydrogen nuclei number density, $n$, calculated from a simulation of molecular cloud formation from initially atomic gas in the solar neighborhood. Heating processes are depicted in red and orange, cooling processes in dark and light blue, processes that in principle can either add or remove heat from the gas component of the ISM are shown in green. Figure adapted from \citet{GloverClark2012a}
\label{fig:heating-cooling}}
\end{figure}

\subsubsection{Dynamical processes}

Finally, hydrodynamical and magneto-hydrodynamical (MHD) effects can also lead to significant heating. In subsonic, gravitationally collapsing regions, such as low-mass prestellar cores, adiabatic compression ($PdV$ heating) can be a major source of heat and can actually be more important in the overall thermal balance of the core than cosmic ray heating. In less quiescent environments, where the gas flow is supersonic, turbulent dissipation in shocks or regions of strong shear is another major heat source. The same is true for magnetic reconnection and other non-ideal MHD processes in magnetized media \cite[e.g.][]{momferratos14}. The rate at which turbulent kinetic energy dissipates in regions of supersonic turbulence is reasonably well established \citep{maclow98, Stone98,maclow99}. The energy dissipation rate within a cloud of mass $M$  can be written to within a factor of order unity as \citep{maclow99}
\begin{equation}
\dot{E}_{\rm kin} \sim - M k_{\rm d} \sigma_{v}^{3},
\end{equation}
where $k_{\rm d}$ is the wavenumber on which energy is injected into the system and where $\sigma_{v}$ is the velocity dispersion at this scale. We assume that it is comparable to the size of the cloud \citep[see e.g.][]{brunt09}. Furthermore, we adopt Larson's relations between the size of the cloud and its velocity dispersion, $\sigma_v \propto L^\alpha$ with $\alpha \sim 0.5$, and number density, $n \propto L^\beta$ with $\beta \sim 1$. We note that normalization and slope are both quite uncertain  \citep{larson81,HennebelleFalgarone2012}. Put together, we arrive at an average turbulent heating rate \citep{pp09}
\begin{equation}
\Gamma_{\rm turb} = 3 \times 10^{-27} \left(\frac{L}{1 \: {\rm pc}} \right)^{0.2} n \: {\rm erg \: s^{-1} \: cm^{-3}}.
\end{equation}
While dominating on large scales, this heating rate can become comparable to the cosmic ray heating rate on small scales, in more quiescent regions deeply embedded regions of the cloud such as low-mass protostellar cores.  We also note that turbulent heating is highly intermittent \citep{pp09}. This means that in much of the cloud, the influence of turbulent dissipation is small, while in small, localized regions, very high heating rates can be produced \citep[see e.g.][]{fal95,god09}. We provide a more detailed account of properties of turbulence in \S\ref{sec:turb}. 

Finally, note that the physical nature of the heating depends upon the strength of the magnetic field within the gas. If the field is weak, energy dissipation occurs mostly through shocks, whereas if the field is strong, a substantial amount of energy is dissipated via non-ideal MHD processes such as ambipolar diffusion, i.e.\ the drift between the neutral and charged constituents of the ISM \citep{pzn00,lmm12}.

\subsubsection{Heating and cooling in the nearby ISM}

Figure \ref{fig:heating-cooling} provides an overview of the most important heating and cooling processes for the solar neighborhood ISM.  The rates are plotted as a function of the hydrogen nuclei number density, $n$. The figure shows that initially atomic gas exhibits three different regimes. At densities $n < 2000\,$cm$^{-3}$, the gas heating is dominated by photoelectric emission from dust grains (\S\ref{subsec:photoelectric-heating}), while cooling is provided by fine structure emission from C$^+$. In the density regime $2000 < n < 10^5\;$cm$^{-3}$, rotational line emission from CO becomes the main coolant. Photoelectric heating remains the main heat source initially but steadily becomes less effective, owing to the larger visual extinction of the cloud at these densities. Other processes -- adiabatic compression of the gas, dissipation of turbulent kinetic energy in shocks and cosmic ray ionization heating -- become more important at $n \sim 6000 \: {\rm cm^{-3}}$ and above. Finally, at densities above about $10^5\;$cm$^{-3}$, the gas couples to the dust (Section \ref{subsubsec:gas-grain-transfer}), which acts as a thermostat and provides most of the cooling power. Weak shocks and adiabatic compressions together dominate the gas heating in this regime, each contributing close to half of the total heating rate \cite[for a more detailed discussion, see][]{GloverClark2012a}.

\section{Thermal structure of the ISM}
\label{sec:thermal-structure}

To understand the ISM it is crucial to know its temperature distribution. As mentioned in the previous section, the ISM consists of two atomic phases, namely the cold neutral medium and the warm neutral medium. The latter has a typical density of about 0.5 cm$^{-3}$ and a temperature of the order of 8000\,K. The former is approximately 100 times denser and cooler. We first describe in short the cooling and heating processes that are relevant for the atomic gas, and we then discuss the principle of thermal instability.

\subsection{Thermal balance}
\label{Sec:Thermal}

\begin{figure}
    \centering
    \includegraphics[width=0.8\textwidth]{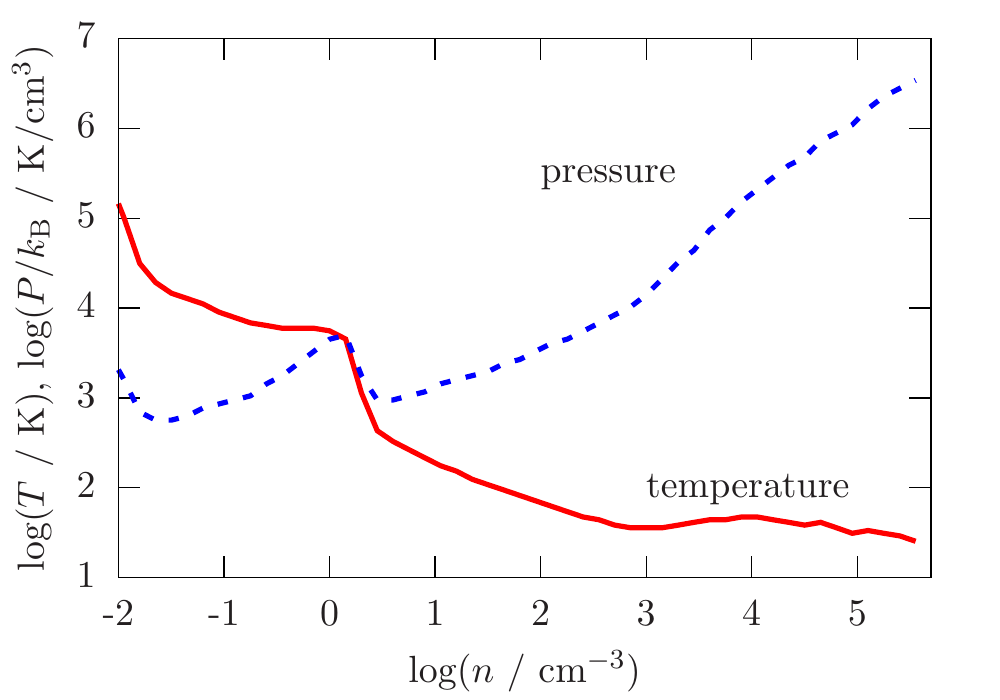}
    \caption{Phase diagram of the ISM including the most relavant heating and cooling processes. Computed using the heating and cooling processes described in \citet{GloverClark2012b} implemented in the SILCC ISM simulations \citep{WalchEtAl2015, GirichidisEtAl2016b, GirichidisEtAl2018b}.} 
    \label{fig:equilibrium}
\end{figure}

The balance of cooling and heating must be computed to provide the temperature as a function of density. The detailed analysis for the thermal equilibrium state in the neutral atomic phase can be found, for example, in \citet{wolfire2003}. The dominant heating mechanisms are photoelectric emission from small grains and PAHs, ionization by cosmic rays and soft X-rays, and the formation and photodissociation of H$_2$. The FUV field is close to the Habing's value, $G_0=1.7$. The most important cooling processes are line emission from H, C, O, Si, and Fe, and rovibrational lines from H$_2$ and CO. Collisions of atomic and molecules with dust grains and the resulting thermal emission from the collisionally heated dust also contribute to cooling.

At high density, typically $10^3 \, {\rm cm^{-3}}<n<10^6 \, {\rm cm^{-3}}$, the cooling depends on the column density and UV radiation field and molecular cooling dominates. How well the radiation field is shielded, in relative importance of H$_2$ self-shielding and dust shielding is duscussed in \citet{sternberg2014}. The chemistry and cooling for a broad range of density, column density, metallicity, and radiation fields are discussed by \citet{koyama2000}, \citet{GloverClark2012a} and \citet{gong2017}.

Figure \ref{fig:equilibrium} displays the phase diagram, i.e. temperature and pressure as a function of density. The interstellar radiation field in this analysis is 1.7 times the Habing field ($G_0=1.7$, \citealt{Habing1968}) and is attenuated in regions of high optical depth \citep{WalchEtAl2015,GirichidisEtAl2016b}. The  At high column density, the main cooling process is due to CO molecules up to about  10$^5$ cm$^{-3}$, above which dust cooling dominates. Under these conditions, the most important heating mechanism is usually cosmic rays.

\subsection{Thermal Instability}
\label{sec:TI}

\begin{figure}
  \centering
  \includegraphics[width=0.7\textwidth]{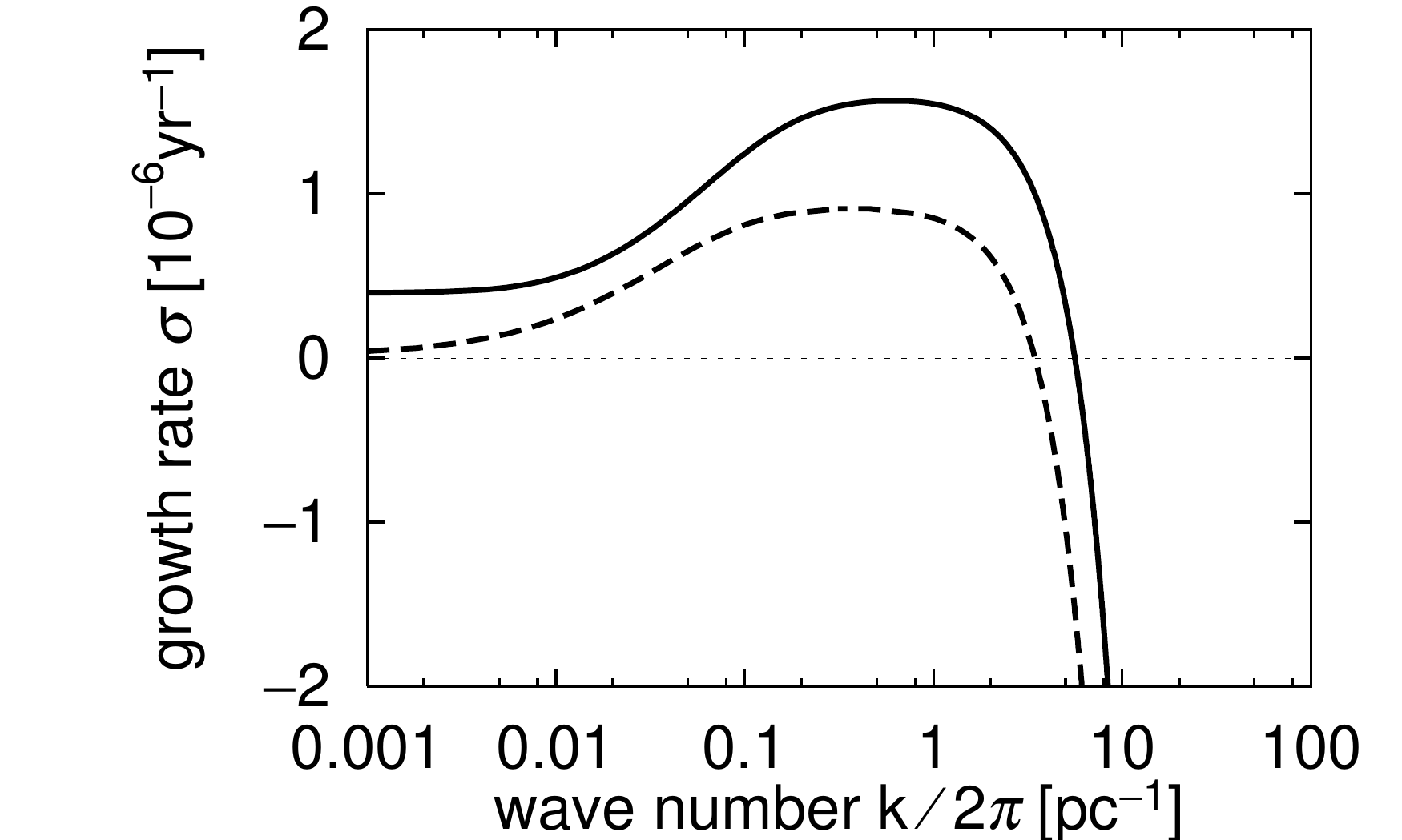}
  \caption{ The dispersion relation for the condensation mode
            of thermal instability analyzed in \citet{koyama2000} for a temperature of $T=10^3\,\mathrm{K}$ and a pressure of $P/k_\mathrm{B}\approx2000\,\mathrm{K\,cm}^{-3}$.
            The dashed curve denotes the result
            for a system at rest in thermal equilibrium.
            The solid curve denotes the case of an isobarically
            contracting core \citep{koyama2000}, \textcopyright AAS. Reproduced with permission.}
            \label{Fig:DR}
\end{figure}

Thermal instability is believed to play an important role in the ISM. It stems from the fact that cooling is proportional to the square of the gas density (because cooling elements get excited through two-body collisions) while the heating function in some range of temperatures is approximately proportional to the density. Quantitatively the existence of the thermal instability can be related to the slope of heat-loss function, $\mathcal{L}=\rho \Lambda-\Gamma$, where $\rho \Lambda$ is the cooling function per unit volume with the density $\rho$ and the cooling rate $\Lambda$ in units of erg\,cm$^{3}$\,s$^{-1}$, and $\Gamma$ is the heating function in units of erg\,g\,s$^{-1}$. The stability conditions of a uniform medium subject to heating and cooling have been studied in great detail by \citet{Field1965}, who investigated the impact of isochoric, isentropic and isobaric perturbations on the thermal equilibrium. In the ISM the unstable isobaric perturbation is believed to be the most relevant one and can be expressed as
\begin{equation}
       \left( \PD{\mathcal{L}}{T} \right)_{\rm P} < 0 \Leftrightarrow
\left( { \partial P \over \partial \rho} \right) _{\mathcal{L}} < 0.
\end{equation}
The physical meaning of this expression is straightforward. Consider thermal equilibrium, i.e. $\mathcal{L} =0$. If the pressure decreases as the density increases the support against a further increase in density reduces. This thermal state is unstable because any small over-density corresponds to a local pressure minimum and will therefore be amplified. Typically this instability appears for in the temperature regime between $\sim$ 100 and $\sim$ 5000\,K. Figure \ref{Fig:DR} displays the growth rate of thermal instability as a function of wave number. The dashed curve shows the dispersion relation in at thermal equilibrium for a system with uniform density at rest. The solid line indicates the counterpart for an isobarically contracting sphere. Thermal conduction is also taken into account in this calculation, which is why the growth rate vanishes at large wave numbers. As discussed by \citet{Field1965}, thermal conduction can counteract thermal instability because it smooths out small scale temperature perturbations. The critical wavelength, i.e. the one at which the growth rate becomes zero, is now called the ``Field length'' and is given by
\begin{equation}
   \lambda_{\rm F}\equiv\sqrt{\frac{KT}{\rho^2 \Lambda}}.
\end{equation}
Here $K$ denotes the coefficient of thermal conduction. The peak of the growth rate occurs at a wavelength that is several times larger than the Field length. Therefore, thermal instability leads to structures larger than the Field length.

The effect of magnetic fields on the linear growth of thermal instability was studied in detail by \citet{ames1973}. A sufficiently strong magnetic field suppresses motions perpendicular to the field lines assuming flux freezing. Consider for instance a slab geometry, in which the magnetic pressure is simply proportional to the density squared (since the magnetic field is proportional to the density) and therefore an increase in the magnetic pressure can compensate for a decrease in thermal pressure. However, perturbations in the direction along the magnetic field are not suppressed and remain unstable if the cooling function satisfies the instability criteria.

\subsection{Thermal front propagation and pressure regulation}

The non-linear development of thermal instability has been studied by various authors \citep[e.g.][]{henne1999,koyama2000,piontek2004,inoue2007,vanloo2007,choi2012}, see also \citet{ElphickRegevShaviv1992} and \citet{ShavivRegev1994} for the dynamics of fronts in thermally bistable media. Typically, after the linear phase, a non-linear structure develops and eventually a cloud of dense gas embedded in the warm surrounding medium settles into an equilibrium. The dense cloud is then connected to the WNM through thermal fronts, whose thicknesses are given by the Field length \citep{Field1965, BegelmanMcKee1990, StoneZweibel2010, KimKim2013}. Note that the Field length varies from about 10$^{-3}$~pc in the CNM to 0.1~pc in the WNM and the fronts have a thickness that can be approximated by the geometric mean of the Field length in the cold and warm medium \citep[e.g.][]{KimKim2013}. The fronts themselves are not in thermal equilibrium, since the denser part is cooling while the more diffuse one is heating. Thus, except for a particular value of the pressure for which the two contributions are equal (also called saturation pressure), it is generally the case that either the cloud evaporates or condenses. If the ambient pressure is lower than the saturation pressure the thermal front becomes an evaporation front. Conversely, if the ambient pressure is larger than the saturation pressure the front leads to a condensation \citep[e.g.][]{InoueInutsukyKoyama2006}.

\section{Turbulence}
\label{sec:turb}
\subsection{Introduction}
Fluid flows are often divided into two sharply different categories: quiet smooth flows known as {\em laminar}, and {\em turbulent} flows in which the fluid velocity exhibits chaotic fluctuations in both space and time over a wide range of length and time scales. The turbulent velocity field is unpredictable because small variations in the initial conditions produce large changes to the subsequent motion. This irregular state of motion is a truly remarkable feature of the governing Navier-Stokes (NS) equation, describing the rate of change of velocity in a viscous fluid,
\begin{equation}
\partial_t\bm u+(\bm u\bm\cdot\bm\nabla)\bm u=-\bm\nabla p/\rho+\nu\bm\nabla^2\bm u+\bm f/\rho,
\label{ns}
\end{equation}
under suitable initial and boundary conditions. This equation 
is simply Newton's law for a fluid, but it includes a seemingly innocent nonlinear inertial term, $(\bm u\bm\cdot\bm\nabla)\bm u$, where the fluid velocity, $\bm u(\bm x,t)$, appears in a quadratic form. It is this nonlinearity in the deterministic NS equation that brings about a source of chaos with it. 
The velocity must further satisfy a simplified form of the continuity equation $\bm\nabla\bm\cdot\bm u=0$,  if the density of the fluid is constant, $\rho=\rho_0$. Finally, $\nu$ is the kinematic viscosity, and $\bm f(\bm x, t)$ is a random external force, which is usually referred to as driving.

The emergence of turbulence only occurs at sufficiently large Reynolds numbers, $Re=UL/\nu$, which measures the ratio of the inertial term to the  viscous term in Eq. (\ref{ns}), assuming $U$ is the characteristic (e.g. root mean squared) velocity of energy-containing eddies\footnote{In fluid dynamics, an eddy is understood as the swirling current of a fluid or a `blob of vorticity', but there is a good tradition in hydrodynamic turbulence of avoiding any formal definition of a turbulent eddy, see page 52 in \citet{davidson04}.} of size, $L$. 

Turbulent fluids are not, however, completely random and unpredictable. Instead their statistics are reproducible and strictly obey certain scaling laws, which can be derived from the NS equation under simplifying assumptions about the symmetries of the underlying problem. 

\subsection{Basic probabilistic tools}

The scale-dependent correlations of fluid variables in turbulent flows are traditionally described in terms of structure functions, correlation functions, and their Fourier counterparts (e.g. power spectra). These standard probabilistic tools are widely used in studies of incompressible turbulence in which the velocity field $\bm u(\bm x,t)$ fully describes the system. The velocity of a turbulent fluid at a given point in space and in time can be treated as a vector-valued centered random variable (zero mean value $\langle\bm u\rangle=0$), as it would still be a function of the initial conditions.

Let us consider the velocity increment, $\delta\bm u(\bm x,\bm r)=\bm u(\bm x+\bm r)-\bm u(\bm x)$, between two points, $\bm x$ and $\bm x+\bm r$, separated by the lag, $\bm r$. In homogeneous systems, the statistics of $\delta\bm u(\bm x,\bm r)$ only depends on the lag, $\bm r$, and not on the position, $\bm x$. If turbulence is also isotropic, the statistics of velocity increments would depend only on the magnitude of the lag, $r=|\bm r|$, not its orientation. Homogeneity and isotropy imply that spatial translations, rotations, and reflections of the original system of coordinate axes $(x_1,x_2,x_3)$ do not change the distribution functions of physical variables.\footnote{In realistic situations, both assumptions are usually not satisfied at large scales. However, if they are valid at small scales and far from boundaries of the flow or its other special regions, this general theoretical framework still remains useful.} It is convenient then to define the longitudinal velocity increment as $\delta u_{\parallel}(r)=\delta\bm u(\bm r)\bm\cdot\bm r/r$, using the natural coordinate frame with the direction of one of the coordinate axes aligned with $\bm r$. The two remaining transverse components, $\delta u_{\perp}(r)$, are equal in the isotropic case. The longitudinal and transverse velocity structure functions of order $p$ are defined, respectively, as $S^{\parallel}_p(r)=\langle[\delta u_{\parallel}(r)]^p\rangle$ and $S^{\perp}_p(r)=\langle[\delta u_{\perp}(r)]^p\rangle$, where $\langle\cdot\rangle$ denotes averaging over an ensemble of point pairs\footnote{Relevant  methods of taking averages of random functions of position and time in homogeneous and ergodic systems are discussed, for instance, in section 2.1 of \citet{batchelor53} and in section 4.4 of \citet{frisch95}.} with the fixed lag magnitude $r$. Note that the velocity structure functions retain Galilean invariance because they depend on $\delta\bm u$. 

The two-point velocity autocorrelation function is defined as $R_{\bm u\bm u}(\bm r)=\langle\bm u(\bm x)\bm\cdot\bm u(\bm x+\bm r)\rangle$, also assuming homogeneity. One can readily show that the velocity autocorrelation function is related to the second order structure function as $R_{\bm u\bm u}(\bm r)=\langle \bm u^2\rangle-\langle[\delta\bm u(r)]^2\rangle/2$ and also that $R_{\bm u\bm u}(\bm 0)=\langle\bm u^2\rangle$. In addition, the three-dimensional power spectrum of velocity, $P_{3D}(\bm u,\bm k)\equiv|\widehat{\bm u}(\bm k)|^2=\widehat{R_{\bm u\bm u}}(\bm k)$, is the Fourier transform of the autocorrelation function, as follows from the Wiener-Khinchin formula \citep{wiener30,khintchine34} (here $\widehat{\ldots}$ denotes the Fourier transform\footnote{Note that the Fourier transforms of homogeneous random functions are, generally, random distributions, i.e.\ not ordinary functions of their argument, $\bm k$. One way to deal with this mathematical difficulty is to replace the ordinary integrals with generalized stochastic Fourier-Stieltjes integrals \citep{batchelor53}; another way is to use low- or high-pass filtering (e.g.\ coarse-graining), which allows one to deal lusively with ordinary functions \citep{frisch95}.} and $\bm k$ is the wave vector). If the turbulence is also isotropic then the 3D power spectrum depends only on the magnitude of the wave vector, $k=|\bm k|$, and hence the one-dimensional (angle-integrated) spectrum can be written as $P(\bm u,k)\equiv\int P_{3D}(\bm u,\bm k)d\Omega_{\bm k}=4\pi k^2P_{3D}(\bm u,k)$. From Parseval's theorem we further get $\int_0^{\infty}P(\bm u,k)dk=\int P_{3D}(\bm u,\bm k)d\bm k=\langle\bm u^2\rangle=R_{\bm u,\bm u}(0)$, which relates various second order moments introduced above to the mean turbulent kinetic energy, $E=\rho_0\langle\bm u^2\rangle/2$, in the incompressible case. Another useful quantity, the velocity dispersion $\sigma_u(r)$ as a function of scale $r=2\pi/\kappa$, can be easily computed from the power spectrum since the variance $\sigma_u^2(r)=\int^{\infty}_{\kappa}P(\bm u,k)dk$, if $\langle\bm u\rangle=0$, and interpreted as a result of high-pass filtering operation.

\subsection{Energy cascade and the four-fifths law}
In the classical phenomenology of turbulence, kinetic energy is supplied at the largest scales (e.g.\ by a stochastic forcing mechanism). Non-linear advection coupled with fluid instabilities then generates motions on progressively smaller and smaller scales. This energy transfer process continues until molecular transport becomes dominant and dissipates the energy as heat. In such scale-by-scale energy cascade\footnote{Envisioned by Lewis F.~Richardson in 1922 \citep{richardson22}.} powered by vortex stretching in three dimensions, fluid motions progressively lose information about (non-universal) details of large-scale energy injection, leading to presumably self-similar (universal) fluid behavior at small scales. This self-similarity leads to a power-law energy spectrum determined solely by the magnitude of the energy flux if we (heuristically) assume that the cascade interactions are local, i.e.\ only comparable spatial scales interact with one another. The locality assumption does not hold universally true, though, as there are non-local scale interactions in Fourier space (see, e.g., section 7.3 in \citet{frisch95} and section 5.5 in \citet{diamond..10}).

\begin{figure}
  \includegraphics[scale=0.8]{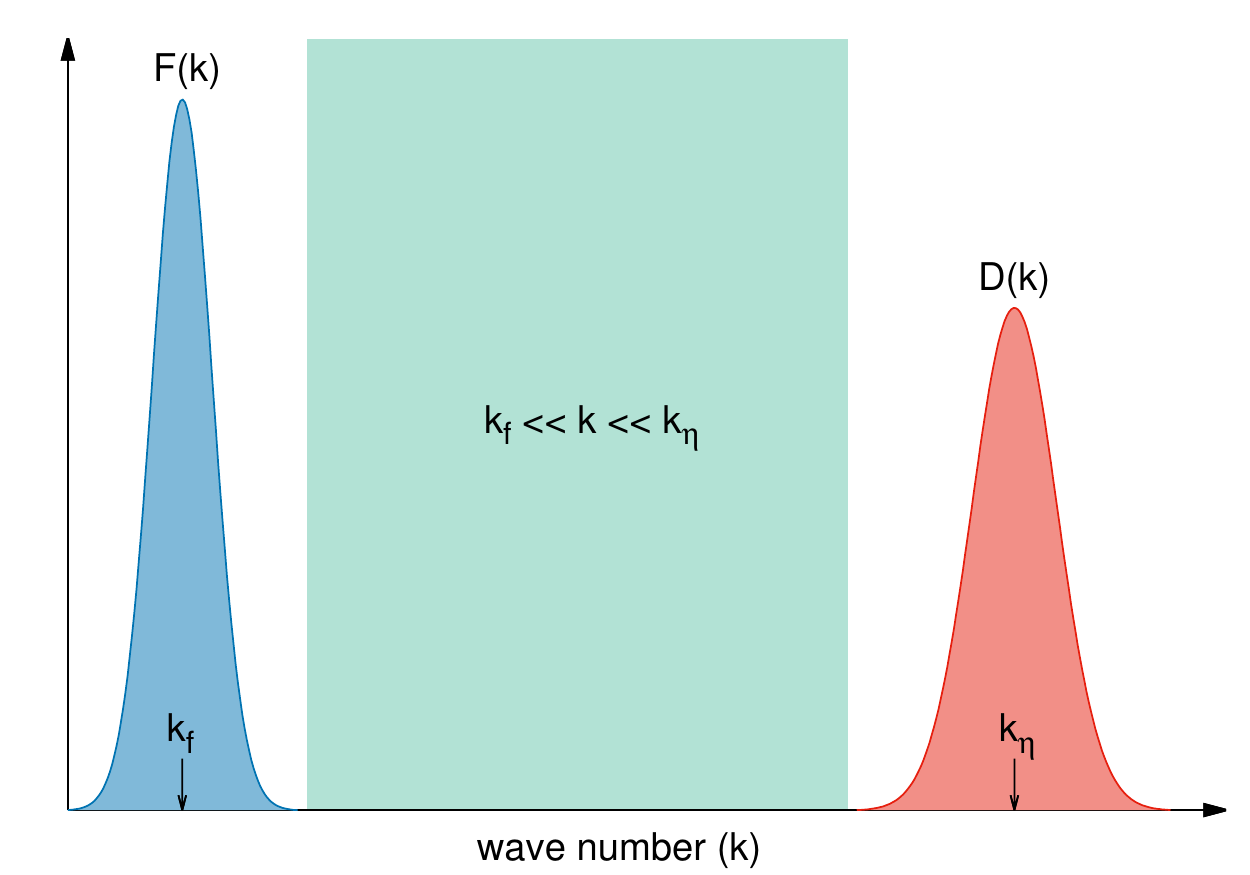}
\caption{Sketch of scale-by-scale energy budget in spectral space. $F(k)$ indicates the forcing range centered around $k_f$, $D(k)$ shows the dissipation range around $k_{\eta}=2\pi/\eta$; the inertial range of scales in between, $k_f\ll k\ll k_{\eta}$, is shown in green.}
\label{fig:t1} 
\end{figure}

To describe the scale-by-scale energy balance, one can write the NS equation in Fourier space and use Fourier integrals to describe turbulent flow in a bounded region of an infinite space, $\mathbb{R}^3$. The resulting equation takes the form \citep[e.g.][]{frisch95} $\partial_t E(k)=T(k)-D(k)+F(k)$, where $\partial_t$ denotes the partial time derivative, $T(k)$ is the transfer function describing energy transfer to scale $k$ due to nonlinear interaction of velocity fluctuations at all different scales, $D(k)=2\nu k^2E(k)$ describes the viscous dissipation of kinetic energy into heat, and $F(k)$ accounts for the energy supply to the system due to the work of an external force. If turbulence is statistically stationary, $E(k)$ does not change with time at all wave numbers, $k$, and hence $T(k)-D(k)+F(k)=0$. In this case, energy injection and dissipation rates are balanced overall. Note that energy supply is mostly concentrated at large scales (small $k$), while energy dissipation occurs at small scales (large $k$), see Fig.~\ref{fig:t1}. Balance between energy injection and dissipation implies that the areas under $F(k)$ and $D(k)$ curves are equal. Since in Fig.~\ref{fig:t1} the energy is supplied on the left and removed on the right, it should be transferred across the range $k\in[k_f,k_{\eta}]$ where $F(k)=D(k)=0$ and hence $T(k)=0$, i.e.\ all energy incoming to $k$ from larger scales gets transferred to smaller scales. The range of scales that are sufficiently distant from both $k_f$ and $k_{\eta}$, where we expect self-similar behavior of fluctuations, is called the inertial range. We can also define the kinetic energy contained at all scales smaller than a given scale as $\int_k^{\infty}E(\kappa)d\kappa$ and the cross-scale energy flux as $\Pi(k)=\int_k^{\infty}T(\kappa)d\kappa$; similarly,  $\Phi(k)=\int_k^{\infty}F(\kappa)d\kappa$ and $\Delta(k)=\int_k^{\infty}D(\kappa)d\kappa$. In statistically stationary turbulence, these quantities are related: $\Pi(k)-\Delta(k)+\Phi(k)=0$. In the inertial range (shown in green in Fig.~\ref{fig:t1}), the energy transfer rate is independent of scale, $\Pi(k)=const$. It is also equal to the energy dissipation rate, $\epsilon$, and to the energy injection rate, $\langle\bm u\bm\cdot\bm f\rangle$, associated with the driving force: $\Pi(k)=\Delta(0)\equiv\epsilon=\Phi(0)=\langle\bm u\bm\cdot\bm f\rangle$.

Suppose that eddies of size $r$ participating in this energy transfer process in the inertial range have a characteristic velocity, $u_r$, and break up on the eddy turn-over time, $\tau_r\equiv r/u_r$. If the energy flux, $\Pi(r),$ along the cascade is constant (we consider a statistically stationary situation) and equal to the energy dissipation rate, $\epsilon$, then $\Pi(r)\sim u_r^2/\tau_r\sim u_r^3/r\sim\epsilon$. In terms of the first-order velocity structure function, this can be written approximately as $S_1(r) = C_1(\epsilon r)^{1/3}$, where $C_1$ is a dimensionless constant, which is usually called the Kolmogorov-Obukhov law \cite{kolmogorov41a,obukhov41}. It was pointed out by \citet{onsager49} that the $r^{1/3}$ law for velocity increments may reflect the lack of smoothness of the velocity field, which is only H\"older continuous of exponent $1/3$ (see, e.g., section 1.3 in \citet{feireisl..16} for the definition of H\"older continuous functions). Onsager was perhaps the first to raise the issue of singularities in turbulent fluid flows \citep{frisch95,onsager.06}. Using dimensional arguments, the Kolmogorov-Obukhov law can be extended to structure functions of arbitrary order $p$ as $S_p(r) = C_p (\epsilon r)^{p/3}$. Similarly, the energy spectrum can be written as $E(k)=2\pi k^2|\widehat{\bm u}(\bm k)|^2 = C_K\epsilon^{2/3}k^{-5/3}$, where $C_K$ is the Kolmogorov constant. Since the dissipation scale, $\eta$ (also called the Kolmogorov scale), depends only on the dissipation rate, $\epsilon$, and on the viscosity, $\nu$, the Kolmogorov scale can be readily written as $\eta=(\nu^3/\epsilon)^{1/4}$ using dimensional arguments \citep[see Eq. (6.66) in][]{frisch95}.

It is important to emphasize that most of the scaling relations introduced above are approximations since they were obtained phenomenologically. The only exception is the exact expression  for the third order longitudinal velocity structure function, $S^{\parallel}_3(r)=-\frac{4}{5}\epsilon r$, derived by Kolmogorov from the NS equations \cite{kolmogorov41b}. This result, known as the four-fifths law, is valid for homogeneous isotropic stationary turbulence in the inertial range. A primitive form of the four-fifths law for isotropic turbulence, which served as a stepping stone for the derivation, combines longitudinal and vector velocity increments, $\langle\delta u_{\parallel}(r) |\delta\bm u(r)|^2\rangle=-\frac{4}{3}\epsilon r$, and is known as the four-thirds law or the von K\'arm\'an-Howarth relation \citep{dekarman.38}.\footnote{A similar relation, $\langle\delta u_{\parallel}(r) [\delta\theta(r)]^2\rangle=-\frac{4}{3}\epsilon_{\theta} r$, was obtained by \citet{yaglom49} for temperature fluctuations in turbulent flows. Here, $\delta\theta(r)$ is the temperature increment and $\epsilon_{\theta}$ is the mean dissipation rate of temperature fluctuations. An anisotropic generalization of this relation exploited in \cite{galtier.11}, $\bm\nabla\bm\cdot\langle|\delta\bm u(\bm r)|^2\delta\bm u(\bm r)\rangle=-4\epsilon$, avoids projection onto the $\parallel$ direction and is sometimes called the von K\'arm\'an-Howarth-Monin relation \citep{frisch95,antonia...97}.}

The third-order moment of velocity increments is special, since it depends on the mean dissipation rate,  $\epsilon=\langle\varepsilon\rangle$. Note that the local dissipation rate, $\varepsilon$, is not a constant but rather a random variable with its own probability density function (PDF). Moments of order $p\ne3$ depend on $\langle\varepsilon^{p/3}\rangle\ne\epsilon^{p/3}$, and hence their scaling relations cannot be universal functions of $\epsilon$ and $\nu$ as they depend on the detailed structure of the $\varepsilon$-PDF, which may be different in different turbulent flow realizations. This lack of universality was first noticed by Landau in his famous remark made at a seminar in Kazan in early 1942, where Kolmogorov has summarized his 1941 work \citep{landau.87,frisch95}.

As noted by Uriel Frisch \citep[Section 6.2]{frisch95}, the four-fifths law ``is one of the most important results in fully developed turbulence, because it is both exact and non-trivial. It thus constitutes a kind of `boundary condition' on theories of turbulence: such theories, to be acceptable, must either satisfy the four-fifths law or explicitly violate the assumptions made in deriving it.''

\subsection{Effects of compressibility}
Turbulence in the ISM, like in many other astrophysical environments (e.g.\ solar wind and stellar winds, supernova collapse, black hole accretion flows) and in numerous terrestrial applications (e.g.\ supersonic mixing in scramjets, hypersonic turbulent boundary layers in high speed aerodynamics, shock-turbulence interaction, inertial confinement fusion, volcanic eruptions, etc.) is highly compressible and hence the formalism developed for incompressible fluids does not directly apply. 

In incompressible turbulence, the pressure is not an autonomous thermodynamic variable, but acts as an enslaved Lagrange multiplier connected to the solenoidal constraint on the velocity. Hence the full solution is contained in the solenoidal  (divergence-free) velocity field, see Section 2.1.3 in \cite{sagaut.18}.

The next level of complexity is represented by a family of compressible barotropic models where the pressure is a function of density only, $p=p(\rho)$. The specific internal energy (or rather the specific Gibbs free energy) in barotropic flows is interpreted as $e\equiv P(\rho)$, where the pressure potential $P(\rho)\equiv-\int_{\rho_0}^{\rho}p(\rho)d(1/\rho)$. The continuity and NS equations in this case form a closed system and the changes to internal energy in such barotropic flows are interpreted in a purely mechanical sense as resulting from the work done on a volume element of a fluid. Two simple examples include an isothermal closure, $p(\rho)=c_{\rm s}^2\rho$, where the sound speed $c_{\rm s}=const$, and a family of polytropic models, $p(\rho)=a\rho^{\gamma}$, where $\gamma>1$ is the adiabatic constant and $a =const>0$ \citep{batchelor67,feireisl04}. In both cases, compressibility brings along a new scalar field, $\rho$, and a new vector field of dilatational (irrotational) velocity, $\bm u_{\rm d}$. This compressible velocity component is responsible for pressure-dilatation effects and for the new channel of dilatational dissipation. Note that, while ideal barotropic models provide a mathematically more tractable description for molecular fluids, inclusion of molecular transport effects in the momentum equation is problematic (e.g.\ \cite{chandrasekhar51a,eyink.18}, see also \cite{batchelor67}).

In non-barotropic fluids, strong coupling of momentum with thermodynamics becomes important due to pressure-dilatation work \citep{aluie13} and baropycnal work \citep{lees.19}. This brings into play fundamentally different energy transfer pathways, missing in the incompressible turbulence \citep{sagaut.18}. Moreover, various constitutive equations may bring new physical nonlinearities in addition to the usual geometric nonlinearity  associated with the Eulerian description \citep{feireisl04}.

To better understand the differences between turbulent fluctuations in incompressible and compressible fluids, it is instructive to consider linearized fluctuation modes on a uniform background. Linear decomposition \citep{kovasznay53} yields three modes of fluctuations: (i) the vortical mode is purely  incompressible, includes solenoidal velocity, $\bm u_{\rm sv}$ ($\bm\nabla\bm\cdot\bm u_{\rm sv}=0$), and no pressure or density fluctuations; (ii) the acoustic mode, which includes only dilatational velocity, $\bm u_{\rm da}$ ($\bm\nabla\bm\times\bm u_{\rm da}=0$), and isentropic pressure and density fluctuations; and (iii) the entropy mode, which is a wave-like linear solution with entropy fluctuations, no pressure fluctuations, and purely dilatational velocity, $\bm u_{\rm de}$. The motions in the entropy mode are induced by the viscous effects, hence in the inviscid case $\bm u_{\rm de}=0$.

In the context of \citet{helmholtz1858} decomposition for the velocity field, $\bm u=\bm u_{\rm s} + \bm u_{\rm d}$, one obtains $\bm u_{\rm s}=\bm u_{\rm sv}$ and $\bm u_{\rm d}=\bm u_{\rm da}+\bm u_{\rm de}$ for the viscous case. As can be seen, the solenoidal velocity component does not include the acoustic waves, but the dilatational field is not limited to acoustic phenomena and generally includes convective effects of heat transfer. It is worth noting that the purely kinematic Helmholtz decomposition is exact and does not rely on any small parameter expansion, see Chapters 2 and 3 in \cite{sagaut.18} for an in-depth discussion of various decomposition techniques for compressible flows.

The linear modes of a uniform state we discussed above are decoupled in the first order in amplitude but become fully coupled already in the second order due to mean gradients and nonlinear inertia terms in the dynamical equations \citep{moyal52,chu.57}. It should be emphasized that such modal decomposition would fail as a general approach to deal with genuinely nonlinear dynamics of turbulence, since corresponding solutions of NS equations cannot be consistently expanded in linear modes. Even though for second-order moments at relatively low turbulent Mach numbers ($M\equiv \sqrt{\langle u^2\rangle}/\langle c_{\rm s}\rangle = 0.1-0.6$) the predictions of the  linear approximation are accurate at a few percent level, the errors inevitably become large for higher-order moments and as fluctuations get stronger at higher Mach numbers \citep{donzis.13,eyink.18}. 

The solenoidal and dilatational Helmholtz projections of the velocity vector $\bm u$ are locally orthogonal in Fourier space (since $\bm k\bm\cdot\widehat{\bm u_{\rm s}}(\bm k)=0$ and $\bm k\bm\times\widehat{\bm u_{\rm d}}(\bm k)=0$) and hence the variance $\langle\bm u\bm\cdot\bm u\rangle$ can be split into a sum of the dilatational and solenoidal parts, regardless of the strength of the fluctuations.\footnote{The same splitting based on $\bm u_{\rm s}$ and $\bm u_{\rm d}$ does not work for the Reynolds stress $R_{ij}\equiv\langle\rho u_iu_j\rangle$, since in addition to solenoidal and dilatational stresses, there is also non-zero cross Reynolds stress \citep{lele94}.} A useful diagnostic of compressibility levels in the energy containing range of fully developed homogeneous turbulence is the ratio,  $\chi\equiv\langle\bm u_{\rm d}\bm\cdot\bm u_{\rm d}\rangle/\langle\bm u_{\rm s}\bm\cdot\bm u_{\rm s}\rangle$. Together with $Re$ and $M$, the compressive ratio, $\chi$, determines the scaling  \cite{donzis.19}. At low Mach numbers ($M<1$) the solenoidal and dilatational modes are mostly decoupled and the statistics of turbulence depend either on $\chi_{\rm i}$, corresponding to initial conditions (in decaying turbulence), or on $\chi_f$, reflecting the mode mixture in the forcing (in stationary turbulence). The lack of mode coupling allows drastically different turbulence regimes with vortical and acoustic modes mixed in different proportions controlled by external factors. 

The idea of universal scaling classes at low Mach numbers was explored in \citep{donzis.19} with a diverse set of direct numerical simulations (DNS) (including homogeneous isotropic turbulence and homogeneous shear turbulence). In DNS of homogeneous turbulence with purely solenoidal forcing at $M\ge0.4$, the dilatational part of the specific kinetic energy is dominated by the acoustic component. If $M<0.4$, small scales are dominated by the pseudo-sound component\footnote{There are hydrodynamic and acoustic pressure fluctuations. The pseudo-sound represents vortical pressure fluctuations advected with the fluid velocity, while acoustic waves propagate at the speed of sound. In the pseudo-sound component, the dilatational velocity field is in equilibrium with the solenoidal pressure. Both types of fluctuations can be measured by the observer.} associated with hydrodynamic eddies \citep{wang..17a}. In all cases with solenoidal forcing, the fraction of specific kinetic energy in dilatational modes is small, $\chi<0.1$. When dilatational forcing is used at $M<0.25$, the compressive ratio $\chi$ can be as high as $\approx 2.5$. At $M\sim 0.6$, the compressive ratio, $\chi$, still can be as high as $\sim1$, but the nonlinear mode coupling would limit its further growth \citep{donzis.19}. 

The energy exchange between vortical and acoustic modes further leads to equipartition at $M>1$, as predicted by \citet{kraichnan55} for the inviscid case with weak excitation based on Liouville's theorem. Since the acoustic energy includes two equal parts (kinetic $+$ potential) and these are in detailed equilibrium across scales \citep{sarkar...91,donzis.13}, the dilatational kinetic energy associated with acoustic waves represents only one half of the full acoustic energy. Thus the expected asymptotic equilibrium value, reflecting energy equipartition between hydrodynamic (eddies) and acoustic (waves) parts of supersonic turbulence, is $\chi_{\infty}=1/2$, which corresponds to the so-called `natural' mix of the modes reproduced in simulations of supersonic turbulence \citep{kritsuk...10}. 

A somewhat lower value of the compressive ratio, $\chi\approx0.3$ was measured in a simulation of supernova driven magnetized molecular cloud turbulence \citep{pan...16}. The energy injection mechanism in this simulation can be described as stochastic thermal forcing, which is different from the standard stochastic large-scale acceleration approach.\footnote{Thermal forcing can also be facilitated through a generalized cooling function (a volumetric source in the energy conservation law  \citep{kritsuk.02}) or through a large-scale injection of internal energy \citep{wang.....19}.} The resulting compressive ratio, nevertheless, is comparable to the value found in isothermal simulations with random solenoidal forcing at $M\approx9$ and Alfv\'enic Mach numbers from 3 to 5 \citep{kritsuk...10} and deviates from $1/2$, as expected, due to different equipartition constraints in magnetohydrodynamic (MHD) turbulence.

Thus, in fully developed compressible turbulence, the Mach number and the compressive ratio, $\chi$, define the scaling universality classes in subsonic regimes, while in supersonic turbulence the compressive ratio assumes an asymptotic value, $\chi_{\infty}$, which only depends on the nature of the system (e.g.\ the magnetization level) and does not have to depend on details of the energy injection mechanism. 

If this phenomenology is correct, in an idealized case of homogeneous supersonic turbulence in a periodic box with large-scale forcing, the energy cascade proceeds from the injection scale with dilatational and solenoidal modes tightly coupled until the sonic scale.\footnote{The sonic scale, $k_{\rm s}$,  is defined as a scale at which the root mean squared velocity fluctuations are equal to the mean sound speed, $\int_{k_{\rm s}}^{\infty}E(k)dk=\langle c_{\rm s}\rangle^2$.} At the sonic scale, the compressive ratio still remains close to its asymptotic `natural' value; at scales below the sonic scale the hydrodynamic and acoustic cascades fully decouples and proceeds independently without active energy exchange.

Observational measurements of $\chi$ in molecular clouds are difficult, primarily due to projection effects. Therefore, methods developed for nearby molecular clouds for which high-sensitivity, high spatial dynamic range spectral line observations are available, have to rely on the assumption of statistical isotropy and hence cannot account for anisotropies caused by the magnetic fields or large-scale energy injection \citep{brunt..10}. Another limiting factor is the emission-weighted nature of observational data, which implies the use of Helmholtz decomposition of the momentum density $\bm j\equiv\rho\bm u$ instead of the velocity $\bm u$ \citep{brunt.14}. It is remarkable that the global compressive ratio measured this way for the Orion B molecular cloud with a mean ${\cal M}\sim6$, using $^{13}$CO$(J = 1-0)$ data in the $11\times6$~pc$^2$ area encompassing the cloud, yields $\chi\lesssim 0.4$ \citep{okrizs...17} -- in reasonable agreement with the prediction based on the energy equipartition conjecture discussed above.

\subsection{Scaling relations and energy cascades in compressible turbulence}

\subsubsection{Numerical simulations}
Most of what we know about turbulence dynamics in the local interstellar medium comes from computer simulations. In a very abstract way such simulations provide us with at least a 12-dimensional data hyperspace (position, time, density, pressure, velocity and magnetic field components), while observations usually yield severely reduced volumes of information, resulting from line-of-sight projection effects and intricate convolutions (e.g.\ two-dimensional emission maps or 3D position-position-velocity cubes). Because of this dramatic information loss, the only sensible approach to test models is through systematic synthetic observations that mimic the involved convolutions in a realistic way (e.g. \citet{okrizs...17}).

Interstellar turbulence regimes of interest for star formation studies involve sonic rms Mach numbers from ${\cal M}\lesssim 1$ through $\sim15$ and Alfv\'enic Mach numbers from $\sim0.5$ through $\sim5$ \citep{kritsuk..17}. These conditions cannot be reproduced in the laboratory, with perhaps a few exceptions \citep[e.g.][]{white-ea-19}. Hence to obtain insights into the physics of interstellar turbulence one has to pursue theoretical research and numerical experiments. One has to keep in mind that realism of star formation simulations is limited by the power of available computational resources and can be improved by advances in numerical methods.

Over the past decade, well-resolved DNS of subsonic regimes with a primary focus on engineering applications have become substantially more mature, as they now routinely use high-order accurate and computationally efficient numerical methods \citep{wang....10}, grids up to $2048^3$ and Taylor microscale Reynolds numbers up to $R_{\lambda}=430$. Emphasis is being placed on creating a database with various simulation cubes populating the vast parameter space \citep[e.g.][]{donzis.13,chen...15,jagannathan.16,donzis.19,wang..17a,wang..17b,wang..17c}. 

For comparison, simulations of homogeneous isotropic incompressible turbulence in a periodic box with  pseudo-spectral methods aimed at improving our understanding of turbulence small-scale structure have reached a resolution of $8192^3$ and $R_{\lambda}=1300$ in 2015 \citep{yeung..15}. The next steps involve a new pseudo-spectral code that scales up to a problem size of $18,432^3$ on the GPU-based  supercomputer {\em Summit} at ORNL and will continue to focus on intermittency in incompressible turbulence \citep{yeung19}. 

The biggest supersonic turbulence simulation to date rely on second-order accurate finite-volume shock-capturing method implemented in the {\sc flash} code, did not include the viscous terms,\footnote{This type of modeling is usually called an implicit large-eddy simulation (ILES) \citep{sytine....00} or a coarse DNS since the dissipation is of purely numerical origin and depends on the numerical method used.} and boosted the grid resolution to $10,048^3$ in 2016 \citep{federrath...16}. This should be compared to the first $1024^3$ simulation of decaying transonic turbulence with the piecewise-parabolic method (PPM) carried out in 1998 \citep{porter..98} to appreciate the stunning improvement in availability of computational resources achieved over the past two decades.

\subsubsection{Scaling in compressible turbulence}

Early simulations of stationary transonic turbulence at ${\cal M}\sim1$ revealed a Kolmogorov-like scaling of velocity in terms of both structure functions and power spectra \citep{porter..02}, thereby demonstrating the compatibility of the $k^{-5/3}$ spectrum with a mild compressibility at transonic Mach numbers. Isothermal simulations exploring supersonic regimes followed closely, but up to mid-2000s did not have enough resolution to measure the absolute scaling exponents. Instead, they had to rely on the so-called extended self-similarity hypothesis \citep{benzi...95} to boost the extent of the scaling range by plotting $S_p(r)$ against $S_3(r)$, which was expected to scale linearly with the lag $r$, as in the four-fifths law. This early work indicated that scaling exponents of the velocity statistics may change gradually between ${\cal M}\sim1$ and $\sim3$ as the Hausdorff dimension of the most singular dissipative structures transitions from $D\sim1$ (vortex filaments) to $D\sim2$ (shock waves) \citep{padoan...04}. Thus the change in the scaling was attributed to intermittency. Higher resolution simulations showed that the velocity spectra indeed get steeper and density spectra get shallower as the turbulent Mach numbers enter the hypersonic range \citep{kim.05,kritsuk...06a}. However, measurements of the absolute exponent $\zeta_3$ of the third-order velocity structure function $S_3(r)\propto r^{\zeta_3}$ at ${\cal M}>3$ returned $\zeta_3\approx1.25$ \citep{kritsuk...06,boldyrev..02},\footnote{A similar slope was also obtained in simulations by W.-C. M\"uller (2005, private communication).} indicating that the observed change in scaling could not occur due to intermittency alone.

This was a clear indication that the energy transfer in supersonic regime is mediated by density fluctuations and hence ignoring the density velocity correlations would lead to large errors in the kinetic energy flux across scales \citep{weizsacker51,lighthill55,henriksen91,fleck96}. A very na\"ive check to make was to consider the third-order structure function of a density-weighted velocity, $\bm v\equiv\rho^{1/3}\bm u$, which happened to be linear \citep{kritsuk...07a}. The linear scaling at high Mach numbers was then independently confirmed in \citet{kowal.07,schmidt..08,schwarz...10,zrake.12}. These results were consistent with simple Richardson-Kolmogorov-like energy cascade phenomenology modified to account for density fluctuations, $\Pi(r)\sim\rho_ru_r^3/r\sim\rho_0\epsilon$, and hence triggered a quest for exact relations describing the energy cascade in compressible turbulence.

Note that the density-velocity correlations are also present at subsonic turbulent Mach numbers, when the compressibility is weak, but their effects are so small that the power spectra of $\bm u$ and $\bm v$ are practically indistinguishable \citep{kritsuk...07b,wang.....12}. At the same time, physics of highly compressible turbulence in molecular clouds stimulated numerical experiments at high Mach numbers, which eventually led to the discovery of new fundamental scaling laws.

The first non-trivial step needed to extend Kolmogorov's phenomenology to compressible turbulence was the interpretation of the four-fifth law; namely, whether the velocity, $\bm u$, should be carried over to the compressible case as velocity or as momentum, $\bm j\equiv\rho \bm u$ (or as both). 

Taking the latter approach, \citet{falkovich..10} derived a series of new exact relations for fluxes and densities of conserved variables. A particular case relevant to the cascading of mean squared momentum, $\langle|\bm j|^2/2\rangle$, injected by external forcing, produced the following exact law: $\bm\nabla_r\bm\cdot\langle(\bm j\bm\cdot\bm j')\bm u'+c_{\rm s}^2\rho'\bm j\rangle=-\langle\rho\rho'\rangle\epsilon$. While this relation reduces to the four-fifths law in the incompressible limit, it does not yield a desired universal scaling law for compressible turbulence. Indeed, it can be readily seen that scaling of the density autocorrelation function in the r.h.s. varies with the Mach number, as does the slope of the density power spectrum \citep{kim.05}, see also \citet{wagner...12,kritsuk..13}. One can also argue that $\bm j^2$ is not a conserved quantity and hence plays no obvious dynamically important role in compressible turbulence \citep{eyink.18}.  Quite a variety of anomalous balance relations for compressible turbulence can be derived. Even though they recover the von K\'arm\'an-Howarth-Monin relation in the incompressible limit, most of them may appear physically irrelevant.

\subsubsection{Total energy as an ideal invariant}

A more traditional interpretation of the four-fifths law in terms of the kinetic energy transfer poses its own challenges when an extension to the compressible case is attempted. First, the kinetic energy is not an ideal invariant of the compressible system, but instead the total energy, $E=K+U=\langle\rho\bm u\bm\cdot\bm u/2+\rho e\rangle$, is invariant. Second, both kinetic, $K$, and internal, $U$, energy terms are not quadratic. Finally, in the isothermal case, the specific free energy per unit mass $e=c_{\rm s}^2\ln(\rho/\rho_0)$ is not sign-definite. Note that including the fluctuations of magnetic field, $\bm b$, would further add the magnetic energy, $M$, to the invariant, $E=K+M+U=\langle\rho\bm u\bm\cdot\bm u/2+\bm b\bm\cdot\bm b/8\pi+\rho e\rangle$, and this new term is quadratic. Same is true for including the effects of self-gravity represented by the fluctuations of the free-fall acceleration, $\bm g$, which adds a negative quadratic contribution of the potential energy, $-W$, to the total energy, $E=K+M+U-W=\langle\rho\bm u\bm\cdot\bm u/2+\bm b\bm\cdot\bm b/8\pi+\rho e-\bm g\bm\cdot\bm g/8\pi G\rangle$. The full self-gravitating MHD system would also include the magnetic induction equation and the Poisson equation, which do not add new nonlinearities.

\subsubsection{Kinetic energy spectra and correlation functions}

Let us first consider the kinetic energy, $K=\langle\rho\bm u\bm\cdot\bm u\rangle/2$, which can be decomposed into solenoidal, dilatational, and mean components by introducing a new variable, $\bm w\equiv\sqrt{\rho}\bm u$, and applying Helmholtz decomposition, $\bm w=\bm w_s+\bm w_d+\bm w_o$ \citep{kida.90,kida.92,miura.95}. The kinetic energy is then the sum of these three components $K=\tilde{K}_s+\tilde{K}_d+\tilde{K}_o$, where $\tilde{K}_s=\langle\bm w_s^2\rangle/2$, $\tilde{K}_d=\langle\bm w_d^2\rangle/2$, and $\tilde{K}_o=\langle\bm w_o^2\rangle/2$. The spectrum of kinetic energy can be defined in terms of the Fourier transform of $\bm w$ as $\tilde{K}(\bm k)=|\widehat{\bm w}(\bm k)|^2/2$, and in the isotropic case $\tilde{K}(k)=4\pi k^2\tilde{K}(\bm k)$. The total kinetic energy spectrum then includes two familiar components, $\tilde{K}(k)=\tilde{K}_s(k)+\tilde{K}_d(k)$, and $K=\int_0^{\infty}\tilde{K}(k)dk+\tilde{K}_o$. This $\bm w$-based decomposition is not unique \citep{sagaut.18}, but convenient since it enforces the positive-definiteness of the compressive and rotational kinetic energies. For that reason, $\bm w$ has been often used  to compute the kinetic energy spectrum in compressible turbulence  \citep[e.g.][]{cook.02,wang.....13,grete....17,mittal.19}. Empirically, it was noticed that in supersonic turbulence $\tilde{K}(k)$ shows a very strong ``bottleneck'' (present in both $\tilde{K}_s(k)$ and $\tilde{K}_d(k)$ spectra), which can be traced to shock fronts \citep[section 3.5 in][]{kritsuk...07a}. While the linear momentum, $\bm j$, is continuous across shock fronts, any mass-weighted velocity, $\rho^{\alpha}\bm u$, with $\alpha\ne1$ is discontinuous. It is therefore likely that strong small-scale autocorrelation of $\bm w$ caused by the fractional weight $\rho^{\alpha}$ with $\alpha=1/2$ is responsible for the ``bottleneck'' in $\tilde{K}(k)$. Note that the  compression factors across isothermal shocks can be very high ($\rho_2/\rho_1\sim {\cal M_{\rm shock}}^2$) and therefore the effect can be strong in supersonic regimes. In contrast, in subsonic turbulence, both $\bm u$- and $\bm w$-based decompositions yield similar results, making mass weighting impractical.

Similar effects of symmetric mass weighting appear in the scaling of third order moments \citep{kritsuk..13}. While the mixed structure function $\langle(\delta\bm j\bm\cdot\delta\bm u)\delta\bm u_{\parallel}\rangle$ showed a clean extended linear scaling range, $\langle|\delta\bm v_{\parallel}|^3\rangle$ displayed only an approximately linear scaling overall with no clear linear range \citep[fig.~6 in][]{kritsuk..13}. The failure of the $\bm v$-based approach to capture the desired universal scaling in simulations of supersonic turbulence suggests that fractional mass weighting should be avoided. 

Generally speaking, there is no fundamental {\em a priori} reason to favor one definition of the kinetic energy correlation function over another. The problem we are facing here is similar to the problem of statistical averaging in variable density fluid turbulence \citep[see  e.g. chapter 5 in][]{chassaing....02} and it is not yet clear what will eventually represent the best way forward. However, some guidance can be found in numerical simulations.

Similar issues arise when coarse-graining is used instead of point-splitting\footnote{In this context, point-splitting regularization refers to the use of two-point statistics (such as correlation functions, structure functions, and power spectra) when the products of fields at the same spatial (temporal) location are not mathematically well-defined.} as a convenient regularization that removes short-distance divergences. The coarse-graining approach was pursued by  \citet{aluie11,aluie..12,aluie13} and more recently by \citet{eyink.18} also to yield an analogue to Kolmogorov's four-fifths law for compressible fluid turbulence. The Favre scale-decomposition framework \citep{favre83} separates the ``resolved'' kinetic energy, $\frac{1}{2}\langle\bm j\rangle_r^2/\langle\rho\rangle_r\leq\frac{1}{2}\langle\bm j\bm\cdot\bm u\rangle_r$, from the ``unresolved'' or ``subscale'' kinetic energy represented by a second-order Favre cumulant, $\frac{1}{2}[\langle\bm j\bm\cdot\bm u\rangle_r-\langle\bm j\rangle_r^2/\langle\rho\rangle_r]$.  When the coarse-graining operation\footnote{The coarse-graining operation is a simple convolution $\langle\bm a\rangle_r(\bm x,t)=\int\phi_r(\bm y)\bm a(\bm x+\bm y,t)d\bm y$ with a smooth mollifier $\phi_r(\bm y)=\phi(\bm y/r)/r^3$ such that $\int\phi_r(\bm y)d\bm y=1$. This type of smooth filtering with compact support in space is used to single out the large scale component of a field variable $\bm a(\bm x,t)$ corresponding to length scales $>r$.} $\langle\cdot\rangle_r$  is applied to scale $r$ that falls in the inertial range, the cumulant (or the so-called ``subscale stress'') is interpreted as an effective inertial range dissipation of kinetic energy \citep{eyink.18}.

However, the mass-weighted Favre filtering is not the only way to decompose scales and regularize the equations. Alternatives include (but are not limited to): $\frac{1}{2}\langle\rho\rangle_r\langle\bm u\rangle_r^2$, $\frac{1}{2}\langle\bm w\rangle_r^2$,  $\frac{1}{2}\langle\bm j\rangle_r\bm\cdot\langle\bm u\rangle_r$,  $\frac{1}{2}\langle\rho\rangle_r\langle|\bm u|^2\rangle_r$, or linear combinations thereof. Based on the so-called inviscid criterion \citep{aluie13}, \citet{zhao.18} showed that the first two decompositions (Reynolds- and $\bm w$-based) fail to capture the inertial range, if density variations are significant. The key idea behind the inviscid criterion is that the scale-decomposition of momentum and kinetic energy should guarantee that viscous contributions are negligible at large enough length-scales, which is necessary for the study of inertial range dynamics. The numerical evidence presented by \citet{zhao.18} is consistent with similar indications from point-splitting approach that we discussed above. This can explain the difficulty of $\bm w$-based point-splitting analysis in \citet{grete....17} and \citet{schmidt.19} to capture a reasonably extended inertial range. 

Note that the so-called transport-selected regrouping  \citep[or mixed-weighted decomposition introduced by][]{chassaing85}, using Reynolds averaging for the transporting agent (e.g. the advection velocity) and Favre averaging for the convected function (e.g. a momentum component), $\frac{1}{2}\langle\bm j\rangle_r\bm\cdot\langle\bm u\rangle_r$, has yet to be evaluated with respect to the inviscid criterion.  However, its point-splitting counterpart is known to yield robust inertial range in supersonic turbulence \citep{kritsuk..13}.

Getting back to the point-splitting approach, let us consider an alternative definition of the kinetic energy correlation function, $K(\bm r)=R_{\bm j\bm u}(\bm r)/2$, using the symmetric part of the cross-covariance of linear momentum density and velocity, $R_{\bm j\bm u}(\bm r)\equiv\langle[\bm j(\bm x)\bm\cdot\bm u(\bm x+\bm r)+\bm j(\bm x+\bm r)\bm\cdot\bm u(\bm x)]\rangle/2$. It was used by \citet{graham..10,galtier.11,banerjee.13,banerjee.14,banerjee.17a,andres.17,banerjee.17b,andres..18,banerjee.18,andres.....18} to derive exact relations for energy transfer in compressible turbulence analogous the von K\'arm\'an-Howarth-Monin relation we introduced earlier. This approach does not suffer from artificial ``bottleneck'' in the kinetic energy cospectrum, $K(\bm k)=\widehat{R_{\bm j\bm u}}(\bm k)/2=\left[\widehat{\bm j}(\bm k)\bm\cdot\widehat{\bm u}^*(\bm k)+\widehat{\bm j}^*(\bm k)\bm\cdot\widehat{\bm u}(\bm k)\right]/4$ since linear momentum decouples from the velocity at shock fronts.\footnote{Here, $^*$ denotes the complex conjugate and we used the convolution theorem to cast the Fourier transform of the correlation function $R_{\bm j\bm u}(\bm r)$ using the Fourier transforms of $\bm j$ and $\bm u$. Using the symmetric cross-covariance makes sure that the spectral kinetic energy density, $K(\bm k)$, is real. Finally, it follows from Parseval's theorem that $K=\int K(\bm k)d\bm k=R_{\bm j\bm u}(0)/2$.} It also recovers the detailed small-scale magnetic-kinetic energy equipartition, $K(k)\approx M(k)$,  in an $1024^3$ MHD turbulence simulation at ${\cal M}\approx10$ and plasma $\beta_0=2$, previously plagued by a large excess in the kinetic energy spectrum, $\tilde{K}(k)$, computed using the Fourier transform of $\bm w$ \citep[e.g. fig.~3g in][]{kritsuk...09}. 

\subsubsection{Internal energy spectra and correlation functions}

The same diversity of formally allowed definitions equally applies to the two-point correlation function of thermodynamic energy. A set of options to split the non-quadratic combination of $\rho e$ betwen points $\bm x$ and $\bm x'=\bm x+\bm r$ discussed in the literature includes: (i) $\langle\rho e'\rangle$ \citep{galtier.11,banerjee.13,banerjee.17a,andres.17,andres..18,banerjee.18,andres.....18}, (ii) $\langle\rho c_{\rm s}c_{\rm s}'\rangle/\gamma(\gamma-1)$ \citep{banerjee.14}, and (iii) $\langle\sqrt{\rho} c_{\rm s}\sqrt{\rho'}c_{\rm s}'\rangle/\gamma(\gamma-1)$ \citep{schmidt.19,mittal.19}. The most popular option (i) uses point splitting between a conserved variable $\rho$ and the specific thermodynamic energy $e$. Option (ii) mimics the momentum-velocity point splitting for the kinetic energy by replacing the velocity $\bm u$ with the sound speed $c_{\rm s}$. It was suggested for systems with polytropic turbulence and cannot be applied to isothermal fluids where $c_{\rm s}$ is constant. Option (iii) is analogous to the $\bm w$-based splitting of the kinetic energy correlator, where the velocity is also replaced by the sound speed. 

In contrast to the kinetic energy correlation function, where centering of the velocity and momentum ($\langle\bm u\rangle=0$, $\langle\bm j\rangle=0$) can be achieved by a proper choice of the reference frame, the thermodynamic energy requires an appropriate decomposition of turbulent fluctuations from uniform background. This issue, overlooked in the early works on energy transfer in isothermal compressible turbulence, was exposed in \citep{banerjee.17b}, where the thermodynamic energy correlation function was defined as $R_{\rho e}(\bm r)=\langle\rho e\rangle/2+\langle\rho e'+\rho'e\rangle/4$. In the single-point limit, we get $R_{\rho e}(0)=\langle\rho e\rangle$, as needed, and the first term in $R_{\rho e}(\bm r)$ describes a single-point contribution required by the presence of a nontrivial uniform background in the homogeneous case. It can be readily shown with data from numerical simulations that using $\tilde{R}_{\rho e}(\bm r)=\langle\rho e'+\rho'e\rangle/2$ instead of  $R_{\rho e}(\bm r)$ would break the acoustic energy equipartition \cite{sarkar...91} at small scales, where turbulence is strongly dominated by the acoustic mode \citep{falkovich.17,kritsuk19}. Moreover, incorrect uniform background removal adds spurious source terms to the scale-by-scale energy balance equation \citep{kritsuk.20}.

We can now use $R_{\rho e}(\bm r)$ to define the thermodynamic energy spectral density for an isothermal fluid as a cospectrum of $\rho$ and $e$, $U(\bm k)=\widehat{R_{\rho e}}(\bm k)=\left[\widehat{\rho}(\bm k)\widehat{e}^*(\bm k)+\widehat{\rho}^*(\bm k)\widehat{e}(\bm k)\right]/4$. From Parseval's rule, we have for the thermodynamic energy of fluctuations: $U=\int U(\bm k)d\bm k$.

\subsubsection{Scale-by-scale energy balance and energy cascades}

As a simple example, let us consider energy cascade in compressible homogeneous isothermal turbulence. The relevant ideal invariant in this case is the total energy, $E=K+U=\langle\rho\bm u^2/2 + \rho e\rangle$, where $e=c_{\rm s}^2\ln(\rho/\rho_0)$. The energy spectral density of turbulent fluctuations is given by $K(k)$ and $U(k)$ defined above. To describe the scale-by-scale energy balance, one can use the same equation we introduced for the incompressible case, $\partial_tE(k)=T(k)-D(k)+F(k)$, but with the transfer, forcing, and dissipation functions taken from \citet{banerjee.17b}. Similar equations can be written for the kinetic and thermodynamic energy balance, but these will include the energy exchange (cross) terms, $X(k)$, which cancel out in the total energy balance: $\partial_tK(k)=T_K(k)-D(k)+F(k)-X_{K\rightarrow U}(k)$ and $\partial_tU(k)=T_U(k)+X_{K\rightarrow U}(k)$. The total transfer function $T(k)=T_K(k)+T_U(k)$. The cross-scale total energy flux is defined in a familiar way, $\Pi(k)=\int_k^{\infty}T(\kappa)d\kappa$, and also includes two components, $\Pi(k)=\Pi_K(k)+\Pi_U(k)$. 

This formalism can be readily used to analyze data from simulations of stationary homogeneous compressible turbulence  \citep[e.g.][]{falkovich.17,kritsuk19,kritsuk.20} and similar analysis can  also be carried out in the configuration space  \citep[e.g.][]{kritsuk..13,kritsuk..15}. By computing statistics of $T(k)$, $D(k)$, and $F(k)$ from the simulation data, one can get direct access to  detailed picture of energy injection, transfer, and dissipation across scales, including the limits on the inertial range, direction of the cascade, possible coexistence of several independent energy cascades, etc. Unique physical definitions of relevant spectral energy densities (or correlation functions) provide means to explore detailed equipartition between various energy components of turbulent fluctuations. It is worth noting that traditional power spectra of density, velocity, and weighted velocity (e.g.\ $\bm v$ or $\bm w$) do not bear a large fraction of that information.

Including self-gravity of the gas and magnetic fields in this analysis does not pose any major technical challenges since both magnetic and gravitational potential energy components are quadratic and Poisson's equation is linear (see \citet{banerjee.17b,banerjee.18}, where the corresponding formalism is developed). It is worth noting that fluctuations of $\bm g$ and $\bm b$ are correlated with other fluctuating quantities in compressible turbulence, hence it is conceptually incorrect to talk about turbulence, gravity, and magnetic fields as separate factors regulating star formation in turbulent molecular clouds. It should be also mentioned that application of virial theorem \citep{mckee.92} to such clouds (or clumps within them), while ignoring the surface terms \citep{dib....07}, would lead to confusing results \citep{ballesteros06,kritsuk..13b} since this is equivalent to setting $\Pi(r)=0$ at the cloud size scale $r$. Thus the whole concept of virial equilibrium in the context of turbulent molecular clouds or their substructure should be taken with a grain of salt. Instead, the formalism describing energy transfer in self-gravitating MHD turbulence developed in \citep{banerjee.17b,banerjee.18,kritsuk..17} should be used. It shows that the role of self-gravity of the gas can be two-fold: (i) it can provide kinetic energy injection in a wide range of scales (wide-band forcing) and (ii) it can work as a trigger of dynamic gravitational collapse at small scales, where collapsing objects decouple from the general turbulent field and the dilatational velocity component $\bm u_d$ gets locally enslaved by the gravitational acceleration $\bm g$, leading to small-scale equipartition of the kinetic energy and gravitational potential energy of the collapsing material, $K(k)\sim W(k)$ at $k>k_{\rm crit}$ \citep{banerjee.17b}.

\subsection{Bibliographical notes}
Our discussion of interstellar turbulence merely provides an overview of the recent progress achieved in the last decade, while the reader is referred to recent books and review articles on the subject. Excellent reviews on interstellar turbulence can be found in \cite{elmegreen.04} and in Chapter 13 of \cite{lequeux2005};  recent accounts of supersonic turbulence in the star formation context are given in \cite{MacLowKlessen2004,mckee.07,HennebelleFalgarone2012,padoan14,federrath18}; The role of magnetic fields in molecular cloud formation and evolution is reviewed in \cite{HennebelleInutsuka2019}. For general theoretical background on incompressible turbulence, see \cite{frisch95}; for magnetohydrodynamic (MHD) turbulence, see \cite{biskamp03,beresnyak.19}; for an up-to-date overview of compressible turbulence, see Chapters 2, 3, and 13-16 in \cite{sagaut.18} and a technical perspective in \cite{chen...15}. Current status of research in MHD turbulence theory and numerical experiments including supersonic MHD turbulence is covered in \cite{beresnyak19,lazarian......20}; recent work on compressibility effects in molecular cloud and MHD turbulence is reviewed in \cite{galtier.18}. Finally, \citet{alexakis.18} is an excellent introduction to cascades and transitions in turbulent fluid flows which are not exactly homogeneous and isotropic, presenting new opportunities for addressing interesting situations in astrophysics, including large-scale turbulence in galactic disks.

\section{Magnetic fields}

\subsection{Introduction and observational facts}

Galaxies and thus the interstellar medium in galaxies are permeated by magnetic fields \citep{FletcherEtAl2011, Beck2015}. First observational evidence for the magnetisation of the ISM dates back to \citet{Hall1949}, \citet{Hiltner1949}, and \citet{DavisGreenstein1951}, who found that polarization of starlight seems to increase with reddening and to be correlated spatially, in
polarization fraction and even more so angle. More recent observations of the interstellar medium and star forming regions allow to quantify the field strength and geometry \citep{Crutcher1999, BourkeEtAl2001, HeilesCrutcher2005, TrolandCrutcher2008, Crutcher2012, Beck2015, Haverkorn2015, PlanckXII2018}. Using polarised synchrotron of electrons, thermal dust emission as well as starlight polarization in extinction, we are able to reconstruct the orientation of the magnetic field in the plane of the sky. The strength can be determined using Faraday rotation and the Zeeman effect, where the former one is mainly used on galactic scales and the diffuse ISM, whereas the latter one is applied in determining the field strength in dense clouds. Observations on galactic scales reveal large-scale fields, which follow the spiral structure of galactic arms \citep{Beck2009, FletcherEtAl2011}. The field strength of this coherent magnetic field component ranges from a few up to a few tens of $\mu\mathrm{G}$ \citep{FletcherEtAl2011}. In the diffuse interstellar medium fields with intensities of $0.1$ to $10\,\mu\mathrm{G}$ have been observed \citep{Crutcher2012} with very little correlation of the field strength with gas density. For column densities above $10^{22}\,\mathrm{cm}^{-2}$ the scatter remains large between individual measurements but there is a clear tendency of the maximum field strength to increases up to the $\mathrm{mG}$ regime. The scaling of the maximum field strength is consistent with the field compression in the ideal MHD approximation, which we will further explain below. The formation of stars is thus likely to be tightly connected to the evolution of magnetic fields. Recent reviews by \citet{HennebelleInutsuka2019} and \citet{KrumholzFederrath2019} focus on the impact of magnetic fields from molecular clouds down to the formation of stars and the resulting initial stellar mass function. The review by \citet{WursterLi2019} explicitly discusses magnetized protostellar discs. In the following we provide a theoretical background and some basic implications of how magnetic fields influence the star formation process.

\subsection{Theoretical background}

\subsubsection{Magneto-hydrodynamics}

The basic equations describing electro-magnetism are Maxwell's equations. In Gaussian cgs units their differential form reads
\begin{align}
\label{eq:Maxwell1}
4 \pi \vektor{j}+\partial \vektor{E} / \partial t &=c \boldsymbol{\nabla} \times \vektor{B} \\
\label{eq:Maxwell2}
\partial \vektor{B} / \partial t &=-c \boldsymbol{\nabla} \times \vektor{E} \\
\label{eq:Maxwell3}
\boldsymbol{\nabla} \cdot \vektor{E} &=4 \pi \rho_\mathrm{e} \\
\label{eq:Maxwell4}
\boldsymbol{\nabla} \cdot \vektor{B} &=0 \end{align}
Here, $\vektor{E}$ and $\vektor{B}$ are the electric and magnetic field vector, $\vektor{j}$ is the electrical current density and $\rho_\mathrm{e}$ the charge density. Equation~\eqref{eq:Maxwell1} is Amp\`{e}re's circuital law, which relates the magnetic field around a closed loop to the electric current passing through the loop. Equation~\eqref{eq:Maxwell2} is known as Faraday's law of induction. Gauss' law in equation~\eqref{eq:Maxwell3} describes the charge density as the source of the electric field. And finally, Equation~\eqref{eq:Maxwell4} describes the magnetic field to be source-free.

Many astrophysical systems are strongly electrically conducting, and so is the interstellar medium. In the limit of an infinite conductivity, this results in an effectively vanishing electric field. Any small electric field would immediately result in a strong current until the electric field has vanished. We note that the electric field only vanishes in the comoving frame of the fluid. An observer moving relative to the fluid with a speed $v$ could observe an electric field,
\begin{equation}
\label{eq:electric-field}
\vektor{E}=-\vektor{v} \times \vektor{B} / c.
\end{equation}
The magnetic field ($\vektor{B}$) and inducition ($\vektor{H}$) are related via the dimensionless relative permeability $\mu$, $\vektor{B}=\mu\vektor{H}$. Similarly, the electric field ($\vektor{E}$) and the electric displacement ($\vektor{D}$) are connected by relative permittivity $\epsilon$, so $\vektor{D}=\epsilon\vektor{E}$. In most astrophysical applications and so in the ISM both proportionality parameters are very close to unity, which allows to ignore the distinction between magnetic field strength and magnetic induction as well as between electric field and electric displacement.

Combining equations \eqref{eq:Maxwell2} and \eqref{eq:electric-field} yields the induction equation
\begin{equation}
\label{eq:induction-eq}
\frac{\partial \vektor{B}}{\partial t}=\boldsymbol{\nabla} \times(\vektor{v} \times \vektor{B}),
\end{equation}
which describes the changes of the magnetic field in the presence of a velocity field $\vektor{v}$ in a perfectly conducting fluid. This approximations is known as \emph{ideal} MHD.

In this limit of ideal MHD we can picture the magnetic field lines to be frozen in the gas flow. This allows the gas and the magnetic field to dynamically interact and transfer momentum and energy. The Lorentz force per unit volume reads
\begin{align}
\vektor{F}_{\mathrm{L}}&=\frac{1}{c} \vektor{j} \times \vektor{B}\\
&=\frac{1}{4 \pi}(\boldsymbol{\nabla} \times \vektor{B}) \times \vektor{B}\\
&=\underbrace{-\frac{1}{8\pi}\boldsymbol{\nabla} \vektor{B}^2}_{\text{pressure term}} + \underbrace{\frac{1}{4\pi}(\vektor{B\cdot\boldsymbol{\nabla}})\vektor{B}}_{\text{curvature term}},
\end{align}
which is a force per unit volume of the fluid exerted on a globally electrically neutral, conducting fluid by the magnetic field. The formulation in the last line illustrates the contribution to the Lorentz force by the gradient of the magnetic pressure $(\vektor{B}^2/8\pi)$ and the curvature of the field. The pressure term can be understood in analogy to thermal pressure where the field lines provide a force per area against compression. The magnetic tension is a restoring force that acts to straighten bent field lines.

The equations of motion for ideal MHD compared to hydrodynamics can be described by adding only the magnetic field to the set of equations, i.e. without reference to other components of the Maxwell's equations. In addition, the induction equation is added as a separate evolution equation for $\vektor{B}$,
\begin{align*}
\frac{\partial\rho}{\partial t} + \boldsymbol{\nabla}\cdot\left(\rho\vektor{v}\right) &= 0\\
\frac{\partial\rho\vektor{v}}{\partial t} + \boldsymbol{\nabla}\cdot\left[\rho\vektor{v}\vektor{v}^\mathrm{T}+\underbrace{\left(P_\mathrm{th} + \frac{\vektor{B}^2}{8\pi}\right)}_{P_\mathrm{tot}}\mathsf{I} - \frac{\vektor{B}\vektor{B}^\mathrm{T}}{4\pi}\right] &= \rho\vektor{g}\\
\frac{\partial e}{\partial t} + \boldsymbol{\nabla}\cdot\left[\left(u + \frac{\rho\vektor{v}^2}{2} + \frac{\vektor{B}^2}{8\pi} + \frac{P_\mathrm{th}}{\rho}\right)\vektor{v} - \frac{\vektor{B}\left(\vektor{v}\cdot\vektor{B}\right)}{4\pi}\right] &= \rho \vektor{v}\cdot\vektor{g} \\
\frac{\partial\vektor{B}}{\partial t} - \boldsymbol{\nabla}\times\left(\vektor{v}\times\vektor{B}\right) &= 0,
\end{align*}
with the gas density $\rho$, the gas velocity $\vektor{v}$, and the gravitational acceleration $\vektor{g}$. The thermal energy density is denoted by $u$, the total energy density 
\begin{equation}
    e = u + \frac{\rho\vektor{v}^2}{2} + \frac{\vektor{B}^2}{8\pi},
\end{equation}
and the thermal pressure by $P_\mathrm{th}$. The notation $\vektor{B}\vektor{B}^\mathrm{T}$ is the dyadic product of two vectors. The total pressure is
\begin{align*}
P_\mathrm{tot} &= P_\mathrm{th} + \frac{\vektor{B}^2}{8\pi}
\end{align*}
and we close the system with the equation of state
\begin{align*}
P_\mathrm{th} = (\gamma-1)\,\rho\,e.
\end{align*}

\subsubsection{Non-ideal MHD effects}

The limit of ideal MHD holds for large regions of the interstellar medium, where the gas is partially ionised. The ions gyrate around the magnetic field lines with the cyclotron frequency and are tied to the field lines. Contrary, the neutrals do not experience the Lorentz force and can move independent of the magnetic field. In principle, this allows for relative motions between the ions and neutrals, which is called \emph{ion-neutral drift} or \emph{ambipolar diffusion} with a drift velocity
\begin{equation}
\vektor{v}_{\mathrm{a}} \equiv \vektor{v}_{\mathrm{i}}-\vektor{v}=\frac{\vektor{F}_{\mathrm{L}}}{\gamma \rho \rho_{\mathrm{i}}} = \frac{1}{4 \pi} \frac{(\nabla \times \vektor{B}) \times \vektor{B}}{\gamma \rho \rho_{\mathrm{i}}}.
\end{equation}
Here, $\rho$ and $\rho_\mathrm{i}$ are the density of the neutrals and the ions, and $\gamma$ is a friction coefficient \citep{MouschoviasPaleologou1981,Balbus2009}. However, ions and neutrals are coupled via collisions and an efficient transfer of momentum between them would effectively also couple the neutrals to the field lines and the relative drift can be small or negligible compared to other speeds in the system. How efficient this coupling is depends on the degree of ionisation. From dimensional arguments the typical time reads
\begin{equation}
\tau_\mathrm{a} = \frac{L}{v_\mathrm{a}} \sim \frac{4\pi\gamma\rho\rho_\mathrm{i}L^2}{\vektor{B}^2}
\end{equation}
with the characteristic length at scale $L$. Using a numerical value of $\gamma=3\times10^{13}\,\mathrm{cm^3\,\mathrm{s^{-1}}\,\mathrm{s^{-1}}}$ \citep{DraineRobergeDalgarno1983} and typical values for a star forming core ($L\sim0.1\,\mathrm{pc}$, $\rho\sim10^{-19}\,\mathrm{g\,cm^{-3}}$, $\rho_\mathrm{i}=10^{-23}\,\mathrm{g\,cm^{-3}}$, $B\sim100\,\mu\mathrm{G}$) this time scale estimate yields $\sim100\,\mathrm{Myr}$, which is much longer than the turbulent crossing time $t_\mathrm{turb}=L/v_\mathrm{turb}\sim0.1\,\mathrm{Myr}$ for a turbulent velocity of $1\,\mathrm{km\,s^{-1}}$. For a lower ion density of the ions (e.g. $\rho_\mathrm{i}=10^{-25}\,\mathrm{g\,cm^{-3}}$) and a five times stronger field both time scales become comparable. At what scale in under what conditions ambipolar diffusion becomes relevant is still debated.

Besides ambipolar diffusion we would like to mention two further processes, namely \emph{Hall drift} and \emph{Ohmic resistivity}, that become important only in the densest regions of star formation such as the in protostellar discs. We refer the reader to the review by \citet{WursterLi2019} for details and their importance on scales below $\sim100\,\mathrm{au}$.
The Hall drift described the effect when the massive particles (ions and charged grains) decouple from the magnetic field, whereas the electrons are still tight to the field lines. This results in relative motions between electrons and ions,
\begin{equation}
\vektor{v}_\mathrm{H} = \vektor{v}_\mathrm{e}-\vektor{v}_\mathrm{i} = -\frac{\vektor{j}}{e\,n_\mathrm{e}} =  -\frac{c}{4\pi}\frac{\boldsymbol{\nabla}\times\vektor{B}}{e\,n_\mathrm{e}}
\end{equation}
with the electron charge $e$ and electron number density $n_\mathrm{e}$.

When all charged components (ions, electrons and charged grains) decouple from the magnetic field, Ohmic resistivity, $\eta_\mathrm{O}$ becomes important. The induction equation can thus be extended to \citep[e.g.][]{Spruit2013,WursterLi2019}
\begin{equation}
\frac{\partial \vektor{B}}{\partial t}=\boldsymbol{\nabla} \times\left[\left(\vektor{v}+\vektor{v}_{\mathrm{H}}+\vektor{v}_{\mathrm{a}}\right) \times \vektor{B}-\eta_\mathrm{O} \boldsymbol{\nabla} \times \vektor{B}\right].
\end{equation}

We would like to highlight that all non-ideal MHD effects allow for a drift between magnetic field lines and the gas. During the collapse of a gas cloud the non-ideal MHD effects therefore allow for a weaker field compared to the ideal MHD approximation.

\subsubsection{Weak and strong fields}

In order to investigate the importance of the field in a dynamical system it is useful to compare quantitatively the energy density in the magnetic field with the thermal and kinetic counterpart.

The magnetization of the gas is accompanied by magnetic waves travelling through the medium. The nature of the magnetic field allows for two types of waves, namely Alfv\'{e}n waves and magneto-sonic waves. In order to determine the speeds of the waves one usually considers a uniform magnetic field in a uniform background material and investigates the transport effects of small perturbations. For a mathematical derivation we refer the reader to \citet{ShuAstroGas1992} and \citet{Spruit2013}. Alv\'{e}n waves are transverse and travel along the field lines because of the magnetic tension like waves travel along a string. Alfv\'{e}n waves propagate at a speed of
\begin{equation}
v_\mathrm{A} = \frac{B}{\sqrt{4\pi\rho}}
\end{equation}
and have the property that pressure and density perturbations vanish in the derivation. This means that Alfv\'{e}n waves are incompressible. Their amplitudes are perpendicular to the unperturbed magnetic field as well as to the direction of propagation.

For magneto-sonic waves the perturbations in pressure and density do not vanish. They can be regarded as the compressive counterpart in analogy to thermal sound waves in a non-magnetised fluid. Magneto-sonic waves are characterized by the angle between the magnetic field and the direction of propagation ($\cos\theta$) as well as the ratio of thermal sound speed $c_\mathrm{s}$ to Alfv\'{e}n speed. The solution of the dispersion relation in the perturbation analysis yields two different modes, which are called the fast mode with a speed
\begin{equation}
    u_\mathrm{f} = \left(c_\mathrm{s}^2 + v_\mathrm{A}^2\right)^{1/2}
\end{equation}
and a slow mode with a speed
\begin{equation}
u_{\mathrm{s}}^{2} = \frac{c_{\mathrm{s}}^{2} v_{\mathrm{A}}^{2}}{c_{\mathrm{s}}^{2}+v_{\mathrm{A}}^{2}} \cos ^{2} \theta.
\end{equation}

An often used quantity is the ratio of thermal to magnetic pressure, which is known as plasma-$\beta$
\begin{equation}
\beta = \frac{P_\mathrm{th}}{P_\mathrm{mag}} \propto \frac{c_\mathrm{s}^2}{v_\mathrm{A}^2},
\end{equation}
where $c_\mathrm{s}$ is the speed of sound. The analogue of the sonic Mach number,
\begin{equation}
\mathcal{M} = \frac{c_\mathrm{s}}{v_\mathrm{A}}.
\end{equation}
is the Alfv\'{e}nic Mach number,
\begin{equation}
\mathcal{M}_\mathrm{A} = \frac{v}{v_\mathrm{A}}.
\end{equation}
For $\beta\gg1$ the effects of the magnetic field are small. Depending on the Mach number, the system is either determined by the thermal pressure ($\mathcal{M}\ll1$) or by the kinetic motions ($\mathcal{M}\gg1$). the latter one is also called the kinematic limit.

\subsection{Magnetic fields and gravity}

In the presence of gravity, it is instructive to compare gravitational and magnetic energy. For a uniform cloud with gravitational energy $E_\mathrm{g}=3GM^2/(5R)$ and magnetic energy $E_\mathrm{mag}=\vektor{B}^2V/(8\pi)$ the ratio
\begin{equation}
\frac{E_\mathrm{g}}{E_\mathrm{mag}} \propto \frac{M^2}{\vektor{B}^2R^4} \propto \left(\frac{M}{\Phi}\right)^2,
\end{equation}
is proportional to the square of the mass-to-flux ratio. It is important to stress the importance of flux freezing in this context. If the gas can drift perpendicular to the field lines the mass of a contracting core can increase without changing the magnetic flux. This ratio is therefore most useful in the the ideal MHD approximation. The mass-to-flux ratio is often expressed in units of the critical value
\begin{equation}
\mu = \frac{M/\Phi}{(M/\Phi)_\mathrm{crit}},
\end{equation}
where the critical value is \citep{MouschoviasSpitzer1976}
\begin{equation}
\left(\frac{M}{\Phi}\right)_\mathrm{crit} = \frac{c_1}{3\pi}\left(\frac{5}{G}\right)^{1/2},
\end{equation}
with a numerical dimensionless parameter $c_1\approx0.53$. Structures with $\mu<1$ are called \emph{magnetically subcritical}, whereas $\mu>1$ refers to \emph{supercritical} regions. It is important to note that the mass is a volume quantity, whereas the magnetic flux is a surface quantity, which makes the critical value to be dependent on the geometry of the region under consideration. The above estimate is based on spherical symmetry and different numerical values have been obtained for spheroids \citep{MouschoviasSpitzer1976} and thin sheets \citep{NakanoNakamura1978}.

\subsection{Field amplification and magnetic dynamo}

There are two fundamental processes that can enhance the magnetic field strength: the first being adiabatic compression (assuming flux freezing), the second being the magnetic dynamo. Whereas adiabatic compression in idealized cases can explain the observed field strength going from galactic scales down to star forming regions, it is unlikely to be the only amplification process: In particular it will not work to attain the $\mu\mathrm{G}$ fields at Galactic scales. Given the complicated (turbulent) motions it is likely that field is also amplified by dynamo processes.
 
We estimate the magnetic field strength based on adiabatic compression. Let us assume that the diffuse gas in molecular clouds ($\rho\sim10\,\mathrm{cm}^{-3}$, $B\sim1\,\mu\mathrm{G}$) is compressed to protostellar cloud densities of $10^6\,\mathrm{cm}^{-3}$ in the turbulent ISM due to gravitational forces. In the case of isotropic motions the magnetic field scales as $\rho^{2/3}$, which would enhance the average field strength to $\sim2\,\mathrm{mG}$. This estimate is at the upper end of the observed relation but still consistent it \citep{Crutcher2012}. 

The adiabatic compression can account locally for strong fields, but the magnetic diffusivity is not entirely zero, even if over a typical dynamical time scale the approximation of ideal MHD is valid. Slow but steady diffusion of field lines with respect to the gas flow will eventually reduce the local field strength by evolving towards a low energy configuration. In addition reconnection of the field lines will result in topological changes of the field structure that overall minimize the energy. Therefore, the dynamical interaction of the fluid flow and a resulting field amplification is needed in order to explain the observed fields. The field amplification can be split into an amplification based on simple fluid flows on the one hand and complex turbulent flows on the other hand, where the latter one is typically referred to as turbulent dynamo. In general, the field amplification and its limitations are relatively complex and we refer to \citet{Spruit2013}, \citet{ChiuderiVelli2015} or \citet{BrandenburgEtAl2012} for further reading. Here, we only illustrate the basic principle of the two main models of the magnetic dynamo.

\paragraph{small-scale dynamo}: The small-scale dynamo naturally acts in plasmas, in which magnetic fields are coupled to the fluid flow. Assume a magnetic flux tube with length $l$ and cross section $A$ embedded in gas with density $\rho$. Conservation of mass implies $\rho A l = \mathrm{constant}$; conservation of magnetic flux in the ideal MHD limit forces $BA$ to remain constant. If velocity fluctuations cause the flux tube to be stretched, the length increases and if the density does not change perceptibly, the cross section decreases, which in turn causes the field strength to increase. This amplification process can continue as long as the time scale for the diffusion of field lines ($\tau_\mathrm{diff}\sim l^2_\mathrm{d}/\eta$) is larger than the dynamical time scale for stretching the flux tubes ($\tau_\mathrm{dyn}\sim l/v$). Equating the two time scales yields $l_\mathrm{d}\sim l R_\mathrm{M}^{-1/2}$, where the magnetic Reynolds number is defined as $R_\mathrm{M}=vl/\eta$. For most of the ISM and star forming regions the magnetic Reynolds number is large, so $l_\mathrm{d}\ll l$, so we expect efficient dynamo action in the ISM \citep{BrandenburgEtAl2012, Subramanian2019}. 

\paragraph{mean field dynamo}:
In order to investigate the evolution of the mean field, we decompose the magnetic field into a mean and a fluctuating component, $\boldsymbol{B}=\overline{\boldsymbol{B}}+\delta\boldsymbol{B}$. The velocity field is decomposed analogously, $\boldsymbol{v}=\overline{\boldsymbol{v}}+\delta\boldsymbol{v}$. The time evolution of the mean field is then given by \citep[e.g.][]{Brandenburg2018,Subramanian2019}
\begin{equation}
\frac{\partial \overline{\boldsymbol{B}}}{\partial t}=\nabla \times(\overline{\boldsymbol{v}} \times \overline{\boldsymbol{B}}+\boldsymbol{\mathcal{E}}-\eta \nabla \times \overline{\boldsymbol{B}}),
\end{equation}
where $\boldsymbol{\mathcal{E}}=\overline{\delta\boldsymbol{v}\times\delta\boldsymbol{B}}$. By choosing an appropriate closure to express $\boldsymbol{\mathcal{E}}$ in terms of the mean field $\overline{\boldsymbol{B}}$ and the mean flow $\overline{\boldsymbol{v}}$, the typical growth times in disc galaxies are of the order of $10^8-10^9\,\mathrm{yr}$, see for example \citet{Subramanian2019}.

\subsection{Effects of magnetic fields in star-forming regions}

The dynamical impact of the magnetic field in the star formation process encompasses several aspects. On galactic scales the strength of the magnetic field is independent of the gas density \citep{TrolandHeiles1986, CrutcherEtAl2010}. The field follows the large scale flows of the galactic rotation and is mainly ordered on scales of hundreds of parsecs \citep{Beck2009, FletcherEtAl2011, Beck2012} with field strengths of a few $\mu\mathrm{G}$. Here, the magnetic pressure supplements the thermal pressure against gravitational compression, which slows down the formation of dense and cold gas \citep[e.g.][]{HillEtAl2012}, in particular molecular gas \citep{GirichidisEtAl2018b}. In addition to this delay the gas structures show smoother distributions if they are magnetized \citep[e.g.][]{PardiEtAl2017}. The individual fragments in the ISM, the filaments and clouds are generally more massive compared to the hydro case because the field reduces the degree of fragmentation from the diffuse gas down to the first hydrostatic core \citep[e.g.][]{CommerconEtAl2011}. In the low-density regime, in which self-gravity is not dominating, the probability distribution function is broadened in the magnetic compared the hydrodynamic case, albeit with a generally weak global impact \citep{molina2012}. The authors compared turbulence simulations with and without magnetic fields showing that mainly the low-density range of the distribution is affected. At high densities the PDF develops a powerlaw tail \citep{KlessenBurkert2001, SlyzEtAl2005, KainulainenEtAl2009, GirichidisEtAl2014, SchneiderEtAl2015b} due to the strong contraction driven by self-gravity. This range is hardy affected by magnetic fields. The reduced degree of fragmentation is propagated from the scale of GMC down to the scales of protostellar cores \citep{CommerconEtAl2011, PetersEtAl2014}.

\section{Gravity}
Concerning gravitational forces we have to consider several aspects. On galactic scales we have to account for the gravitational attraction towards the galactic midplane, which is mainly caused by the disc as a whole and more specifically the stellar disc. In the Milky Way the stellar surface density is a factor of three larger than the gas surface density. For the hot and the warm diffuse gas, self gravity is not important. Turbulent motions and gravity on galactic scales like spiral density waves generate the seeds for molecular clouds that form close to the midplane in the dense sprial arms. Cooling results in lower thermal pressure support, but the molecular clouds as a whole are mostly still not gravitationally bound. Only the densest structures in the clouds are dynamically dominated by self-gravity and start to collapse if the gravitational compression exceeds the opposing pressure forces such as magnetic and thermal contributions and rotational support. Here we focus on the basic principles of self-gravitating isothermal gas dynamics.

\subsection{Ratio of thermal and gravitational energies}
We start by computing the ratio between the thermal energy,
\begin{equation}
E_{\rm therm}= {M \over (\gamma -1) m_p} k_\mathrm{B} T \propto R^3 P \propto R^3\rho^\Gamma\propto R^{3-3\Gamma}
\end{equation}
and the gravitational counterpart $E_{\rm grav} = -(3/5) M^2 G / R $, with the cloud mass 
$M$, the radius $R$, the mean mass per particle $m_p$, the temperature $T$ and the pressure $P$. The Boltzmann constant is $k_B$ and the adiabatic index $\gamma$ depends on the number of internal degrees of freedom of the gas. Assuming a polytropic gas, the thermal pressure is $P = K \rho^{\Gamma}$ with $\Gamma$ being the effective adiabatic exponent including cooling processes. With this expression we find
\begin{eqnarray}
{ E_{\rm therm} \over E_{\rm grav}} \propto R^{4-3 \Gamma}.
\label{ratio_ener}
\end{eqnarray}
The scaling with $R$ revelas that $\Gamma =4/3$ is a critical case below which thermal pressure is unable to support the cloud against gravitational collapse because of a decreasing ratio of thermal support to gravitational energy with increasing radius. While this is true for the isothermal case, $\Gamma=1$, the gravitational collapse will be halted by thermal pressure as soon as the gas is unable to cool efficiently any more, which occurs when the gas becomes optically thick and the heating doe to compression cannot be radiated away. We note that for a monoatomic gas $\Gamma \simeq \gamma = 5/3$ and for a diatomic one $\Gamma \simeq \gamma = 7/5$.

\subsection{Jeans length, Jeans mass and freefall time}
The Jeans length \citep{Jeans1902,lequeux2005} is obtained via a linear analysis of the self-gravitating fluid equations. For a uniform cloud with density $\rho_0$, radius $R$, and sound speed $c_\mathrm{s}$ a linear analysis leads to the dispersion relation
\begin{eqnarray}
\omega^2 = c_\mathrm{s}^2 k ^2 - 4 \pi G \rho_0, 
\label{eq1}
\end{eqnarray}
We note however, that a self-gravitating isothermal cloud cannot have a strictly uniform density because pressure forces have to compensate for the gravitational attraction. For a wave number, $k$, smaller than $\sqrt{4 \pi G \rho_0} / c_\mathrm{s}$, the waves cannot propagate and perturbations are exponentially amplified. This threshold leads to the Jeans length,
\begin{eqnarray}
\lambda_J =  \sqrt{\pi c_\mathrm{s}^2 \over G \rho_0 },
\end{eqnarray}
with the gravitational constant $G$. Physically this result means that self-gravity induces a contraction on a time scale of $1/\sqrt{G \rho_0}$. Thermal pressure counteracts this contraction by reestablishing a uniform density over the time scale of a sound crossing time, $R / c_\mathrm{s}$. If $1/\sqrt{G \rho_0} < R / c_\mathrm{s}$, the thermal pressure cannot erase fluctuations induced by the gravitational forces before the entire cloud collapses.

The Jeans mass is simply defined as the mass contained in a volume with a radius of the Jeans length, $\lambda_J$, and reads 
\begin{eqnarray}
\nonumber
M_J &=& 4 \pi /3 \rho _0 (\lambda_J/2)^3 \\
&=& {\pi ^{5/2} \over 6}  {c_\mathrm{s}^{3} \over \left( G ^3 \rho_0  \right)^{1/2} }.
\label{jeans_mass}
\end{eqnarray}
We note that there is no fundamental justification for this choice within a factor of a few.

Equation~\eqref{jeans_mass} indicates that the Jeans mass decreases with increasing density, assuming an isothermal equation of state. Consequently, during the collapse of a region of a given mass the number of Jeans masses increases as the collapse proceeds. \citet{Hoyle1953} used this argument to propose the concept of recursive fragmentation by which a cloud continues to fragment into more and more condensations as the density increases. However, as shown by eq.~\eqref{eq1}, the growth rate of the gravitational instability decreases with increasing $k$, (since $\omega^2<0$). This means that perturbations at large scales evolve faster than their small scale counterparts. As a result, the recursive fragmentation scenario suffers a timescale problem. A perturbation analysis of the exact solutions of the hydrostatic equilibrium reveals that the growth rate tends to zero for $k \rightarrow 0$ and the fastest growing mode corresponds to a few times the Jeans length (e.g., \citealt{NagaiEtAl1998} for layers or \citealt{FiegePudritz2000} for filaments). This solves the aforementioned time scale issue.

Generally, a solution for a cloud to collapse cannot be computed analytically. However, in the limit of spherical cold cloud of uniform density with a vanishing pressure one can calculate the collapse time exactly, which yields the free-fall time \citep[see e.g.,][]{lequeux2005},
\begin{equation}
\tau _{\rm ff} = \sqrt{ {3 \pi \over 32 G \rho_0 }}.
\end{equation}

\subsection{The smallest Jeans mass in contemporary molecular clouds}
The derivation of the Jeans mass in eq.~(\ref{jeans_mass}) assumes a barotropic equation of state (i.e. the pressure is solely a function of density), which includes isothermal fluids as a particular case, and neglects explicit heating and cooling processes. The hierarchy of fragmentation as mentioned above in (close to) isothermal conditions will halt if the gas is so optically thick or the collapse so fast that the $PdV$ work released during the contraction cannot be radiated away any more. This limit determines the value of the smallest Jeans mass. We follow the approach of \citet{Rees1976} and \citet{WhitworthEtAl2007} to compute this mass limit. Two conditions must be fulfilled. The first requires the size of the condensation, $R$, to be of the order of the Jeans length as explained above. The condition $R \simeq \lambda_J$ leads to
\begin{eqnarray}
R \simeq {6 \over \pi ^2} {G \over c_\mathrm{s}^2} M _J.
\label{cond1}
\end{eqnarray}
The second condition requires the energy released through gravitational contraction to be efficiently radiated. If this condition is not fulfilled the effective adiabatic index, $\Gamma$, will be larger than $4/3$ and thermal pressure will halt the collapse. The heating rate is given by the work of the thermal pressure per unit time $P dV/dt$. For a collpse in approximately freefall, we find $v = dR/dt \simeq \sqrt{2GM/R}$ and therefore
\begin{eqnarray}
-P \frac{dV}{dt} = -\rho c_\mathrm{s}^2 {d \over dt} \left( {4 \pi \over 3} R^3 \right) \simeq {3 c_\mathrm{s}^2 M \over R} \sqrt{GM \over R}.
\end{eqnarray}
The cooling due to radiative losses in the optically thick regime is given by
\citep[e.g.][]{MihalasMihalas1984, HansenEtAl2004, WhitworthEtAl2007}
\begin{eqnarray}
\mathcal{L} = \frac{4 \pi R^2 \sigma T^4}{\tau_\mathrm{eff}}
\end{eqnarray}
where $\tau_\mathrm{eff}$ is the effective optical depth and $\sigma = 2 \pi^5 k_\mathrm{B}^4 / 15 h^3 c^2$ the Stefan-Boltzmann constant. In order to be ravitationally unstable heating and cooling need to balance,
\begin{eqnarray}
-P \frac{dV}{dt} \simeq {3 c_\mathrm{s}^2 M \over R} \sqrt{GM \over R} \simeq 
\mathcal{L} = \frac{4 \pi R^2 \sigma T^4}{\tau_\mathrm{eff}}.
\end{eqnarray}
Solving for the radius yields
\begin{eqnarray}
R  \simeq \left( {3^4 5^2 \over 2^{8} \pi^{12}} \right)^{1/7} \left( {G h^6 c^4 \over c_\mathrm{s}^{12} m_p^8 } \right)^{1/7}
\tau_\mathrm{eff}^{2/7} M^{3/7},
\label{cond2}
\end{eqnarray}
where we assume the mean molecular weight to be unity in the sound speed. Combining the conditions stated by eqs.~(\ref{cond1}) and (\ref{cond2}), we find a characteristic mass of
\begin{align}
M &\simeq \left(5^2 \pi^2 \over 2^{15} 3^3 \right)^{1/4} \left( {h c \over G}\right)^{3/2}
m_p^{-2} \left( {c_\mathrm{s} \over c} \right)^{1/2} \tau_\mathrm{eff}^{1/2}\\
&\simeq {m_\mathrm{Planck}^3 \over m_p^2} \left( {c_\mathrm{s} \over c} \right)^{1/2} \tau_\mathrm{eff}^{1/2}, 
\end{align}
where $m_\mathrm{Planck} = \sqrt{\hbar c /G}$ is the Planck mass. The exact numerical factor depends on the detailed assumptions. For an optical depth $\tau_\mathrm{eff} \simeq 1$, the minimum mass that can collapse of order a few Jupiter masses. \citet{WhitworthStamatellos2006} and \citet{MasunagaInutsuka2000} discuss that $\tau \ge 1$ is not a necessary condition for collapse although it appears reasonable in this context.

\subsection{Equilibrium configurations}
In equilibrium configurations the pressure forces compensate gravitational forces. These static solutions of the fluid equations are of interest in order to test numerical codes and perform more rigorous stability analyses than the Jeans analysis. In some cases they can directly be compared to observations. The equations of hydrostatic equilibrium assuming an isothermal equation of state and the Poisson equation read
\begin{equation}
\label{eq2}
-c_\mathrm{s}^2 \partial _X \rho + \rho \partial_X \phi = 0,
\end{equation}
\begin{equation}
\label{eq3}
{1 \over X ^{D-1}} \partial _X (X^{D-1} \partial_X \phi)  = - 4 \pi G \rho.
\end{equation}
Combining these two equations yields the so-called Lane-Emden equation, 
\begin{equation}
{1 \over X ^{D-1}} \partial _X \left(X^{D-1} {\partial_X \rho \over \rho} \right)  
= - {4 \pi G \over c_\mathrm{s}^2} \rho, 
\label{eq4}
\end{equation}
with the dimension $D$ and the spatial coordinate $X$.  

The one-dimensional configuration ($D=1$, $X$ is the usual Cartesian coordinate, $z$) represents the plane-parallel geometry. The analytical solution of self-gravitating layer has been investigated by \citet{Spitzer1942}.
In the two-dimensional cases corresponding to cylindrical coordinates with $D=2$ and $X$ being the cylindrical radius, the analytical solution of a self-gravitating filament has been obtained by \citet{Ostriker1964}. Both solutions are characterized by a flat density profile near $X=0$. At large $X$ the two solutions differ. Whereas \citet{Spitzer1942} presents an exponential decrease of the density at large $z$, \citet{Ostriker1964} finds a profile decreasing as $r^{-4}$.

In spherical geometry, where $D=3$ and $X$ is the spherical radius, $r$, the solutions of eq.~\eqref{eq4} are called Bonnor-Ebert spheres \citep{Ebert55,Bonnor1956}. In general, the equations cannot be solved analytically but must be obtained numerically. A noticeable exception is the singular isothermal sphere (SIS) whose density profile is given by $\rho_{\rm SIS} = c_\mathrm{s}^2 / (2 \pi G r^2) $. The density profile of the Bonnor-Ebert sphere is flat in the central part and asymptotes toward the profile of the singular isothermal sphere at large radii. In order to obtain a finite radius, the solutions are obtained up to an arbitrarily defined radius, assuming pressure equilibrium diffuse and warm medium outside the cloud. A whole family of equilibrium solutions is obtained, which are parameterized by the density contrast between the central core and the integration edge. Performing a analysis reveals that solutions with a density contrast smaller than about 14 are stable, and unstable otherwise.  

Stability analyses of self-gravitating layers and filaments have been performed (e.g. \citealt{Larson1985}, \citealt{FiegePudritz2000}). Both configurations are unstable to perturbations with a wavelengths of order the Jeans length. This suggests that periodically distributed cores or filaments could form through gravitational instability within self-gravitating filaments and layers. Since the interstellar medium is not in steady state equilibrium but stirred supersonic motions, it is difficult to address this paradigm quantitatively. However, qualitatively, spatially approximately periodically distributed cores and filaments are observed \citep{DutreyEtAl1991, TakahashiEtAl2013, PalauEtAl2018, LuEtAl2018}.

\subsection{Gravitational collapse}
The gravitational collapse of a spherical cloud has been investigated both analytically and numerically. Analytical models have mainly focused on self-similar solutions \citep[e.g.][]{Larson1969, Penston1969, Shu1977, WhitworthSummers1985} which allow the reduction of the non-linear equations to simpler ordinary differential equations. These solutions provide easily time-dependent density and velocity fields and help to understand the physics of the collapse. There are two main types of solutions. The one by \citet{Larson1969} and \citet{Penston1969} describes supersonic infall velocities at large radii ($\simeq 3.3 c_\mathrm{s}$). Contrary, in the solution by \citet{Shu1977} the gas is initially at rest and collapse evolves inside-out. Starting from the centre in which the protostar forms, A rarefaction wave propagates outwards at the sound speed. All self-similar solutions share the property of a constant accretion rate, which is equal to a few up to several times $c_\mathrm{s}^3 / G$. We note that all solutions are characterized by a density field proportional to $r^{-2}$ in the outer part of the cloud and a $r^{-3/2}$ profile in the central region which has been reached by the rarefaction wave. Finally, the density in the Larson-Penston solution is about 8 times larger than the one in the Shu solution at infinite radius. 

There are also a few noticeable numerical solutions. Starting from an spherical cloud with initially uniform density \citet{Larson1969} calculates the gas contraction up to the formation of the protostar including a simplified model of radiative transfer \citep[see also][]{MasunagaInutsuka2000}. The first accretion shock in his simulation forms at the edge of the thermally supported core. This core forms when the dust becomes opaque to its own radiation, i.e. at a density of about $10^{-13}$\,g\,cm$^{-3}$, and is sometimes called the first Larson core. A second accretion shock develops at the edge of the protostar at significantly higher density ($\simeq 10^{-2}$ g cm$^{-3}$). The model by \citet{FosterChevalier1993} starts with a marginally unstable Bonnor-Ebert sphere. In their solution the collapse proceeds very slowly in the outer part of the cloud and only develop subsonic infall motions. Supersonic motions only appear in the inner part of the cloud, which indeed converges towards the Larson-Penston solution. In the outer part of the envelope, the density profile remains close to solution of the SIS. Simulations of triggered collapse have also been investigated by various authors \citep[e.g.][]{HennebelleEtAl2003}. There, typically faster infall velocities are reached and densities a few times higher than the ones of the SIS. Contrary to the self-similar solution, the triggered numerical models all develop strongly varying accretion rate varies over time.

\section{Overview of Stellar Feedback}

Stars impact their environment through a range of energetic processes including radiation, magnetically launched outflows, winds, and supernova explosions. This stellar feedback powers a variety of cosmic processes including heavy element production, evolution of galaxies, reionization of the Universe, formation of planetary systems and ultimately the prevalence of life.

Stellar feedback acts over a broad range of physical scales, carrying mass, momentum and energy from stellar scales ($\sim$ au) up to galactic scales ($\sim$ kpc). Unlike the fundamental processes described above, feedback is not a single physical process but a heterogeneous set of effects that arise from the messy and energetic life cycle of stars. 

In this section, we begin with the feedback from individual stars -- protostellar outflows, radiation and winds -- and then discuss the collective and multiplicative effect of feedback when many stars act together.  It is instructive to visualize this progression as a ``feedback ladder,"  with the various sources ordered based on their energy and scale of influence \citep{Bally2011}. We will begin with the lowest ``rung" of the ladder: protostellar outflows.

\subsection{Protostellar Outflows}

Stars begin to shape their environment during formation. The process of accretion is surprisingly violent, producing significant radiation  (see \S\ref{sec:feedbackrad}) and flinging mass at velocities of 10s-100s of km/s out to $\sim$0.1-1 pc from the forming star in an outflow. Outflows form as a result of rotating gas that winds up the magnetic field lines. Mass coupled to the field is redirected outwards, carrying away angular momentum and thereby facilitating accretion of lower angular momentum material. Here we use the term outflow to refer to the phenomena of collimated mass-loss from young stars, while jet refers to a very narrowly collimated outflow. The first protostellar outflows were discovered in molecular emission in 1980 \citep{Snell+1980,Rodriguez+1980} (see \cite{Bally2016} for a recent detailed review of protostellar outflows). Molecular outflows, predominately observed in $^{12}$CO and $^{13}$CO are typically associated with the youngest and most embedded stage of star formation when the young outflow entrains a significant amount of the surrounding core envelope as it travels away from the star. The result is that molecular outflows are relatively slow moving with velocities of a few to 10s of km\,s$^{-1}$  and gas temperatures of 10s of K. Older, less embedded young stellar objects are associated with highly collimated, optical jets, which achieve velocities of 200 km\,s$^{-1}$ and commonly exhibit atomic line and maser emission \citep{ReipurthBally2001}.

It is now accepted that outflows are a fundamental part of the star formation process and a by-product of the formation of stars ranging from brown dwarfs \citep{Lee+2009,Whelan+2009} to massive stars \citep{ShepherdChurchwell1996,Zhang+2005,Cyganowski+2008}. The ubiquity of outflows suggests that a universal launching mechanism is at work.

Obscuration by dust and gas during the early stages of star formation have frustrated high-resolution observational studies of the expected $\sim au$ launching region. Consequently, analytic models and numerical simulations have provided the primary insights into the launching mechanism. Seminal theoretical work by \cite{BlandfordPayne1982} proposed a general theory to describe outflow launching in hydro-magnetic accreting systems, ranging from protostars to active galactic nuclei. In this model, outflows are caused by magneto-rotational coupling, whereby magnetic fields anchored to rotating accreting gas can centrifugally redirect the gas outwards along open field lines, accelerating it to high velocities. They derived a minimum critical angle of 30\degree between the poloidal component of the magnetic field and the rotation axis for jet launching to occur. Successive work built on these principles to propose models describing outflows from young accreting stars. The ``X-Wind" model developed a formulation for outflows in which the magnetic fields are anchored near the magnetospherical truncation radius of the disk, i.e.\ where the magnetic pressure balances the ram pressure of the accreting material, which is otherhwise known as the ``X-point" \citep{Shu+1988,Shu+1995}. Due to the small launching radius, the predicted velocities are expected to be comparable to Keplerian velocity at the stellar surface, $v_w = \sqrt{GM_*/r_*} \sim 100-200$ km/s with $\sim 30$\% of accreting material being redirected into a well-collimated jet. In contrast, in the ``disk-wind" model, the magnetic field lines are anchored within the disk. and material is launched over a wider range of radii \citep{PelletierPudrtiz1992}. Disk-winds are slower, less well collimated and expected to eject 10\% of accreting material. However, the two mechanisms are not mutually exclusive. High-resolution ALMA observations find both a wide-angle component and a well-collimated, episodic component \citep{Zhang+2016,Zhang+2019}, which suggests both launching mechanisms may be active simultaneously. 

Numerical simulations are necessary to progress beyond idealized, stationary axisymmetric MHD outflow models and to explore the impact of initial conditions \citep{BanerjeePudritz2006,HennebelleFromang2008,Machida+2008,Commercon+2010,Tomida+2010,Price+2012,MachidaHosokawa2013}. However, the launching velocity is sensitive to the simulation resolution \citep{Seifried+2012}, where $\sim$ au resolution is required to produce even slow outflows of 10s of km/s, while $R_\odot$ resolution is required to resolve the launching of the highest velocity material. High-resolution calculations that follow the collapse until the formation of the second core (protostar) find that outflow launching ensues as soon as a compact rotating structure forms, leading to a slow outflow of a few km/s even before the protostar forms \citep{Tomida+2010,Price+2012}. Further advances in computing power are necessary to achieve higher resolution, multi-physics simulations and be able to follow the outflow evolution over star formation timescales of a few $\sim$0.1 Myr.

Outflows impact the star formation process and molecular clouds in a variety of ways. They play an important role in setting the efficiency of star formation \citep{Wang+2010,Hansen+2012,Federrath+2014,Tanaka+2017}, clearing the natal envelope \citep{OffnerArce2014,Zhang+2016}, driving turbulence both locally \citep{OffnerArce2014,OffnerChaban2017} and globally \citep{Nakamura+2007,Carroll+2009,Wang+2010,Hansen+2012} and transporting angular momentum \citep{Bai+2016}. 

\subsection{Radiation} \label{sec:feedbackrad}

Stars emit a significant amount of radiation while forming. Radiation influences the surrounding gas in three main ways: heating, ionization and dynamics, namely via radiation pressure. The latter two effects are relevant only for high-mass protostars ($M >10\,\msun$). We discuss the origin of the radiation and the scope of the impact below.

{\it Origin.} During the earliest stages of protostar formation, nuclear processes have not yet started and gravitational contraction is the source of the energy ultimately emitted as radiation.  During accretion, gas accelerates as it falls into the gravitational potential well of the star. It slams to a halt on the stellar surface in a strong accretion shock. The gas kinetic energy is converted to heat, most of which is radiated away. The resulting accretion luminosity can be expressed as \citep{Stahler+1980,OffnerMcKee2011}:
\begin{equation}
L_{\rm acc} = f_{\rm acc} \frac{GM_* \dot M_*}{R_*} \simeq 6.2 f_{\rm acc} \left( \frac{M_*}{0.25\,\msun} \right) \left(\frac{\dot M_*}{2 \times 10^{-6}\, \msun\,{\rm yr}^{-1}} \right) \left( \frac{2.5\,\rsun}{R_*} \right)\,\lsun,
\end{equation}
where $M_*$ and $R_*$ are the stellar mass and radius, respectively, $\dot M_*$ is the accretion rate, and $f_{\rm acc}$ is an efficiency factor that reflects how efficiently heat is radiated away. This factor is not well-constrained, since it depends on the properties of the shock and accretion flow \citep{OstrikerShu1995}, and it is often assumed that $f_{\rm acc}\simeq$1 \citep{Hartmann+2016}. However, this factor is central to the details of protostellar evolution, radii and inferred ages \citep{Baraffe+2009,Hosokawa+2011}. 

As protostars contract along the Hayashi track (the luminosity-temperature relationship followed by pre-main sequence stars on the Hertzsprung-Russell diagram)  towards the main sequence, they also radiate according to internal processes, including  gravitational contraction and deuterium burning. For low-mass stars, $L_{\rm int}$ is generally negligible compared to the accretion luminosity. Once nuclear processes begin and accretion starts to decline the intrinsic luminosity contributes an increasing larger fraction of the total luminosity. 

Low-mass stars do not begin to fuse hydrogen until reaching the main sequence, but they fuse deuterium once their central temperatures reach $T \simeq 10^6$\,K. In contrast, high-mass stars join the main sequence and begin burning hydrogen while still vigorously accreting. For stars with $M > 10 \msun$, the intrinsic stellar luminosity exceeds the accretion luminosity even at high accretion rates of $\dot M = 10^{-4} \msun$\,yr$^{-1}$ \citep{Krumholz+2009}.

{\it Heating.}  The emitted radiation heats the surrounding gas and the gas temperature declines are
\begin{equation}
T = \left( \frac{L_*}{4 \pi \sigma r^2} \right)^{1/4} \simeq 70 \left( \frac{L_*}{1\, \lsun} \right)^{1/4} \left( \frac{r}{100\,{\rm au}} \right)^{-1/2}\,K,    
\end{equation}
where $r$ is the distance from the emitting source. The radiation may be beamed more asymmetrically in the presence of an accretion disk and outflow cavity, in which radiation is preferentially escapes in the polar direction where the optical depth is lower \citep{YorkeSonnhalter2002,Krumholz2005,Robitaille2011}. 
While the extent of the heating is relatively modest for low-mass stars it is sufficient to increase the stability of the accretion disk, ultimately shaping the stellar initial mass function \citep{offner2014}.

\subsection{Stellar Winds}

Stars of all masses and ages continuously shed mass in a high-velocity wind. The high mass-loss rates 
of winds launched by massive OB-type stars inject significant momentum and energy into the surrounding gas \citep{Churchwell+2006}, aid in cloud dispersal \citep{RogersPittard2013} 
and help to power galaxy evolution \citep{VanDerKruitFeeman2011,Hopkins+2018}. 
The strength and character of stellar winds vary as a function of stellar type and evolutionary phase, spanning a broad range of physical mechanisms. Winds are typically described by two fundamental parameters: $\dot M_w$, the mass-loss rate or mass-loss per unit time, and $v_{\infty}$, the terminal velocity or wind velocity far from the star. We review each of the three main wind-driving mechanisms below. 

{\it Gas-Pressure Driven Winds.} Coronal winds, such as that of our Sun, are powered by gas pressure. The wind is launched by an outwardly increasing temperature gradient $\sim10^4-$a few $10^6$\,K in the solar photosphere that lifts mass from the surface \citep{Parker1958}. 
All cool stars with effective temperatures $T_{\rm eff}\lesssim 6,500$\,K and a sub-surface convection zone have winds driven by gas pressure \citep{LamersCassinelli1999}.  This includes main-sequence stars with spectral types later than F5V or post-main sequence stars with types F5IV-K1III \citep{LamersCassinelli1999}. Although coronal winds achieve terminal velocities of hundreds of km\,s$^{-1}$, the mass-loss rates are quite low, $\dot M <10^{-10}\,\msun$yr\e, so the net wind momentum and impact on the surroundings is small.

{\it Radiation-Pressure Driven Winds.} O, B, and A-type main-sequence stars, giants and supergiants emit prodigious amounts of radiation with net photon momentum that drives mass-loss rates of $\dot M \sim 10^{-9}-10^{-4} \msun$\,yr\e and wind velocities up to 2,000\,km\,s\e \citep{Vink+2001}. Such winds are known as ``line-driven winds," since they are
mediated by optically thick spectral lines. The Doppler effect plays a key role: a velocity gradient between the photosphere and outer stellar atmosphere allows redshifted photons to be absorbed in the outer layers without undergoing significant attenuation by the intervening material thereby contributing to accelerating the outer layers.
 
 The momentum in the wind can be related to the stellar luminosity,
 \begin{equation}
\dot M v_{\infty} = (L_*/c) \tau_w,     
 \end{equation}
 where $\tau_w$ is the wind optical depth \citep{LamersCassinelli1999}. In the limit that all photons are absorbed or scattered by the wind, $\tau_w =1$. In reality the wind is driven by a finite number of optically thick lines, $N_{\rm eff}$, such that the efficiency of momentum transfer is \citep{LamersCassinelli1999}:
 \begin{equation}
\eta_{\rm mom} = \dot M v_{\infty}/ (L_*/c) \simeq N_{\rm eff} v_{\infty}/c,     
 \end{equation}
which depends on the typical optical depth of the lines. Then the terminal velocity is $v_\infty \simeq c/N_{\rm eff}$ for $N_{\rm eff}$ non-overlapping lines. Observed velocities of $v_\infty \sim 10^3$\,km\,s\e attest to the large number of spectral lines contributing to wind acceleration. 
 
 Line-driven winds are sensitive to the stellar metallicity, $Z$, since wind acceleration is powered by the absorption and re-emission of UV photons by heavy ions such as C, N, O, and Fe. Wind mass-loss rates scale as $\dot M \propto Z^m$, where $m\simeq0.7-0.8$ \citep{Vink+2001,Smith2014}. 
 The efficiency of the acceleration also depends on the fractional abundance of metal ions, line optical depths and the amount of small-scale density inhomogeneities, which leads to lower mass-loss rates in early-O and late-type B stars \citep{LucySolomon1970,Smith2014}.
 
Luminous, cool stars, namely cool supergiants, exhibit high mass-loss rates; however, these stars do not have coronae and their winds are not driven by gas pressure like those of other cool stars. Instead, radiation pressure acts on small dust grains that form in relatively cool regions above the photosphere ($T\lesssim1500$\,K) \citep{Smith2014}. The dust opacities are significantly higher than spectral line opacities, allowing efficient acceleration. While the mass-loss rates from such cool stars are relatively high, wind velocities are slow, reaching only a few 10s of km\,s\e.

{\it Magnetically Driven Winds.} Low-mass stars, such as F-type and later, have convective outer envelopes that enable the production of magnetic waves and magnetic energy transfer. Magnetic fields in concert with rotation can boost wind mass-loss rates, momentum and energy via ``magnetic rotator" winds. One side effect of magnetically driven winds is that stellar angular momentum is carried away by the wind, such that stars spin-down over time \citep{WeberDavis1967}. The Sun and other low-mass stars have appreciably lower rotation rates compared to their younger counterparts due to this angular momentum transport mechanism \citep{Kraft1967,Bouvier+2014}. Magnetic rotator winds can also act in conjunction with the wind acceleration mechanisms described above, enhancing the wind strength, especially for young fast-rotating stars (``fast magnetic rotators") \citep{LamersCassinelli1999}.

The wind mass-loss rate depends on the magnetic field strength and the stellar rotation rate, which vary with stellar type and age. Typically, magnetically-driven wind velocities are a few hundred km\,s\e.

\subsection{Supernovae}
Supernovae can be classified into two main types.
In thermonuclear
supernovae, a white dwarf experiences 
runaway nuclear burning after being driven 
over the Chandrasekhar mass by accretion from
a companion star (single-degenerate case) or merging
with another white dwarf \citep[double-degenerate case,][]{Churea14,Diehlea14a,JMS19}.
In core collapse supernovae, the core of a 
massive star collapses after its main sequence 
life \citep{Smartt09}. In both cases, 
typical explosion energies are of the order
$10^{51}$~erg, as inferred from the typically
ejected mass ($1 \msun$) and velocity 
(10,000~km~s$^-1$) of the ejecta. According 
to a widely used historical classification
dating back to \citet{Mink41} and subsequently 
refined, thermonuclear supernovae are mainly of
spectral type~Ia (no hydrogen lines), 
whereas core collapse supernovae
are classified as type~II (strong hydrogen lines). 

Massive stars ($>8-9 M_\odot$) explode as core collapse
supernovae, but not all stars explode \citep{Muellea16,Ebingea19}.
The theoretical models indicate certain mass ranges and possibly
dependence on other parameters that determine if a star 
explodes or collapses to a black hole without explosion.

Stars with initially several ten solar masses that do explode
may sometimes have high explosion energies, up to 
several times $10^{52}$~erg \citep{Mazea14,Heesea15}

\subsection{Cosmic Rays}

Cosmic rays (CRs) are high-energy particles with non-thermal spectral energy distributions \citep{StrongMoskalenkoPtuskin2007, GrenierBlackStrong2015}. Strong shocks can accelerate particles from the thermal bath to super-thermal energies via diffusive shock acceleration \citep{AxfordEtAl1977, Krymskii1977, BlandfordOstriker1978, Bell1978, CaprioliSpitkovsky2014a, MarcowithEtAl2016}, which makes them the most promising site for CR production. The majority of Galactic CRs are produced by shocks in SN remnants, where approximately $5-10$ percent of the SN energy can be converted to the high-energy component \citep{AharonianEtAl2004, AharonianEtAl2006, AharonianEtAl2007, AbdoEtAl2011, AckermannEtAl2013}. This fraction has been constrained using Gamma ray emission, which is emitted during the interactions of CRs with the gas in the ISM. Stellar winds also provide appropriate conditions for CR acceleration. However, the total amount of CRs produced by stellar winds is certainly less than that by SNe \citep{WebbAxfordForman1985}, and the exact number is still a matter of debate. The composition of CRs reflects to first order the abundances in the ISM with protons occupying the largest fraction \citep[e.g.][]{GrenierBlackStrong2015}. The energy distribution ranges from the thermal bath up to particle energies of $10^{20}\,\mathrm{eV}$ \citep[e.g.][]{Zweibel2013}. Most of the energy is in CR protons at an energy of a few GeV. Particles with higher energies are much less abundant, the low-energy counterpart does not carry enough energy per particle. As a result the GeV protons are the CRs that are dynamically relevant because of two reasons. They have comparable energy densities to the magnetic, thermal and kinetic one in the ISM \citep{ferriere01}. In addition the GeV CRs efficiently interact with the magnetic field in the ISM via the streaming instability \citep{KulsrudPearce1969}, which heats the ISM \citep{WienerZweibelOh2013} and transfers enough momentum to drive galactic winds \citep{Zweibel2017}. It is important to note that the cross section of CRs with the gas in the ISM is very small. The momentum transfer by direct interactions that lead to Gamma ray emission are negligible. The dynamical interactions are primarily transferred via the magnetic field, except for CR energies $\lesssim\,\mathrm{MeV}$. Concerning the transport and the detailed impact of CRs in the star formation process including the impact of low-energy CRs on the chemistry we refer the reader to the chapter on CRs in this volume. Here, we only give a brief outline. Being coupled to the magnetic field results in highly anisotropic transport where the particles mainly diffuse or stream along the magnetic fields \citep[see e.g.][]{SkillingStrong1976, CesarskyVolk1978, Chandran2000, PadovaniGalli2011, Padovani2013}. The lack of direct collisions also allows for CRs to move relatively freely with respect to the gas. Consequently, the energy can be redistributed independently of the local gas motions. 

With regard to the impact of CRs it is necessary to distinguish between particles with sub-relativistic energies ($\sim\,\mathrm{MeV}$), the GeV range, and higher energies. Low-energy CRs are an important source of heating and ionization in molecular clouds because their cross section with the thermal gas increases perceptibly \citep{Padovani2009}. They can penetrate deeply into star-forming regions and provide an effective temperature floor. In addition they directly influence chemical reaction chains and alter observables in star-forming cores \citep{IndrioloMcCall2013}. GeV CRs provide dynamical impact on scales of $\gtrsim10-100\,\mathrm{pc}$ by accelerating via their interaction with the magnetic field and driving galactic outflows. Energies above $\sim10^2-10^3\,\mathrm{GeV}$ do not dynamically alter the star formation process.

\begin{acknowledgements}
We thank the staff of the International Space Science Institute (ISSI) for their generous hospitality and for creating a stimulating collaborative environment.
We thank the reviewer for carefully reading the manuscript as well as for her/his valuable comments.
P.G. acknowledges funding from the European Research Council under ERC-CoG grant CRAGSMAN-646955.
S.S.R.O. acknowledges funding from NSF Career grant AST-1650486.
J.M.D.K. gratefully acknowledges funding from the German Research Foundation (DFG) in the form of an Emmy Noether Research Group (grant number KR4801/1-1) and a DFG Sachbeihilfe Grant (grant number KR4801/2-1), from the European Research Council (ERC) under the European Union's Horizon 2020 research and innovation programme via the ERC Starting Grant MUSTANG (grant agreement number 714907), and from Sonderforschungsbereich SFB 881 ``The Milky Way System'' (subproject B2) of the DFG.
R.S.K.\ acknowledges support from the Deutsche Forschungsgemeinschaft via the SFB 881 “The Milky Way System” (subprojects B1, B2, and B8) as well as funding from the Heidelberg Cluster of Excellence STRUCTURES in the framework of Germany’s Excellence Strategy (grant EXC-2181/1 - 390900948).
A.G.K. acknowledges support from the NASA ATP Grant No. 80NSSC18K0561 and NASA TCAN Grant No.  NNH17ZDA001N. 
M.P. acknowledges funding from the INAF PRIN-SKA 2017 program 1.05.01.88.04.
\end{acknowledgements}

\bibliographystyle{spbasic}
%\bibliography{allrefs.bib}

\begin{thebibliography}{469}
\providecommand{\natexlab}[1]{#1}
\providecommand{\url}[1]{{#1}}
\providecommand{\urlprefix}{URL }
\expandafter\ifx\csname urlstyle\endcsname\relax
  \providecommand{\doi}[1]{DOI~\discretionary{}{}{}#1}\else
  \providecommand{\doi}{DOI~\discretionary{}{}{}\begingroup
  \urlstyle{rm}\Url}\fi
\providecommand{\eprint}[2][]{\url{#2}}

\bibitem[{{Abdo} et~al.(2011){Abdo}, {Ackermann}, {Ajello}, {Allafort},
  {Baldini}, {Ballet}, {Barbiellini}, and {Baring}}]{AbdoEtAl2011}
{Abdo} AA, {Ackermann} M, {Ajello} M, et~al. (2011) {Observations of the Young
  Supernova Remnant RX J1713.7-3946 with the Fermi Large Area Telescope}. \apj
  734:28

\bibitem[{{Abgrall} et~al.(2000){Abgrall}, {Roueff}, and {Drira}}]{abg00}
{Abgrall} H, {Roueff} E, and {Drira} I (2000) {Total transition probability and
  spontaneous radiative dissociation of B, C, B' and D states of molecular
  hydrogen}. \aaps 141:297--300

\bibitem[{{Ackermann} et~al.(2013){Ackermann}, {Ajello}, {Allafort}, {Baldini},
  {Ballet}, {Barbiellini}, {Baring}, {Bastieri}, {Bechtol}, {Bellazzini},
  {Blandford}, {Yang}, and {Zimmer}}]{AckermannEtAl2013}
{Ackermann} M, {Ajello} M, {Allafort} A, et~al. (2013) {Detection of the
  Characteristic Pion-Decay Signature in Supernova Remnants}. Science
  339:807--811

\bibitem[{{Aharonian} et~al.(2004){Aharonian}, {Akhperjanian}, {Aye},
  {Bazer-Bachi}, {Beilicke}, {Benbow}, {Berge}, {Berghaus}, {Bernl{\"o}hr},
  {Bolz}, {Boisson}, {Borgmeier}, {Breitling}, {Brown}, {Bussons Gordo},
  {Chadwick}, {Chitnis}, {Chounet}, {Cornils}, {Costamante}, {Degrange},
  {Djannati-Ata{\"\i}}, {O'C. Drury}, {Ergin}, {Espigat}, {Feinstein},
  {Fleury}, {Fontaine}, {Funk}, {Gallant}, {Giebels}, {Gillessen}, {Goret},
  {Guy}, {Hadjichristidis}, {Hauser}, {Heinzelmann}, {Henri}, {Hermann},
  {Hinton}, {Hofmann}, {Holleran}, {Horns}, {de Jager}, {Jung}, {Kh{\'e}lifi},
  {Komin}, {Konopelko}, {Latham}, {Le Gallou}, {Lemoine}, {Lemi{\`e}re},
  {Leroy}, {Lohse}, {Marcowith}, {Masterson}, {McComb}, {de Naurois}, {Nolan},
  {Noutsos}, {Orford}, {Osborne}, {Ouchrif}, {Panter}, {Pelletier}, {Pita},
  {Pohl}, {P{\"u}hlhofer}, {Punch}, {Raubenheimer}, {Raue}, {Raux}, {Rayner},
  {Redondo}, {Reimer}, {Reimer}, {Ripken}, {Rivoal}, {Rob}, {Rolland},
  {Rowell}, {Sahakian}, {Saug{\'e}}, {Schlenker}, {Schlickeiser}, {Schuster},
  {Schwanke}, {Siewert}, {Sol}, {Steenkamp}, {Stegmann}, {Tavernet},
  {Th{\'e}oret}, {Tluczykont}, {van der Walt}, {Vasileiadis}, {Vincent},
  {Visser}, {V{\"o}lk}, and {Wagner}}]{AharonianEtAl2004}
{Aharonian} F, {Akhperjanian} AG, {Aye} KM, et~al. (2004) {Very high energy
  gamma rays from the direction of Sagittarius A$^{*}$}. \aap 425:L13--L17

\bibitem[{{Aharonian} et~al.(2006){Aharonian}, {Akhperjanian}, {Bazer-Bachi},
  {Beilicke}, {Benbow}, {Berge}, {Bernl{\"o}hr}, {Boisson}, {Bolz}, {Borrel},
  {Braun}, {Breitling}, {Brown}, {Chadwick}, {Chounet}, {Cornils},
  {Costamante}, {Degrange}, {Dickinson}, {Djannati-Ata{\"\i}}, {Drury},
  {Dubus}, {Emmanoulopoulos}, {Espigat}, {Feinstein}, {Fontaine}, {Fuchs},
  {Funk}, {Gallant}, {Giebels}, {Gillessen}, {Glicenstein}, {Goret},
  {Hadjichristidis}, {Hauser}, {Heinzelmann}, {Henri}, {Hermann}, {Hinton},
  {Hofmann}, {Holleran}, {Horns}, {Jacholkowska}, {de Jager}, {Kh{\'e}lifi},
  {Komin}, {Konopelko}, {Latham}, {Le Gallou}, {Lemi{\`e}re},
  {Lemoine-Goumard}, {Leroy}, {Lohse}, {Martin}, {Martineau-Huynh},
  {Marcowith}, {Masterson}, {McComb}, {de Naurois}, {Nolan}, {Noutsos},
  {Orford}, {Osborne}, {Ouchrif}, {Panter}, {Pelletier}, {Pita},
  {P{\"u}hlhofer}, {Punch}, {Raubenheimer}, {Raue}, {Raux}, {Rayner}, {Reimer},
  {Reimer}, {Ripken}, {Rob}, {Rolland}, {Rowell}, {Sahakian}, {Saug{\'e}},
  {Schlenker}, {Schlickeiser}, {Schuster}, {Schwanke}, {Siewert}, {Sol},
  {Spangler}, {Steenkamp}, {Stegmann}, {Tavernet}, {Terrier}, {Th{\'e}oret},
  {Tluczykont}, {Vasileiadis}, {Venter}, {Vincent}, {V{\"o}lk}, and
  {Wagner}}]{AharonianEtAl2006}
{Aharonian} F, {Akhperjanian} AG, {Bazer-Bachi} AR, et~al. (2006) {The H.E.S.S.
  Survey of the Inner Galaxy in Very High Energy Gamma Rays}. \apj
  636(2):777--797

\bibitem[{{Aharonian} et~al.(2007){Aharonian}, {Akhperjanian}, {Bazer-Bachi},
  {Behera}, {Beilicke}, {Benbow}, {Berge}, {Bernl{\"o}hr}, {Boisson}, {Bolz},
  {Borrel}, {Boutelier}, {Braun}, {Brion}, {Brown}, {B{\"u}hler},
  {B{\"u}sching}, {Bulik}, {Carrigan}, {Chadwick}, {Clapson}, {Chounet},
  {Coignet}, {Cornils}, {Costamante}, {Degrange}, {Dickinson},
  {Djannati-Ata{\"\i}}, {Domainko}, {Drury}, {Dubus}, {Dyks}, {Egberts},
  {Emmanoulopoulos}, {Espigat}, {Farnier}, {Feinstein}, {Fiasson},
  {F{\"o}rster}, {Fontaine}, {Funk}, {Funk}, {F{\"u}{\ss}ling}, {Gallant},
  {Giebels}, {Glicenstein}, {Gl{\"u}ck}, {Goret}, {Hadjichristidis}, {Hauser},
  {Hauser}, {Heinzelmann}, {Henri}, {Hermann}, {Hinton}, {Hoffmann}, {Hofmann},
  {Holleran}, {Hoppe}, {Horns}, {Jacholkowska}, {de Jager}, {Kendziorra},
  {Kerschhaggl}, {Kh{\'e}lifi}, {Komin}, {Kosack}, {Lamanna}, {Latham}, {Le
  Gallou}, {Lemi{\`e}re}, {Lemoine-Goumard}, {Lenain}, {Lohse}, {Martin},
  {Martineau-Huynh}, {Marcowith}, {Masterson}, {Maurin}, {McComb}, {Moderski},
  {Moulin}, {de Naurois}, {Nedbal}, {Nolan}, {Olive}, {Orford}, {Osborne},
  {Ostrowski}, {Panter}, {Pedaletti}, {Pelletier}, {Petrucci}, {Pita},
  {P{\"u}hlhofer}, {Punch}, {Ranchon}, {Raubenheimer}, {Raue}, {Rayner},
  {Renaud}, {Ripken}, {Rob}, {Rolland}, {Rosier-Lees}, {Rowell}, {Rudak},
  {Ruppel}, {Sahakian}, {Santangelo}, {Saug{\'e}}, {Schlenker}, {Schlickeiser},
  {Schr{\"o}der}, {Schwanke}, {Schwarzburg}, {Schwemmer}, {Shalchi}, {Sol},
  {Spangler}, {Stawarz}, {Steenkamp}, {Stegmann}, {Superina}, {Tam},
  {Tavernet}, {Terrier}, {van Eldik}, {Vasileiadis}, {Venter}, {Vialle},
  {Vincent}, {Vivier}, {V{\"o}lk}, {Volpe}, {Wagner}, {Ward}, and
  {Zdziarski}}]{AharonianEtAl2007}
{Aharonian} F, {Akhperjanian} AG, {Bazer-Bachi} AR, et~al. (2007) {An
  Exceptional Very High Energy Gamma-Ray Flare of PKS 2155-304}. \apjl
  664(2):L71--L74

\bibitem[{{Alexakis} and {Biferale}(2018)}]{alexakis.18}
{Alexakis} A and {Biferale} L (2018) {Cascades and transitions in turbulent
  flows}. Phys Reports 767:1--101

\bibitem[{{Aluie}(2011)}]{aluie11}
{Aluie} H (2011) {Compressible Turbulence: The Cascade and its Locality}. \prl
  106(17):174502

\bibitem[{{Aluie}(2013)}]{aluie13}
{Aluie} H (2013) {Scale decomposition in compressible turbulence}. Physica D
  Nonlinear Phenomena 247(1):54--65

\bibitem[{{Aluie} et~al.(2012){Aluie}, {Li}, and {Li}}]{aluie..12}
{Aluie} H, {Li} S, and {Li} H (2012) {Conservative Cascade of Kinetic Energy in
  Compressible Turbulence}. \apjl 751(2):L29

\bibitem[{{Ames}(1973)}]{ames1973}
{Ames} S (1973) {Magneto-Gravitational and Thermal Instability in the Galactic
  Disk}. \apj 182:387--404

\bibitem[{{Andr{\'e}s} and {Sahraoui}(2017)}]{andres.17}
{Andr{\'e}s} N and {Sahraoui} F (2017) {Alternative derivation of exact law for
  compressible and isothermal magnetohydrodynamics turbulence}. \pre
  96(5):053205

\bibitem[{{Andr{\'e}s} et~al.(2018{\natexlab{a}}){Andr{\'e}s}, {Galtier}, and
  {Sahraoui}}]{andres..18}
{Andr{\'e}s} N, {Galtier} S, and {Sahraoui} F (2018{\natexlab{a}}) {Exact law
  for homogeneous compressible Hall magnetohydrodynamics turbulence}. \pre
  97(1):013204

\bibitem[{{Andr{\'e}s} et~al.(2018{\natexlab{b}}){Andr{\'e}s}, {Sahraoui},
  {Galtier}, {Hadid}, {Dmitruk}, and {Mininni}}]{andres.....18}
{Andr{\'e}s} N, {Sahraoui} F, {Galtier} S, et~al. (2018{\natexlab{b}}) {Energy
  cascade rate in isothermal compressible magnetohydrodynamic turbulence}.
  Journal of Plasma Physics 84(4):905840404

\bibitem[{{Antonia} et~al.(1997){Antonia}, {Ould-Rouis}, {Anselmet}, and
  {Zhu}}]{antonia...97}
{Antonia} RA, {Ould-Rouis} M, {Anselmet} F, et~al. (1997) {Analogy between
  predictions of Kolmogorov and Yaglom}. Journal of Fluid Mechanics
  332(1):395--409

\bibitem[{{Arzoumanian} et~al.(2011){Arzoumanian}, {Andr{\'e}}, {Didelon},
  {K{\"o}nyves}, {Schneider}, {Men'shchikov}, {Sousbie}, {Zavagno}, {Bontemps},
  {di Francesco}, {Griffin}, {Hennemann}, {Hill}, {Kirk}, {Martin}, {Minier},
  {Molinari}, {Motte}, {Peretto}, {Pezzuto}, {Spinoglio}, {Ward-Thompson},
  {White}, and {Wilson}}]{ArzoumanianEtAl2011}
{Arzoumanian} D, {Andr{\'e}} P, {Didelon} P, et~al. (2011) {Characterizing
  interstellar filaments with Herschel in IC 5146}. \aap 529:L6

\bibitem[{Atkins and Friedman(2011)}]{atkins11}
Atkins PW and Friedman RS (2011) Molecular quantum mechanics. Oxford university
  press

\bibitem[{{Axford} et~al.(1977){Axford}, {Leer}, and
  {Skadron}}]{AxfordEtAl1977}
{Axford} WI, {Leer} E, and {Skadron} G (1977) {The acceleration of cosmic rays
  by shock waves}. International Cosmic Ray Conference 11:132--137

\bibitem[{{Bacchini} et~al.(2019){Bacchini}, {Fraternali}, {Pezzulli},
  {Marasco}, {Iorio}, and {Nipoti}}]{BacchiniEtAl2019}
{Bacchini} C, {Fraternali} F, {Pezzulli} G, et~al. (2019) {The volumetric star
  formation law in the Milky Way}. \aap 632:A127

\bibitem[{{Bai} et~al.(2016){Bai}, {Ye}, {Goodman}, and {Yuan}}]{Bai+2016}
{Bai} XN, {Ye} J, {Goodman} J, et~al. (2016) {Magneto-thermal Disk Winds from
  Protoplanetary Disks}. The Astrophysical Journal 818(2):152

\bibitem[{{Bakes} and {Tielens}(1994)}]{bt94}
{Bakes} ELO and {Tielens} AGGM (1994) {The Photoelectric Heating Mechanism for
  Very Small Graphitic Grains and Polycyclic Aromatic Hydrocarbons}. \apj
  427:822

\bibitem[{{Balbus}(2009)}]{Balbus2009}
{Balbus} SA (2009) {Magnetohydrodynamics of Protostellar Disks}. arXiv e-prints
  arXiv:0906.0854

\bibitem[{{Ballesteros-Paredes}(2006)}]{ballesteros06}
{Ballesteros-Paredes} J (2006) {Six myths on the virial theorem for
  interstellar clouds}. \mnras 372(1):443--449

\bibitem[{{Bally}(2011)}]{Bally2011}
{Bally} J (2011) {Observations of Winds, Jets, and Turbulence Generation in
  GMCs}. In: {Alves} J, {Elmegreen} BG, {Girart} JM, et~al. (eds) Computational
  Star Formation, Proceedings of the International Astronomical Union, vol 270,
  pp 247--254

\bibitem[{{Bally}(2016)}]{Bally2016}
{Bally} J (2016) {Protostellar Outflows}. Annual Review of Astronomy and
  Astrophysics 54:491--528

\bibitem[{{Banerjee} and {Pudritz}(2006)}]{BanerjeePudritz2006}
{Banerjee} R and {Pudritz} RE (2006) {Outflows and Jets from Collapsing
  Magnetized Cloud Cores}. \apj 641(2):949--960

\bibitem[{{Banerjee} and {Galtier}(2013)}]{banerjee.13}
{Banerjee} S and {Galtier} S (2013) {Exact relation with two-point correlation
  functions and phenomenological approach for compressible magnetohydrodynamic
  turbulence}. \pre 87(1):013019

\bibitem[{{Banerjee} and {Galtier}(2014)}]{banerjee.14}
{Banerjee} S and {Galtier} S (2014) {A Kolmogorov-like exact relation for
  compressible polytropic turbulence}. Journal of Fluid Mechanics 742:230--242

\bibitem[{{Banerjee} and {Galtier}(2017)}]{banerjee.17a}
{Banerjee} S and {Galtier} S (2017) {An alternative formulation for exact
  scaling relations in hydrodynamic and magnetohydrodynamic turbulence}.
  Journal of Physics A Mathematical General 50(1):015501

\bibitem[{{Banerjee} and {Kritsuk}(2017)}]{banerjee.17b}
{Banerjee} S and {Kritsuk} AG (2017) {Exact relations for energy transfer in
  self-gravitating isothermal turbulence}. \pre 96(5):053116

\bibitem[{{Banerjee} and {Kritsuk}(2018)}]{banerjee.18}
{Banerjee} S and {Kritsuk} AG (2018) {Energy transfer in compressible
  magnetohydrodynamic turbulence for isothermal self-gravitating fluids}. \pre
  97(2):023107

\bibitem[{{Baraffe} et~al.(2009){Baraffe}, {Chabrier}, and
  {Gallardo}}]{Baraffe+2009}
{Baraffe} I, {Chabrier} G, and {Gallardo} J (2009) {Episodic Accretion at Early
  Stages of Evolution of Low-Mass Stars and Brown Dwarfs: A Solution for the
  Observed Luminosity Spread in H-R Diagrams?} \apjl 702(1):L27--L31

\bibitem[{{Batchelor}(1953)}]{batchelor53}
{Batchelor} GK (1953) {The Theory of Homogeneous Turbulence}. Cambridge
  University Press

\bibitem[{{Batchelor}(1967)}]{batchelor67}
{Batchelor} GK (1967) {An Introduction to Fluid Dynamics}. Cambridge University
  Press

\bibitem[{{Beck}(2009)}]{Beck2009}
{Beck} R (2009) {Galactic and extragalactic magnetic fields - a concise
  review}. Astrophysics and Space Sciences Transactions 5:43--47

\bibitem[{{Beck}(2012)}]{Beck2012}
{Beck} R (2012) {Magnetic Fields in Galaxies}. \ssr 166:215--230

\bibitem[{{Beck}(2015)}]{Beck2015}
{Beck} R (2015) {Magnetic fields in spiral galaxies}. \aapr 24:4

\bibitem[{{Begelman} and {McKee}(1990)}]{BegelmanMcKee1990}
{Begelman} MC and {McKee} CF (1990) {Global Effects of Thermal Conduction on
  Two-Phase Media}. \apj 358:375

\bibitem[{{Bell}(1978)}]{Bell1978}
{Bell} AR (1978) {The acceleration of cosmic rays in shock fronts. I}. \mnras
  182:147--156

\bibitem[{{Benzi} et~al.(1995){Benzi}, {Ciliberto}, {Baudet}, and
  {Chavarria}}]{benzi...95}
{Benzi} R, {Ciliberto} S, {Baudet} C, et~al. (1995) {On the scaling of
  three-dimensional homogeneous and isotropic turbulence}. Physica D Nonlinear
  Phenomena 80(4):385--398

\bibitem[{{Beresnyak}(2019)}]{beresnyak19}
{Beresnyak} A (2019) {MHD turbulence}. Living Reviews in Computational
  Astrophysics 5:2

\bibitem[{{Beresnyak} and {Lazarian}(2019)}]{beresnyak.19}
{Beresnyak} A and {Lazarian} A (2019) {Turbulence in Magnetohydrodynamics}

\bibitem[{{Bigiel} et~al.(2008){Bigiel}, {Leroy}, {Walter}, {Brinks}, {de
  Blok}, {Madore}, and {Thornley}}]{BigielEtAl2008}
{Bigiel} F, {Leroy} A, {Walter} F, et~al. (2008) {The Star Formation Law in
  Nearby Galaxies on Sub-Kpc Scales}. \aj 136:2846--2871

\bibitem[{{Biskamp}(2003)}]{biskamp03}
{Biskamp} D (2003) {Magnetohydrodynamic Turbulence}

\bibitem[{{Black}(1981)}]{black81}
{Black} JH (1981) {The physical state of primordial intergalactic clouds}.
  \mnras 197:553--563

\bibitem[{{Black} and {Dalgarno}(1977)}]{bd77}
{Black} JH and {Dalgarno} A (1977) {Models of interstellar clouds. I. The Zeta
  Ophiuchi cloud.} \apjs 34:405--423

\bibitem[{{Blandford} and {Ostriker}(1978)}]{BlandfordOstriker1978}
{Blandford} RD and {Ostriker} JP (1978) {Particle acceleration by astrophysical
  shocks}. \apjl 221:L29--L32

\bibitem[{{Blandford} and {Payne}(1982)}]{BlandfordPayne1982}
{Blandford} RD and {Payne} DG (1982) {Hydromagnetic flows from accretion disks
  and the production of radio jets.} Monthly Notices of the Royal Astronomical
  Society 199:883--903

\bibitem[{{Boldyrev} et~al.(2002){Boldyrev}, {Nordlund}, and
  {Padoan}}]{boldyrev..02}
{Boldyrev} S, {Nordlund} {\r{A}}, and {Padoan} P (2002) {Scaling Relations of
  Supersonic Turbulence in Star-forming Molecular Clouds}. \apj 573(2):678--684

\bibitem[{{Bonnell} et~al.(2001){Bonnell}, {Bate}, {Clarke}, and
  {Pringle}}]{BonnellEtAl2001}
{Bonnell} IA, {Bate} MR, {Clarke} CJ, et~al. (2001) {Competitive accretion in
  embedded stellar clusters}. \mnras 323(4):785--794

\bibitem[{{Bonnor}(1956)}]{Bonnor1956}
{Bonnor} WB (1956) {Boyle's Law and gravitational instability}. \mnras 116:351

\bibitem[{{Bourke} et~al.(2001){Bourke}, {Myers}, {Robinson}, and {Hyland
  }}]{BourkeEtAl2001}
{Bourke} TL, {Myers} PC, {Robinson} G, et~al. (2001) {New OH Zeeman
  Measurements of Magnetic Field Strengths in Molecular Clouds}. \apj
  554(2):916--932

\bibitem[{{Bouvier} et~al.(2014){Bouvier}, {Matt}, {Mohanty}, {Scholz},
  {Stassun}, and {Zanni}}]{Bouvier+2014}
{Bouvier} J, {Matt} SP, {Mohanty} S, et~al. (2014) {Angular Momentum Evolution
  of Young Low-Mass Stars and Brown Dwarfs: Observations and Theory}.
  Protostars and Planets VI pp 433--450

\bibitem[{{Brandenburg}(2018)}]{Brandenburg2018}
{Brandenburg} A (2018) {Advances in mean-field dynamo theory and applications
  to astrophysical turbulence}. Journal of Plasma Physics 84(4):735840404

\bibitem[{{Brandenburg} et~al.(2012){Brandenburg}, {Sokoloff}, and
  {Subramanian}}]{BrandenburgEtAl2012}
{Brandenburg} A, {Sokoloff} D, and {Subramanian} K (2012) {Current Status of
  Turbulent Dynamo Theory. From Large-Scale to Small-Scale Dynamos}. \ssr
  169(1-4):123--157

\bibitem[{{Bressert} et~al.(2010){Bressert}, {Bastian}, {Gutermuth}, {Megeath},
  {Allen}, {Evans}, {Rebull}, {Hatchell}, {Johnstone}, {Bourke}, {Cieza},
  {Harvey}, {Merin}, {Ray}, and {Tothill}}]{BressertEtAl2010}
{Bressert} E, {Bastian} N, {Gutermuth} R, et~al. (2010) {The spatial
  distribution of star formation in the solar neighbourhood: do all stars form
  in dense clusters?} \mnras 409:L54--L58

\bibitem[{{Bron} et~al.(2014){Bron}, {Le Bourlot}, and {Le Petit}}]{bron14}
{Bron} E, {Le Bourlot} J, and {Le Petit} F (2014) {Surface chemistry in the
  interstellar medium. II. H$_{2}$ formation on dust with random temperature
  fluctuations}. \aap 569:A100

\bibitem[{{Brunt} and {Federrath}(2014)}]{brunt.14}
{Brunt} CM and {Federrath} C (2014) {An observational method to measure the
  relative fractions of solenoidal and compressible modes in interstellar
  clouds}. \mnras 442(2):1451--1469

\bibitem[{{Brunt} et~al.(2009){Brunt}, {Heyer}, and {Mac Low}}]{brunt09}
{Brunt} CM, {Heyer} MH, and {Mac Low} MM (2009) {Turbulent driving scales in
  molecular clouds}. \aap 504(3):883--890

\bibitem[{{Brunt} et~al.(2010){Brunt}, {Federrath}, and {Price}}]{brunt..10}
{Brunt} CM, {Federrath} C, and {Price} DJ (2010) {A method for reconstructing
  the variance of a 3D physical field from 2D observations: application to
  turbulence in the interstellar medium}. \mnras 403(3):1507--1515

\bibitem[{{Burton} et~al.(1990){Burton}, {Hollenbach}, and {Tielens}}]{bht90}
{Burton} MG, {Hollenbach} DJ, and {Tielens} AGGM (1990) {Line Emission from
  Clumpy Photodissociation Regions}. \apj 365:620

\bibitem[{{Caprioli} and {Spitkovsky}(2014)}]{CaprioliSpitkovsky2014a}
{Caprioli} D and {Spitkovsky} A (2014) {Simulations of Ion Acceleration at
  Non-relativistic Shocks. I. Acceleration Efficiency}. \apj 783(2):91

\bibitem[{{Carroll} et~al.(2009){Carroll}, {Frank}, {Blackman}, {Cunningham},
  and {Quillen}}]{Carroll+2009}
{Carroll} JJ, {Frank} A, {Blackman} EG, et~al. (2009) {Outflow-Driven
  Turbulence in Molecular Clouds}. The Astrophysical Journal 695(2):1376--1381

\bibitem[{{Caselli} et~al.(1998){Caselli}, {Walmsley}, {Terzieva}, and
  {Herbst}}]{cas98}
{Caselli} P, {Walmsley} CM, {Terzieva} R, et~al. (1998) {The Ionization
  Fraction in Dense Cloud Cores}. \apj 499(1):234--249

\bibitem[{{Cen}(1992)}]{cen92}
{Cen} R (1992) {A Hydrodynamic Approach to Cosmology: Methodology}. \apjs
  78:341

\bibitem[{{Cen} and {Fang}(2006)}]{cf06}
{Cen} R and {Fang} T (2006) {Where Are the Baryons? III. Nonequilibrium Effects
  and Observables}. \apj 650(2):573--591

\bibitem[{{Cesarsky} and {Volk}(1978)}]{CesarskyVolk1978}
{Cesarsky} CJ and {Volk} HJ (1978) {Cosmic Ray Penetration into Molecular
  Clouds}. \aap 70:367

\bibitem[{{Ceverino} et~al.(2010){Ceverino}, {Dekel}, and {Bournaud}}]{g10}
{Ceverino} D, {Dekel} A, and {Bournaud} F (2010) {High-redshift clumpy discs
  and bulges in cosmological simulations}. \mnras 404(4):2151--2169

\bibitem[{{Chabrier}(2003)}]{Chabrier2003}
{Chabrier} G (2003) {Galactic Stellar and Substellar Initial Mass Function}.
  \pasp 115:763--795

\bibitem[{{Chandran}(2000)}]{Chandran2000}
{Chandran} BDG (2000) {Confinement and Isotropization of Galactic Cosmic Rays
  by Molecular-Cloud Magnetic Mirrors When Turbulent Scattering Is Weak}. \apj
  529(1):513--535

\bibitem[{Chandrasekhar(1951)}]{chandrasekhar51a}
Chandrasekhar S (1951) The fluctuations of density in isotropic turbulence.
  Proceedings of the Royal Society of London Series A, Mathematical and
  Physical Sciences 210(1100):18--25,
  \urlprefix\url{http://www.jstor.org/stable/98946}

\bibitem[{{Chassaing}(1985)}]{chassaing85}
{Chassaing} P (1985) {An alternative formulation of the equations of turbulent
  motion for a fluid of variable density}. Journal de Mecanique Theorique et
  Appliquee 4(3):375--389

\bibitem[{{Chassaing} et~al.(2002){Chassaing}, {Antonia}, {Anselmet}, {Joly},
  and {Sarkar}}]{chassaing....02}
{Chassaing} P, {Antonia} R, {Anselmet} F, et~al. (2002) {Variable Density Fluid
  Turbulence}, Fluid Mechanics and its Applications, vol~69. Kluwer Academic
  Ppublishers

\bibitem[{Chen et~al.(2015)Chen, Xia, Wang, and Yang}]{chen...15}
Chen S, Xia Z, Wang J, et~al. (2015) Recent progress in compressible
  turbulence. Acta Mechanica Sinica 31(3):275

\bibitem[{{Chevance} et~al.(2019){Chevance}, {Kruijssen}, {Hygate}, {Schruba},
  {Longmore}, {Groves}, {Henshaw}, {Herrera}, {Hughes}, {Jeffreson}, {Lang},
  {Leroy}, {Meidt}, {Pety}, {Razza}, {Rosolowsky}, {Schinnerer}, {Bigiel},
  {Blanc}, {Emsellem}, {Faesi}, {Glover}, {Haydon}, {Ho}, {Kreckel}, {Lee},
  {Liu}, {Querejeta}, {Saito}, {Sun}, {Usero}, and {Utomo}}]{ChevanceEtAl2019}
{Chevance} M, {Kruijssen} JMD, {Hygate} APS, et~al. (2019) {The lifecycle of
  molecular clouds in nearby star-forming disc galaxies}. \mnras~in~press p
  arXiv:1911.03479

\bibitem[{{Chiuderi} and {Velli}(2015)}]{ChiuderiVelli2015}
{Chiuderi} C and {Velli} M (2015) {Basics of Plasma Astrophysics}

\bibitem[{{Choi} and {Stone}(2012)}]{choi2012}
{Choi} E and {Stone} JM (2012) {The Effect of Anisotropic Conduction on the
  Thermal Instability in the Interstellar Medium}. \apj 747:86

\bibitem[{{Chu} and {Kovasznay}(1957)}]{chu.57}
{Chu} BT and {Kovasznay} LSG (1957) {Non-linear interactions in a viscous
  heat-conducting compressible gas}. Journal of Fluid Mechanics 3:494--514

\bibitem[{{Churazov} et~al.(2014){Churazov}, {Sunyaev}, {Isern},
  {Kn{\"o}dlseder}, {Jean}, {Lebrun}, {Chugai}, {Grebenev}, {Bravo}, {Sazonov},
  and {Renaud}}]{Churea14}
{Churazov} E, {Sunyaev} R, {Isern} J, et~al. (2014) {Cobalt-56
  {\ensuremath{\gamma}}-ray emission lines from the type Ia supernova 2014J}.
  \nat 512(7515):406--408

\bibitem[{{Churchwell} et~al.(2006){Churchwell}, {Povich}, {Allen}, {Taylor},
  {Meade}, {Babler}, {Indebetouw}, {Watson}, {Whitney}, {Wolfire}, {Bania},
  {Benjamin}, {Clemens}, {Cohen}, {Cyganowski}, {Jackson}, {Kobulnicky},
  {Mathis}, {Mercer}, {Stolovy}, {Uzpen}, {Watson}, and
  {Wolff}}]{Churchwell+2006}
{Churchwell} E, {Povich} MS, {Allen} D, et~al. (2006) {The Bubbling Galactic
  Disk}. \apj 649(2):759--778

\bibitem[{{Clark} and {Glover}(2014)}]{clark14}
{Clark} PC and {Glover} SCO (2014) {On column density thresholds and the star
  formation rate}. \mnras 444(3):2396--2414

\bibitem[{{Clark} et~al.(2011){Clark}, {Glover}, {Klessen}, and
  {Bromm}}]{ClarkEtAl2011b}
{Clark} PC, {Glover} SCO, {Klessen} RS, et~al. (2011) {Gravitational
  Fragmentation in Turbulent Primordial Gas and the Initial Mass Function of
  Population III Stars}. \apj 727(2):110

\bibitem[{{Clark} et~al.(2013){Clark}, {Glover}, {Ragan}, {Shetty}, and
  {Klessen}}]{clark13}
{Clark} PC, {Glover} SCO, {Ragan} SE, et~al. (2013) {On the Temperature
  Structure of the Galactic Center Cloud G0.253+0.016}. \apjl 768(2):L34

\bibitem[{{Commer{\c c}on} et~al.(2011){Commer{\c c}on}, {Hennebelle}, and
  {Henning}}]{CommerconEtAl2011}
{Commer{\c c}on} B, {Hennebelle} P, and {Henning} T (2011) {Collapse of Massive
  Magnetized Dense Cores Using Radiation Magnetohydrodynamics: Early
  Fragmentation Inhibition}. \apjl 742:L9

\bibitem[{{Commer{\c{c}}on} et~al.(2010){Commer{\c{c}}on}, {Hennebelle},
  {Audit}, {Chabrier}, and {Teyssier}}]{Commercon+2010}
{Commer{\c{c}}on} B, {Hennebelle} P, {Audit} E, et~al. (2010) {Protostellar
  collapse: radiative and magnetic feedbacks on small-scale fragmentation}.
  \aap 510:L3

\bibitem[{{Congiu} et~al.(2009){Congiu}, {Matar}, {Kristensen}, {Dulieu}, and
  {Lemaire}}]{con09}
{Congiu} E, {Matar} E, {Kristensen} LE, et~al. (2009) {Laboratory evidence for
  the non-detection of excited nascent H$_{2}$ in dark clouds}. \mnras
  397(1):L96--L100

\bibitem[{Cook and Zhou(2002)}]{cook.02}
Cook AW and Zhou Y (2002) Energy transfer in rayleigh-taylor instability. Phys
  Rev E 66:026312,
  \urlprefix\url{https://link.aps.org/doi/10.1103/PhysRevE.66.026312}

\bibitem[{{Crutcher}(1999)}]{Crutcher1999}
{Crutcher} RM (1999) {Magnetic Fields in Molecular Clouds: Observations
  Confront Theory}. \apj 520(2):706--713

\bibitem[{{Crutcher}(2012)}]{Crutcher2012}
{Crutcher} RM (2012) {Magnetic Fields in Molecular Clouds}. \araa 50:29--63

\bibitem[{{Crutcher} et~al.(2010){Crutcher}, {Wandelt}, {Heiles}, {Falgarone},
  and {Troland}}]{CrutcherEtAl2010}
{Crutcher} RM, {Wandelt} B, {Heiles} C, et~al. (2010) {Magnetic Fields in
  Interstellar Clouds from Zeeman Observations: Inference of Total Field
  Strengths by Bayesian Analysis}. \apj 725:466--479

\bibitem[{{Cyganowski} et~al.(2008){Cyganowski}, {Whitney}, {Holden}, {Braden},
  {Brogan}, {Churchwell}, {Indebetouw}, {Watson}, {Babler}, {Benjamin},
  {Gomez}, {Meade}, {Povich}, {Robitaille}, and {Watson}}]{Cyganowski+2008}
{Cyganowski} CJ, {Whitney} BA, {Holden} E, et~al. (2008) {A Catalog of Extended
  Green Objects in the GLIMPSE Survey: A New Sample of Massive Young Stellar
  Object Outflow Candidates}. The Astronomical Journal 136(6):2391--2412

\bibitem[{{Daddi} et~al.(2010){Daddi}, {Elbaz}, {Walter}, {Bournaud}, {Salmi},
  {Carilli}, {Dannerbauer}, {Dickinson}, {Monaco}, and {Riechers}}]{daddi10}
{Daddi} E, {Elbaz} D, {Walter} F, et~al. (2010) {Different Star Formation Laws
  for Disks Versus Starbursts at Low and High Redshifts}. \apjl 714:L118--L122

\bibitem[{{Dale} et~al.(2013){Dale}, {Ercolano}, and
  {Bonnell}}]{DaleErcolanoBonnell2013}
{Dale} JE, {Ercolano} B, and {Bonnell} IA (2013) {Ionizing feedback from
  massive stars in massive clusters - III. Disruption of partially unbound
  clouds}. \mnras 430(1):234--246

\bibitem[{{Dalgarno} and {Black}(1976)}]{db76}
{Dalgarno} A and {Black} JH (1976) {REVIEW: Molecule formation in the
  interstellar gas}. Reports on Progress in Physics 39(6):573--612

\bibitem[{{Dalgarno} et~al.(1999){Dalgarno}, {Yan}, and {Liu}}]{dyl99}
{Dalgarno} A, {Yan} M, and {Liu} W (1999) {Electron Energy Deposition in a Gas
  Mixture of Atomic and Molecular Hydrogen and Helium}. \apjs 125(1):237--256

\bibitem[{{Davidson}(2004)}]{davidson04}
{Davidson} PA (2004) {Turbulence : an introduction for scientists and
  engineers}

\bibitem[{{Davis} and {Greenstein}(1951)}]{DavisGreenstein1951}
{Davis} J Leverett and {Greenstein} JL (1951) {The Polarization of Starlight by
  Aligned Dust Grains.} \apj 114:206

\bibitem[{{de Avillez} and {Breitschwerdt}(2012)}]{db12}
{de Avillez} MA and {Breitschwerdt} D (2012) {The Diagnostic O VI Absorption
  Line in Diffuse Plasmas: Comparison of Non-equilibrium Ionization Structure
  Simulations to FUSE Data}. \apjl 761(2):L19

\bibitem[{{Diamond} et~al.(2010){Diamond}, {Itoh}, and {Itoh}}]{diamond..10}
{Diamond} PH, {Itoh} SI, and {Itoh} K (2010) {Modern Plasma Physics}

\bibitem[{{Dib} et~al.(2007){Dib}, {Kim}, {V{\'a}zquez-Semadeni}, {Burkert},
  and {Shadmehri}}]{dib....07}
{Dib} S, {Kim} J, {V{\'a}zquez-Semadeni} E, et~al. (2007) {The Virial Balance
  of Clumps and Cores in Molecular Clouds}. \apj 661(1):262--284

\bibitem[{{Diehl} et~al.(2014){Diehl}, {Siegert}, {Hillebrandt}, {Grebenev},
  {Greiner}, {Krause}, {Kromer}, {Maeda}, {R{\"o}pke}, and
  {Taubenberger}}]{Diehlea14a}
{Diehl} R, {Siegert} T, {Hillebrandt} W, et~al. (2014) {Early $^{56}$Ni decay
  gamma rays from SN2014J suggest an unusual explosion}. Science 345:1162--1165

\bibitem[{{Donzis} and {Jagannathan}(2013)}]{donzis.13}
{Donzis} DA and {Jagannathan} S (2013) {Fluctuations of thermodynamic variables
  in stationary compressible turbulence}. Journal of Fluid Mechanics
  733:221--244

\bibitem[{{Donzis} and {Panickacheril}(2019)}]{donzis.19}
{Donzis} DA and {Panickacheril} J (2019) {Universality and scaling in
  compressible turbulence}. arXiv e-prints arXiv:1907.07871

\bibitem[{{Draine}(2011)}]{draine11}
{Draine} BT (2011) {Physics of the Interstellar and Intergalactic Medium}

\bibitem[{{Draine} and {Bertoldi}(1996)}]{db96}
{Draine} BT and {Bertoldi} F (1996) {Structure of Stationary Photodissociation
  Fronts}. \apj 468:269

\bibitem[{{Draine} and {Lee}(1984)}]{dl84}
{Draine} BT and {Lee} HM (1984) {Optical Properties of Interstellar Graphite
  and Silicate Grains}. \apj 285:89

\bibitem[{{Draine} and {Li}(2007)}]{draine07}
{Draine} BT and {Li} A (2007) {Infrared Emission from Interstellar Dust. IV.
  The Silicate-Graphite-PAH Model in the Post-Spitzer Era}. \apj
  657(2):810--837

\bibitem[{{Draine} and {Sutin}(1987)}]{ds87}
{Draine} BT and {Sutin} B (1987) {Collisional Charging of Interstellar Grains}.
  \apj 320:803

\bibitem[{{Draine} et~al.(1983){Draine}, {Roberge}, and
  {Dalgarno}}]{DraineRobergeDalgarno1983}
{Draine} BT, {Roberge} WG, and {Dalgarno} A (1983) {Magnetohydrodynamic shock
  waves in molecular clouds.} \apj 264:485--507

\bibitem[{{Dutrey} et~al.(1991){Dutrey}, {Duvert}, {Castets}, {Langer},
  {Bally}, and {Wilson}}]{DutreyEtAl1991}
{Dutrey} A, {Duvert} G, {Castets} A, et~al. (1991) {Periodically spaced
  fragmentation in Orion A.} \aap 247:L9

\bibitem[{{Ebert}(1955)}]{Ebert55}
{Ebert} R (1955) {{\" U}ber die Verdichtung von H I-Gebieten. Mit 5
  Textabbildungen}. \zap 37:217--+

\bibitem[{{Ebinger} et~al.(2019){Ebinger}, {Curtis}, {Fr{\"o}hlich}, {Hempel},
  {Perego}, {Liebend{\"o}rfer}, and {Thielemann}}]{Ebingea19}
{Ebinger} K, {Curtis} S, {Fr{\"o}hlich} C, et~al. (2019) {PUSHing Core-collapse
  Supernovae to Explosions in Spherical Symmetry. II. Explodability and Remnant
  Properties}. \apj 870(1):1

\bibitem[{{Elmegreen} and {Burkert}(2010)}]{eb10}
{Elmegreen} BG and {Burkert} A (2010) {Accretion-Driven Turbulence and the
  Transition to Global Instability in Young Galaxy Disks}. \apj 712(1):294--302

\bibitem[{{Elmegreen} and {Scalo}(2004)}]{elmegreen.04}
{Elmegreen} BG and {Scalo} J (2004) {Interstellar Turbulence I: Observations
  and Processes}. \araa 42:211--273

\bibitem[{{Elphick} et~al.(1992){Elphick}, {Regev}, and
  {Shaviv}}]{ElphickRegevShaviv1992}
{Elphick} C, {Regev} O, and {Shaviv} N (1992) {Dynamics of Fronts in Thermally
  Bistable Fluids}. \apj 392:106

\bibitem[{{Eyink} and {Drivas}(2018)}]{eyink.18}
{Eyink} GL and {Drivas} TD (2018) {Cascades and Dissipative Anomalies in
  Compressible Fluid Turbulence}. Physical Review X 8(1):011022

\bibitem[{{Eyink} and {Sreenivasan}(2006)}]{onsager.06}
{Eyink} GL and {Sreenivasan} KR (2006) {Onsager and the theory of hydrodynamic
  turbulence}. Reviews of Modern Physics 78(1):87--135

\bibitem[{{Falgarone} et~al.(1995){Falgarone}, {Pineau des Forets}, and
  {Roueff}}]{fal95}
{Falgarone} E, {Pineau des Forets} G, and {Roueff} E (1995) {Chemical
  signatures of the intermittency of turbulence in low density interstellar
  clouds.} \aap 300:870

\bibitem[{{Falkovich} and {Kritsuk}(2017)}]{falkovich.17}
{Falkovich} G and {Kritsuk} AG (2017) {How vortices and shocks provide for a
  flux loop in two-dimensional compressible turbulence}. Physical Review Fluids
  2(9):092603

\bibitem[{{Falkovich} et~al.(2010){Falkovich}, {Fouxon}, and
  {Oz}}]{falkovich..10}
{Falkovich} G, {Fouxon} I, and {Oz} Y (2010) {New relations for correlation
  functions in Navier-Stokes turbulence}. Journal of Fluid Mechanics 644:465

\bibitem[{{Faucher-Gigu{\`e}re} et~al.(2013){Faucher-Gigu{\`e}re}, {Quataert},
  and {Hopkins}}]{fauchergiguere13}
{Faucher-Gigu{\`e}re} CA, {Quataert} E, and {Hopkins} PF (2013)
  {Feedback-regulated star formation in molecular clouds and galactic discs}.
  \mnras 433(3):1970--1990

\bibitem[{{Favre}(1983)}]{favre83}
{Favre} A (1983) {Turbulence: Space-time statistical properties and behavior in
  supersonic flows}. Physics of Fluids 26(10):2851--2863

\bibitem[{{Federrath}(2018)}]{federrath18}
{Federrath} C (2018) {The turbulent formation of stars}. Physics Today
  71(6):38--42

\bibitem[{{Federrath} et~al.(2014){Federrath}, {Schr{\"o}n}, {Banerjee}, and
  {Klessen}}]{Federrath+2014}
{Federrath} C, {Schr{\"o}n} M, {Banerjee} R, et~al. (2014) {Modeling Jet and
  Outflow Feedback during Star Cluster Formation}. The Astrophysical Journal
  790(2):128

\bibitem[{{Federrath} et~al.(2016){Federrath}, {Klessen}, {Iapichino}, and
  {Hammer}}]{federrath...16}
{Federrath} C, {Klessen} RS, {Iapichino} L, et~al. (2016) {The world's largest
  turbulence simulations}. arXiv e-prints arXiv:1607.00630

\bibitem[{Feireisl(2004)}]{feireisl04}
Feireisl E (2004) Dynamics of Viscous Compressible Fluids. Oxford University
  Press

\bibitem[{Feireisl et~al.(2016)Feireisl, Karper, and Pokorn\'y}]{feireisl..16}
Feireisl E, Karper TG, and Pokorn\'y M (2016) Mathematical Theory of
  Compressible Viscous Fluids. Lecture Notes in Mathematical Fluid Mechanics,
  Birkh\"auser Basel

\bibitem[{{Feldmann} et~al.(2011){Feldmann}, {Gnedin}, and
  {Kravtsov}}]{feldmann11}
{Feldmann} R, {Gnedin} NY, and {Kravtsov} AV (2011) {How Universal is the
  \textbackslashSigma \_SFR\{--\}\textbackslashSigma \_H\_2 Relation?} \apj
  732(2):115

\bibitem[{{Ferri{\`e}re}(2001)}]{ferriere01}
{Ferri{\`e}re} KM (2001) {The interstellar environment of our galaxy}. Reviews
  of Modern Physics 73:1031--1066

\bibitem[{{Fiege} and {Pudritz}(2000)}]{FiegePudritz2000}
{Fiege} JD and {Pudritz} RE (2000) {Helical fields and filamentary molecular
  clouds - I}. \mnras 311(1):85--104

\bibitem[{{Field}(1965)}]{Field1965}
{Field} GB (1965) {Thermal Instability.} \apj 142:531

\bibitem[{{Field} et~al.(1969){Field}, {Goldsmith}, and {Habing}}]{fgh69}
{Field} GB, {Goldsmith} DW, and {Habing} HJ (1969) {Cosmic-Ray Heating of the
  Interstellar Gas}. \apjl 155:L149

\bibitem[{{Fleck}(1996)}]{fleck96}
{Fleck} J Robert~C (1996) {Scaling Relations for the Turbulent,
  Non--Self-gravitating, Neutral Component of the Interstellar Medium}. \apj
  458:739

\bibitem[{{Fletcher} et~al.(2011){Fletcher}, {Beck}, {Shukurov}, {Berkhuijsen},
  and {Horellou}}]{FletcherEtAl2011}
{Fletcher} A, {Beck} R, {Shukurov} A, et~al. (2011) {Magnetic fields and spiral
  arms in the galaxy M51}. \mnras 412:2396--2416

\bibitem[{{Flower} et~al.(2003){Flower}, {Le Bourlot}, {Pineau des For{\^e}ts},
  and {Cabrit}}]{flower03}
{Flower} DR, {Le Bourlot} J, {Pineau des For{\^e}ts} G, et~al. (2003) {The
  contributions of J-type shocks to the H$_{2}$ emission from molecular outflow
  sources}. \mnras 341(1):70--80

\bibitem[{{Foster} and {Chevalier}(1993)}]{FosterChevalier1993}
{Foster} PN and {Chevalier} RA (1993) {Gravitational Collapse of an Isothermal
  Sphere}. \apj 416:303

\bibitem[{{Frisch}(1995)}]{frisch95}
{Frisch} U (1995) {Turbulence}

\bibitem[{{Fukui} et~al.(2001){Fukui}, {Mizuno}, {Yamaguchi}, {Mizuno}, and
  {Onishi}}]{fukui02}
{Fukui} Y, {Mizuno} N, {Yamaguchi} R, et~al. (2001) {On the Mass Spectrum of
  Giant Molecular Clouds in the Large Magellanic Cloud}. \pasj 53(6):L41--L44

\bibitem[{{Gaensler} et~al.(2008){Gaensler}, {Madsen}, {Chatterjee}, and
  {Mao}}]{gae08}
{Gaensler} BM, {Madsen} GJ, {Chatterjee} S, et~al. (2008) {The Vertical
  Structure of Warm Ionised Gas in the Milky Way}. \pasa 25(4):184--200

\bibitem[{{Galametz} et~al.(2011){Galametz}, {Madden}, {Galliano}, {Hony},
  {Bendo}, and {Sauvage}}]{gala11}
{Galametz} M, {Madden} SC, {Galliano} F, et~al. (2011) {Probing the dust
  properties of galaxies up to submillimetre wavelengths. II. Dust-to-gas mass
  ratio trends with metallicity and the submm excess in dwarf galaxies}. \aap
  532:A56

\bibitem[{Galtier(2018)}]{galtier.18}
Galtier S (2018) Turbulence in space plasmas and beyond. Journal of Physics A:
  Mathematical and Theoretical 51(29):293001

\bibitem[{{Galtier} and {Banerjee}(2011)}]{galtier.11}
{Galtier} S and {Banerjee} S (2011) {Exact Relation for Correlation Functions
  in Compressible Isothermal Turbulence}. \prl 107(13):134501

\bibitem[{{Geen} et~al.(2015){Geen}, {Rosdahl}, {Blaizot}, {Devriendt}, and
  {Slyz}}]{GeenEtAl2015}
{Geen} S, {Rosdahl} J, {Blaizot} J, et~al. (2015) {A detailed study of feedback
  from a massive star}. \mnras 448:3248--3264

\bibitem[{{Genzel} et~al.(2010){Genzel}, {Tacconi}, {Gracia-Carpio},
  {Sternberg}, {Cooper}, {Shapiro}, {Bolatto}, {Bouch{\'e}}, {Bournaud},
  {Burkert}, {Combes}, {Comerford}, {Cox}, {Davis}, {Schreiber},
  {Garcia-Burillo}, {Lutz}, {Naab}, {Neri}, {Omont}, {Shapley}, and
  {Weiner}}]{genzel10}
{Genzel} R, {Tacconi} LJ, {Gracia-Carpio} J, et~al. (2010) {A study of the
  gas-star formation relation over cosmic time}. \mnras 407:2091--2108

\bibitem[{{Girichidis} et~al.(2011){Girichidis}, {Federrath}, {Banerjee}, and
  {Klessen}}]{GirichidisEtAl2011}
{Girichidis} P, {Federrath} C, {Banerjee} R, et~al. (2011) {Importance of the
  initial conditions for star formation - I. Cloud evolution and morphology}.
  \mnras 413:2741--2759

\bibitem[{{Girichidis} et~al.(2012){Girichidis}, {Federrath}, {Banerjee}, and
  {Klessen}}]{GirichidisEtAl2012a}
{Girichidis} P, {Federrath} C, {Banerjee} R, et~al. (2012) {Importance of the
  initial conditions for star formation - II. Fragmentation-induced starvation
  and accretion shielding}. \mnras 420:613--626

\bibitem[{{Girichidis} et~al.(2014){Girichidis}, {Konstandin}, {Whitworth}, and
  {Klessen}}]{GirichidisEtAl2014}
{Girichidis} P, {Konstandin} L, {Whitworth} AP, et~al. (2014) {On the Evolution
  of the Density Probability Density Function in Strongly Self-gravitating
  Systems}. \apj 781:91

\bibitem[{{Girichidis} et~al.(2016){Girichidis}, {Walch}, {Naab}, {Gatto},
  {W{\"u}nsch}, {Glover}, {Klessen}, {Clark}, {Peters}, {Derigs}, and
  {Baczynski}}]{GirichidisEtAl2016b}
{Girichidis} P, {Walch} S, {Naab} T, et~al. (2016) {The SILCC (SImulating the
  LifeCycle of molecular Clouds) project - II. Dynamical evolution of the
  supernova-driven ISM and the launching of outflows}. \mnras 456:3432--3455

\bibitem[{{Girichidis} et~al.(2018){Girichidis}, {Seifried}, {Naab}, {Peters},
  {Walch}, {W{\"u}nsch}, {Glover}, and {Klessen}}]{GirichidisEtAl2018b}
{Girichidis} P, {Seifried} D, {Naab} T, et~al. (2018) {The SILCC project - V.
  The impact of magnetic fields on the chemistry and the formation of molecular
  clouds}. \mnras 480:3511--3540

\bibitem[{{Glassgold} and {Langer}(1975)}]{gl75}
{Glassgold} AE and {Langer} WD (1975) {The C$^{+}$-CO transition in
  interstellar clouds.} \apj 197:347--350

\bibitem[{{Glassgold} et~al.(2012){Glassgold}, {Galli}, and {Padovani}}]{ggp12}
{Glassgold} AE, {Galli} D, and {Padovani} M (2012) {Cosmic-Ray and X-Ray
  Heating of Interstellar Clouds and Protoplanetary Disks}. \apj 756(2):157

\bibitem[{{Glover}(2003)}]{glover03}
{Glover} SCO (2003) {Comparing Gas-Phase and Grain-catalyzed H$_{2}$
  Formation}. \apj 584(1):331--338

\bibitem[{{Glover} and {Clark}(2012{\natexlab{a}})}]{GloverClark2012b}
{Glover} SCO and {Clark} PC (2012{\natexlab{a}}) {Approximations for modelling
  CO chemistry in giant molecular clouds: a comparison of approaches}. \mnras
  421:116--131

\bibitem[{{Glover} and {Clark}(2012{\natexlab{b}})}]{GloverClark2012a}
{Glover} SCO and {Clark} PC (2012{\natexlab{b}}) {Is molecular gas necessary
  for star formation?} \mnras 421:9--19

\bibitem[{{Glover} and {Clark}(2012{\natexlab{c}})}]{gc12c}
{Glover} SCO and {Clark} PC (2012{\natexlab{c}}) {Star formation in metal-poor
  gas clouds}. \mnras 426(1):377--388

\bibitem[{{Glover} and {Jappsen}(2007)}]{gj07}
{Glover} SCO and {Jappsen} AK (2007) {Star Formation at Very Low Metallicity.
  I. Chemistry and Cooling at Low Densities}. \apj 666(1):1--19

\bibitem[{{Glover} and {Mac Low}(2007)}]{gm07b}
{Glover} SCO and {Mac Low} MM (2007) {Simulating the Formation of Molecular
  Clouds. II. Rapid Formation from Turbulent Initial Conditions}. \apj
  659(2):1317--1337

\bibitem[{{Gnat} and {Ferland}(2012)}]{GnatFerland2012}
{Gnat} O and {Ferland} GJ (2012) {Ion-by-ion Cooling Efficiencies}. \apjs
  199:20

\bibitem[{{Gnedin} and {Hollon}(2012)}]{gh12}
{Gnedin} NY and {Hollon} N (2012) {Cooling and Heating Functions of
  Photoionized Gas}. \apjs 202(2):13

\bibitem[{{Godard} et~al.(2009){Godard}, {Falgarone}, and {Pineau Des
  For{\^e}ts}}]{god09}
{Godard} B, {Falgarone} E, and {Pineau Des For{\^e}ts} G (2009) {Models of
  turbulent dissipation regions in the diffuse interstellar medium}. \aap
  495(3):847--867

\bibitem[{{Goldreich} and {Kwan}(1974)}]{gk74}
{Goldreich} P and {Kwan} J (1974) {Molecular Clouds}. \apj 189:441--454

\bibitem[{{Goldsmith}(2001)}]{Goldsmith01}
{Goldsmith} PF (2001) {Molecular Depletion and Thermal Balance in Dark Cloud
  Cores}. \apj 557:736--746

\bibitem[{{Goldsmith} and {Langer}(1978)}]{goldsmith78}
{Goldsmith} PF and {Langer} WD (1978) {Molecular cooling and thermal balance of
  dense interstellar clouds.} \apj 222:881--895

\bibitem[{{Gong} et~al.(2017){Gong}, {Ostriker}, and {Wolfire}}]{gong2017}
{Gong} M, {Ostriker} EC, and {Wolfire} MG (2017) {A Simple and Accurate Network
  for Hydrogen and Carbon Chemistry in the Interstellar Medium}. \apj 843:38

\bibitem[{{Gould} and {Salpeter}(1963)}]{gs63}
{Gould} RJ and {Salpeter} EE (1963) {The Interstellar Abundance of the Hydrogen
  Molecule. I. Basic Processes.} \apj 138:393

\bibitem[{Graham et~al.(2010)Graham, Cameron, and Schüssler}]{graham..10}
Graham JP, Cameron R, and Schüssler M (2010) {TURBULENT} {SMALL}-{SCALE}
  {DYNAMO} {ACTION} {IN} {SOLAR} {SURFACE} {SIMULATIONS}. The Astrophysical
  Journal 714(2):1606--1616

\bibitem[{{Gredel} et~al.(1987){Gredel}, {Lepp}, and {Dalgarno}}]{gld87}
{Gredel} R, {Lepp} S, and {Dalgarno} A (1987) {The C/CO Ratio in Dense
  Interstellar Clouds}. \apjl 323:L137

\bibitem[{{Gredel} et~al.(1989){Gredel}, {Lepp}, {Dalgarno}, and
  {Herbst}}]{gredel89}
{Gredel} R, {Lepp} S, {Dalgarno} A, et~al. (1989) {Cosmic-Ray--induced
  Photodissociation and Photoionization Rates of Interstellar Molecules}. \apj
  347:289

\bibitem[{{Greif} et~al.(2011){Greif}, {Springel}, {White}, {Glover}, {Clark},
  {Smith}, {Klessen}, and {Bromm}}]{GreifEtAl2011}
{Greif} TH, {Springel} V, {White} SDM, et~al. (2011) {Simulations on a Moving
  Mesh: The Clustered Formation of Population III Protostars}. \apj 737(2):75

\bibitem[{{Grenier} et~al.(2015){Grenier}, {Black}, and
  {Strong}}]{GrenierBlackStrong2015}
{Grenier} IA, {Black} JH, and {Strong} AW (2015) {The Nine Lives of Cosmic Rays
  in Galaxies}. \araa 53:199--246

\bibitem[{{Grete} et~al.(2017){Grete}, {O'Shea}, {Beckwith}, {Schmidt}, and
  {Christlieb}}]{grete....17}
{Grete} P, {O'Shea} BW, {Beckwith} K, et~al. (2017) {Energy transfer in
  compressible magnetohydrodynamic turbulence}. Physics of Plasmas 24(9):092311

\bibitem[{{Gutermuth} et~al.(2011){Gutermuth}, {Pipher}, {Megeath}, {Myers},
  {Allen}, and {Allen}}]{gutermuth11}
{Gutermuth} RA, {Pipher} JL, {Megeath} ST, et~al. (2011) {A Correlation between
  Surface Densities of Young Stellar Objects and Gas in Eight Nearby Molecular
  Clouds}. \apj 739:84

\bibitem[{{Habing}(1968)}]{Habing1968}
{Habing} HJ (1968) {The interstellar radiation density between 912 A and 2400
  A}. \bain 19:421

\bibitem[{{Haffner} et~al.(2009){Haffner}, {Dettmar}, {Beckman}, {Wood},
  {Slavin}, {Giammanco}, {Madsen}, {Zurita}, and {Reynolds}}]{HaffnerEtAl2009}
{Haffner} LM, {Dettmar} RJ, {Beckman} JE, et~al. (2009) {The warm ionized
  medium in spiral galaxies}. Reviews of Modern Physics 81(3):969--997

\bibitem[{{Hall}(1949)}]{Hall1949}
{Hall} JS (1949) {Observations of the Polarized Light from Stars}. Science
  109(2825):166--167

\bibitem[{{Hansen} et~al.(2012){Hansen}, {Klein}, {McKee}, and
  {Fisher}}]{Hansen+2012}
{Hansen} CE, {Klein} RI, {McKee} CF, et~al. (2012) {Feedback Effects on
  Low-mass Star Formation}. The Astrophysical Journal 747(1):22

\bibitem[{{Hansen} et~al.(2004){Hansen}, {Kawaler}, and
  {Trimble}}]{HansenEtAl2004}
{Hansen} CJ, {Kawaler} SD, and {Trimble} V (2004) {Stellar interiors : physical
  principles, structure, and evolution}

\bibitem[{{Hartmann} et~al.(2016){Hartmann}, {Herczeg}, and
  {Calvet}}]{Hartmann+2016}
{Hartmann} L, {Herczeg} G, and {Calvet} N (2016) {Accretion onto
  Pre-Main-Sequence Stars}. \araa 54:135--180

\bibitem[{{Haverkorn}(2015)}]{Haverkorn2015}
{Haverkorn} M (2015) {Magnetic Fields in the Milky Way}. In: {Lazarian} A, {de
  Gouveia Dal Pino} EM, and {Melioli} C (eds) Magnetic Fields in Diffuse Media,
  Astrophysics and Space Science Library, vol 407, p 483

\bibitem[{{Hayward} and {Hopkins}(2017)}]{hayward17}
{Hayward} CC and {Hopkins} PF (2017) {How stellar feedback simultaneously
  regulates star formation and drives outflows}. \mnras 465(2):1682--1698

\bibitem[{{Heesen} et~al.(2015){Heesen}, {Brinks}, {Krause}, {Harwood}, {Rau},
  {Rupen}, {Hunter}, {Chyzy}, and {Kitchener}}]{Heesea15}
{Heesen} V, {Brinks} E, {Krause} MGH, et~al. (2015) {The non-thermal
  superbubble in IC 10: the generation of cosmic ray electrons caught in the
  act.} \mnras 447:L1--L5

\bibitem[{{Heiles} and {Crutcher}(2005)}]{HeilesCrutcher2005}
{Heiles} C and {Crutcher} R (2005) {Magnetic Fields in Diffuse HI and Molecular
  Clouds}. In: {Wielebinski} R and {Beck} R (eds) Cosmic Magnetic Fields,
  Lecture Notes in Physics, Berlin Springer Verlag, vol 664, p 137

\bibitem[{Helmholtz(1858)}]{helmholtz1858}
Helmholtz H (1858) Über integrale der hydrodynamischen gleichungen, welche den
  wirbelbewegungen entsprechen. Journal für die reine und angewandte
  Mathematik 55:25--55, \urlprefix\url{http://eudml.org/doc/147720}

\bibitem[{{Hennebelle} and {Falgarone}(2012)}]{HennebelleFalgarone2012}
{Hennebelle} P and {Falgarone} E (2012) {Turbulent molecular clouds}. \aapr
  20:55

\bibitem[{{Hennebelle} and {Fromang}(2008)}]{HennebelleFromang2008}
{Hennebelle} P and {Fromang} S (2008) {Magnetic processes in a collapsing dense
  core. I. Accretion and ejection}. \aap 477(1):9--24

\bibitem[{{Hennebelle} and {Inutsuka}(2019)}]{HennebelleInutsuka2019}
{Hennebelle} P and {Inutsuka} Si (2019) {The role of magnetic field in
  molecular cloud formation and evolution}. Frontiers in Astronomy and Space
  Sciences 6:5

\bibitem[{{Hennebelle} and {P{\'e}rault}(1999)}]{henne1999}
{Hennebelle} P and {P{\'e}rault} M (1999) {Dynamical condensation in a
  thermally bistable flow. Application to interstellar cirrus}. \aap
  351:309--322

\bibitem[{{Hennebelle} et~al.(2003){Hennebelle}, {Whitworth}, {Gladwin}, and
  {Andr{\'e}}}]{HennebelleEtAl2003}
{Hennebelle} P, {Whitworth} AP, {Gladwin} PP, et~al. (2003) {Protostellar
  collapse induced by compression}. \mnras 340(3):870--882

\bibitem[{{Henriksen}(1991)}]{henriksen91}
{Henriksen} RN (1991) {On Molecular Cloud Scaling Laws and Star Formation}.
  \apj 377:500

\bibitem[{{Herrera-Camus} et~al.(2012){Herrera-Camus}, {Fisher}, {Bolatto},
  {Leroy}, {Walter}, {Gordon}, {Roman-Duval}, {Donaldson}, {Mel{\'e}ndez}, and
  {Cannon}}]{hc12}
{Herrera-Camus} R, {Fisher} DB, {Bolatto} AD, et~al. (2012) {Dust-to-gas Ratio
  in the Extremely Metal-poor Galaxy I Zw 18}. \apj 752(2):112

\bibitem[{{Heyer} and {Dame}(2015)}]{HeyerDame2015}
{Heyer} M and {Dame} TM (2015) {Molecular Clouds in the Milky Way}. \araa
  53:583--629

\bibitem[{{Heyer} et~al.(2009){Heyer}, {Krawczyk}, {Duval}, and
  {Jackson}}]{HeyerEtAl2009}
{Heyer} M, {Krawczyk} C, {Duval} J, et~al. (2009) {Re-Examining Larson's
  Scaling Relationships in Galactic Molecular Clouds}. \apj 699:1092--1103

\bibitem[{{Heyer} et~al.(1998){Heyer}, {Brunt}, {Snell}, {Howe}, {Schloerb},
  and {Carpenter}}]{heyer98}
{Heyer} MH, {Brunt} C, {Snell} RL, et~al. (1998) {The Five College Radio
  Astronomy Observatory CO Survey of the Outer Galaxy}. \apjs 115(2):241--258

\bibitem[{{Hill} et~al.(2012){Hill}, {Joung}, {Mac Low}, {Benjamin}, {Haffner},
  {Klingenberg}, and {Waagan}}]{HillEtAl2012}
{Hill} AS, {Joung} MR, {Mac Low} MM, et~al. (2012) {Vertical Structure of a
  Supernova-driven Turbulent, Magnetized Interstellar Medium}. \apj 750:104

\bibitem[{{Hiltner}(1949)}]{Hiltner1949}
{Hiltner} WA (1949) {On the Presence of Polarization in the Continuous
  Radiation of Stars. II.} \apj 109:471

\bibitem[{{Hirano} et~al.(2014){Hirano}, {Hosokawa}, {Yoshida}, {Umeda},
  {Omukai}, {Chiaki}, and {Yorke}}]{HiranoEtAl2014}
{Hirano} S, {Hosokawa} T, {Yoshida} N, et~al. (2014) {One Hundred First Stars:
  Protostellar Evolution and the Final Masses}. \apj 781(2):60

\bibitem[{{Hocuk} and {Spaans}(2010)}]{hs10}
{Hocuk} S and {Spaans} M (2010) {The impact of X-rays on molecular cloud
  fragmentation and the inital mass function}. \aap 522:A24

\bibitem[{{Hollenbach} and {McKee}(1989)}]{hm89}
{Hollenbach} D and {McKee} CF (1989) {Molecule Formation and Infrared Emission
  in Fast Interstellar Shocks. III. Results for J Shocks in Molecular Clouds}.
  \apj 342:306

\bibitem[{{Hopkins} et~al.(2014){Hopkins}, {Kere{\v s}}, {O{\~n}orbe},
  {Faucher-Gigu{\`e}re}, {Quataert}, {Murray}, and {Bullock}}]{HopkinsEtAl2014}
{Hopkins} PF, {Kere{\v s}} D, {O{\~n}orbe} J, et~al. (2014) {Galaxies on FIRE
  (Feedback In Realistic Environments): stellar feedback explains
  cosmologically inefficient star formation}. \mnras 445:581--603

\bibitem[{{Hopkins} et~al.(2018){Hopkins}, {Wetzel}, {Kere{\v{s}}},
  {Faucher-Gigu{\`e}re}, {Quataert}, {Boylan-Kolchin}, {Murray}, {Hayward},
  {Garrison-Kimmel}, {Hummels}, {Feldmann}, {Torrey}, {Ma},
  {Angl{\'e}s-Alc{\'a}zar}, {Su}, {Orr}, {Schmitz}, {Escala}, {Sanderson},
  {Grudi{\'c}}, {Hafen}, {Kim}, {Fitts}, {Bullock}, {Wheeler}, {Chan},
  {Elbert}, and {Narayanan}}]{Hopkins+2018}
{Hopkins} PF, {Wetzel} A, {Kere{\v{s}}} D, et~al. (2018) {FIRE-2 simulations:
  physics versus numerics in galaxy formation}. \mnras 480(1):800--863

\bibitem[{{Hosokawa} et~al.(2011){Hosokawa}, {Offner}, and
  {Krumholz}}]{Hosokawa+2011}
{Hosokawa} T, {Offner} SSR, and {Krumholz} MR (2011) {On the Reliability of
  Stellar Ages and Age Spreads Inferred from Pre-main-sequence Evolutionary
  Models}. \apj 738(2):140

\bibitem[{{Hoyle}(1953)}]{Hoyle1953}
{Hoyle} F (1953) {On the Fragmentation of Gas Clouds Into Galaxies and Stars.}
  \apj 118:513

\bibitem[{{Hoyle} and {Ellis}(1963)}]{he63}
{Hoyle} F and {Ellis} GRA (1963) {On the Existence of an Ionized Layer about
  the Galactic Plane}. Australian Journal of Physics 16:1

\bibitem[{{Indriolo} and {McCall}(2013)}]{IndrioloMcCall2013}
{Indriolo} N and {McCall} BJ (2013) {Cosmic-ray astrochemistry}. Chemical
  Society Reviews 42:7763--7773

\bibitem[{{Inoue} et~al.(2006){Inoue}, {Inutsuka}, and
  {Koyama}}]{InoueInutsukyKoyama2006}
{Inoue} T, {Inutsuka} Si, and {Koyama} H (2006) {Structure and Stability of
  Phase Transition Layers in the Interstellar Medium}. \apj 652(2):1331--1338

\bibitem[{{Inoue} et~al.(2007){Inoue}, {Inutsuka}, and {Koyama}}]{inoue2007}
{Inoue} T, {Inutsuka} Si, and {Koyama} H (2007) {The Role of Ambipolar
  Diffusion in the Formation Process of Moderately Magnetized Diffuse Clouds}.
  \apjl 658:L99--L102

\bibitem[{{Jagannathan} and {Donzis}(2016)}]{jagannathan.16}
{Jagannathan} S and {Donzis} DA (2016) {Reynolds and Mach number scaling in
  solenoidally-forced compressible turbulence using high-resolution direct
  numerical simulations}. Journal of Fluid Mechanics 789:669--707

\bibitem[{{Jeans}(1902)}]{Jeans1902}
{Jeans} JH (1902) {The Stability of a Spherical Nebula}. Royal Society of
  London Philosophical Transactions Series A 199:1--53

\bibitem[{{Jeffreson} and {Kruijssen}(2018)}]{jeffreson18}
{Jeffreson} SMR and {Kruijssen} JMD (2018) {A general theory for the lifetimes
  of giant molecular clouds under the influence of galactic dynamics}. \mnras
  476:3688--3715

\bibitem[{{Jenkins}(2013)}]{jenk13}
{Jenkins} EB (2013) {The Fractional Ionization of the Warm Neutral Interstellar
  Medium}. \apj 764(1):25

\bibitem[{{Jha} et~al.(2019){Jha}, {Maguire}, and {Sullivan}}]{JMS19}
{Jha} SW, {Maguire} K, and {Sullivan} M (2019) {Observational properties of
  thermonuclear supernovae}. Nature Astronomy 3:706--716

\bibitem[{{Joung} et~al.(2009){Joung}, {Mac Low}, and {Bryan}}]{joung09}
{Joung} MR, {Mac Low} MM, and {Bryan} GL (2009) {Dependence of Interstellar
  Turbulent Pressure on Supernova Rate}. \apj 704(1):137--149

\bibitem[{{Jura}(1975)}]{jura75}
{Jura} M (1975) {Interstellar clouds containing optically thin H$_{2}$.} \apj
  197:575--580

\bibitem[{{Kainulainen} et~al.(2009){Kainulainen}, {Beuther}, {Henning}, and
  {Plume}}]{KainulainenEtAl2009}
{Kainulainen} J, {Beuther} H, {Henning} T, et~al. (2009) {Probing the evolution
  of molecular cloud structure. From quiescence to birth}. \aap 508:L35--L38

\bibitem[{{Kalberla} and {Kerp}(2009)}]{kk09}
{Kalberla} PMW and {Kerp} J (2009) {The Hi Distribution of the Milky Way}.
  \araa 47(1):27--61

\bibitem[{de~Karman and Howarth(1938)}]{dekarman.38}
de~Karman T and Howarth L (1938) On the statistical theory of isotropic
  turbulence. Proc Roy Soc London A 164:192--215

\bibitem[{{Kennicutt} and {Evans}(2012)}]{KennicuttEvans2012}
{Kennicutt} RC and {Evans} NJ (2012) {Star Formation in the Milky Way and
  Nearby Galaxies}. \araa 50:531--608

\bibitem[{{Kennicutt}(1989)}]{kennicutt89}
{Kennicutt} RC Jr (1989) {The star formation law in galactic disks}. \apj
  344:685--703

\bibitem[{{Kennicutt}(1998)}]{KennicuttSchmidt1998}
{Kennicutt} RC Jr (1998) {The Global Schmidt Law in Star-forming Galaxies}.
  \apj 498:541--552

\bibitem[{{Kennicutt} et~al.(2007){Kennicutt}, {Calzetti}, {Walter}, {Helou},
  {Hollenbach}, {Armus}, {Bendo}, {Dale}, {Draine}, {Engelbracht}, {Gordon},
  {Prescott}, {Regan}, {Thornley}, {Bot}, {Brinks}, {de Blok}, {de Mello},
  {Meyer}, {Moustakas}, {Murphy}, {Sheth}, and {Smith}}]{kennicutt07}
{Kennicutt} RC Jr, {Calzetti} D, {Walter} F, et~al. (2007) {Star Formation in
  NGC 5194 (M51a). II. The Spatially Resolved Star Formation Law}. \apj
  671:333--348

\bibitem[{Khintchine(1934)}]{khintchine34}
Khintchine A (1934) Korrelationstheorie der station\"aren stochastischen
  prozesse. Mathematische Annalen 109:604--615,
  \urlprefix\url{http://eudml.org/doc/159698}

\bibitem[{{Kida} and {Orszag}(1990)}]{kida.90}
{Kida} S and {Orszag} SA (1990) {Energy and spectral dynamics in forced
  compressible turbulence}. Journal of Scientific Computing 5:85--125

\bibitem[{{Kida} and {Orszag}(1992)}]{kida.92}
{Kida} S and {Orszag} SA (1992) {Energy and spectral dynamics in decaying
  compressible turbulence}. Journal of Scientific Computing 7:1--34

\bibitem[{{Kim} and {Ostriker}(2015)}]{KimOstriker2015}
{Kim} CG and {Ostriker} EC (2015) {Momentum Injection by Supernovae in the
  Interstellar Medium}. \apj 802:99

\bibitem[{{Kim} and {Ryu}(2005)}]{kim.05}
{Kim} J and {Ryu} D (2005) {Density Power Spectrum of Compressible Hydrodynamic
  Turbulent Flows}. \apjl 630(1):L45--L48

\bibitem[{{Kim} and {Kim}(2013)}]{KimKim2013}
{Kim} JG and {Kim} WT (2013) {Instability of Evaporation Fronts in the
  Interstellar Medium}. \apj 779(1):48

\bibitem[{{Klessen} and {Burkert}(2001)}]{KlessenBurkert2001}
{Klessen} RS and {Burkert} A (2001) {The Formation of Stellar Clusters:
  Gaussian Cloud Conditions. II.} \apj 549:386--401

\bibitem[{{Klessen} and {Glover}(2016)}]{klessen16}
{Klessen} RS and {Glover} SCO (2016) {Physical Processes in the Interstellar
  Medium}. Saas-Fee Advanced Course 43:85

\bibitem[{{Klessen} and {Hennebelle}(2010)}]{klessen10}
{Klessen} RS and {Hennebelle} P (2010) {Accretion-driven turbulence as
  universal process: galaxies, molecular clouds, and protostellar disks}. \aap
  520:A17

\bibitem[{{Kolmogorov}(1941{\natexlab{a}})}]{kolmogorov41a}
{Kolmogorov} A (1941{\natexlab{a}}) {The Local Structure of Turbulence in
  Incompressible Viscous Fluid for Very Large Reynolds Numbers}. Akademiia Nauk
  SSSR Doklady 30:301--305

\bibitem[{{Kolmogorov}(1941{\natexlab{b}})}]{kolmogorov41b}
{Kolmogorov} AN (1941{\natexlab{b}}) {Dissipation of Energy in Locally
  Isotropic Turbulence}. Akademiia Nauk SSSR Doklady 32:16

\bibitem[{Kov\'asznay(1953)}]{kovasznay53}
Kov\'asznay LSG (1953) Turbulence in supersonic flow. Journal of the
  Aeronautical Sciences 20(10):657--674,
  \urlprefix\url{https://doi.org/10.2514/8.2793}

\bibitem[{{Kowal} and {Lazarian}(2007)}]{kowal.07}
{Kowal} G and {Lazarian} A (2007) {Scaling Relations of Compressible MHD
  Turbulence}. \apjl 666(2):L69--L72

\bibitem[{{Koyama} and {Inutsuka}(2000)}]{koyama2000}
{Koyama} H and {Inutsuka} SI (2000) {Molecular Cloud Formation in
  Shock-compressed Layers}. \apj 532:980--993

\bibitem[{{Kraft}(1967)}]{Kraft1967}
{Kraft} RP (1967) {Studies of Stellar Rotation. V. The Dependence of Rotation
  on Age among Solar-Type Stars}. \apj 150:551

\bibitem[{{Kraichnan}(1955)}]{kraichnan55}
{Kraichnan} RH (1955) {On the Statistical Mechanics of an Adiabatically
  Compressible Fluid}. Acoustical Society of America Journal 27(3):438

\bibitem[{{Kreckel} et~al.(2018){Kreckel}, {Faesi}, {Kruijssen}, {Schruba},
  {Groves}, {Leroy}, {Bigiel}, {Blanc}, {Chevance}, {Herrera}, {Hughes},
  {McElroy}, {Pety}, {Querejeta}, {Rosolowsky}, {Schinnerer}, {Sun}, {Usero},
  and {Utomo}}]{kreckel18}
{Kreckel} K, {Faesi} C, {Kruijssen} JMD, et~al. (2018) {A 50 pc Scale View of
  Star Formation Efficiency across NGC 628}. \apjl 863:L21

\bibitem[{Kritsuk(2019)}]{kritsuk19}
Kritsuk AG (2019) Energy Transfer and Spectra in Simulations of Two-dimensional
  Compressible Turbulence, ERCOFTAC Series, vol~26, Springer Nature Switzerland
  AG, pp 61--70

\bibitem[{{Kritsuk} and {Banerjee}(2020)}]{kritsuk.20}
{Kritsuk} AG and {Banerjee} S (2020) {Energy Transfer in Subsonic Isothermal
  Turbulence}. Phys Rev Fluids In preparation

\bibitem[{{Kritsuk} and {Norman}(2002)}]{kritsuk.02}
{Kritsuk} AG and {Norman} ML (2002) {Interstellar Phase Transitions Stimulated
  by Time-dependent Heating}. \apjl 580(1):L51--L55

\bibitem[{{Kritsuk} et~al.(2006{\natexlab{a}}){Kritsuk}, {Norman}, and
  {Padoan}}]{kritsuk...06a}
{Kritsuk} AG, {Norman} ML, and {Padoan} P (2006{\natexlab{a}}) {Adaptive Mesh
  Refinement for Supersonic Molecular Cloud Turbulence}. \apjl 638(1):L25--L28

\bibitem[{{Kritsuk} et~al.(2006{\natexlab{b}}){Kritsuk}, {Wagner}, {Norman},
  and {Padoan}}]{kritsuk...06}
{Kritsuk} AG, {Wagner} R, {Norman} ML, et~al. (2006{\natexlab{b}}) {High
  resolution simulations of supersonic turbulence in molecular clouds},
  Astronomical Society of the Pacific Conference Series, vol 359, p~84

\bibitem[{{Kritsuk} et~al.(2007{\natexlab{a}}){Kritsuk}, {Norman}, {Padoan},
  and {Wagner}}]{kritsuk...07a}
{Kritsuk} AG, {Norman} ML, {Padoan} P, et~al. (2007{\natexlab{a}}) {The
  Statistics of Supersonic Isothermal Turbulence}. \apj 665(1):416--431

\bibitem[{{Kritsuk} et~al.(2007{\natexlab{b}}){Kritsuk}, {Padoan}, {Wagner},
  and {Norman}}]{kritsuk...07b}
{Kritsuk} AG, {Padoan} P, {Wagner} R, et~al. (2007{\natexlab{b}}) {Scaling laws
  and intermittency in highly compressible turbulence}. In: {Shaikh} D and
  {Zank} GP (eds) Turbulence and Nonlinear Processes in Astrophysical Plasmas,
  American Institute of Physics Conference Series, vol 932, pp 393--399

\bibitem[{{Kritsuk} et~al.(2009){Kritsuk}, {Ustyugov}, {Norman}, and
  {Padoan}}]{kritsuk...09}
{Kritsuk} AG, {Ustyugov} SD, {Norman} ML, et~al. (2009) {Simulating supersonic
  turbulence in magnetized molecular clouds}. In: Journal of Physics Conference
  Series, Journal of Physics Conference Series, vol 180, p 012020

\bibitem[{{Kritsuk} et~al.(2010){Kritsuk}, {Ustyugov}, {Norman}, and
  {Padoan}}]{kritsuk...10}
{Kritsuk} AG, {Ustyugov} SD, {Norman} ML, et~al. (2010) {Self-organization in
  Turbulent Molecular Clouds: Compressional Versus Solenoidal Modes},
  Astronomical Society of the Pacific Conference Series, vol 429, p~15

\bibitem[{{Kritsuk} et~al.(2013{\natexlab{a}}){Kritsuk}, {Lee}, and
  {Norman}}]{kritsuk..13b}
{Kritsuk} AG, {Lee} CT, and {Norman} ML (2013{\natexlab{a}}) {A supersonic
  turbulence origin of Larson's laws}. \mnras 436(4):3247--3261

\bibitem[{{Kritsuk} et~al.(2013{\natexlab{b}}){Kritsuk}, {Wagner}, and
  {Norman}}]{kritsuk..13}
{Kritsuk} AG, {Wagner} R, and {Norman} ML (2013{\natexlab{b}}) {Energy cascade
  and scaling in supersonic isothermal turbulence}. Journal of Fluid Mechanics
  729:R1

\bibitem[{{Kritsuk} et~al.(2015){Kritsuk}, {Wagner}, and
  {Norman}}]{kritsuk..15}
{Kritsuk} AG, {Wagner} R, and {Norman} ML (2015) {Scaling in Supersonic
  Isothermal Turbulence}, Astronomical Society of the Pacific Conference
  Series, vol 498, p~16

\bibitem[{{Kritsuk} et~al.(2017){Kritsuk}, {Ustyugov}, and
  {Norman}}]{kritsuk..17}
{Kritsuk} AG, {Ustyugov} SD, and {Norman} ML (2017) {The structure and
  statistics of interstellar turbulence}. New Journal of Physics 19(6):065003

\bibitem[{{Kroupa}(2001)}]{Kroupa01}
{Kroupa} P (2001) {On the variation of the initial mass function}. \mnras
  322:231--246

\bibitem[{{Kruijssen}(2012)}]{Kruijssen2012}
{Kruijssen} JMD (2012) {On the fraction of star formation occurring in bound
  stellar clusters}. \mnras 426:3008--3040

\bibitem[{{Kruijssen} and {Longmore}(2014)}]{KruijssenLongmore2014}
{Kruijssen} JMD and {Longmore} SN (2014) {An uncertainty principle for star
  formation - I. Why galactic star formation relations break down below a
  certain spatial scale}. \mnras 439:3239--3252

\bibitem[{{Kruijssen} et~al.(2012){Kruijssen}, {Maschberger}, {Moeckel},
  {Clarke}, {Bastian}, and {Bonnell}}]{KruijssenEtAl2012}
{Kruijssen} JMD, {Maschberger} T, {Moeckel} N, et~al. (2012) {The dynamical
  state of stellar structure in star-forming regions}. \mnras 419:841--853

\bibitem[{{Kruijssen} et~al.(2018){Kruijssen}, {Schruba}, {Hygate}, {Hu},
  {Haydon}, and {Longmore}}]{kruijssen18}
{Kruijssen} JMD, {Schruba} A, {Hygate} APS, et~al. (2018) {An uncertainty
  principle for star formation - II. A new method for characterizing the
  cloud-scale physics of star formation and feedback across cosmic history}.
  \mnras 479:1866--1952

\bibitem[{{Kruijssen} et~al.(2019){Kruijssen}, {Schruba}, {Chevance},
  {Longmore}, {Hygate}, {Haydon}, {McLeod}, {Dalcanton}, {Tacconi}, and {van
  Dishoeck}}]{KruijssenEtAl2019}
{Kruijssen} JMD, {Schruba} A, {Chevance} M, et~al. (2019) {Fast and inefficient
  star formation due to short-lived molecular clouds and rapid feedback}. \nat
  569(7757):519--522

\bibitem[{{Krumholz} and {Federrath}(2019)}]{KrumholzFederrath2019}
{Krumholz} MR and {Federrath} C (2019) {The Role of Magnetic Fields in Setting
  the Star Formation Rate and the Initial Mass Function}. Frontiers in
  Astronomy and Space Sciences 6:7

\bibitem[{{Krumholz} et~al.(2005){Krumholz}, {McKee}, and
  {Klein}}]{Krumholz2005}
{Krumholz} MR, {McKee} CF, and {Klein} RI (2005) {How Protostellar Outflows
  Help Massive Stars Form}. \apjl 618(1):L33--L36

\bibitem[{{Krumholz} et~al.(2009{\natexlab{a}}){Krumholz}, {Klein}, {McKee},
  {Offner}, and {Cunningham}}]{Krumholz+2009}
{Krumholz} MR, {Klein} RI, {McKee} CF, et~al. (2009{\natexlab{a}}) {The
  Formation of Massive Star Systems by Accretion}. Science 323(5915):754

\bibitem[{{Krumholz} et~al.(2009{\natexlab{b}}){Krumholz}, {McKee}, and
  {Tumlinson}}]{krumholz09b}
{Krumholz} MR, {McKee} CF, and {Tumlinson} J (2009{\natexlab{b}}) {The
  Atomic-to-Molecular Transition in Galaxies. II: H I and H$_{2}$ Column
  Densities}. \apj 693:216--235

\bibitem[{{Krumholz} et~al.(2009{\natexlab{c}}){Krumholz}, {McKee}, and
  {Tumlinson}}]{krumholz09c}
{Krumholz} MR, {McKee} CF, and {Tumlinson} J (2009{\natexlab{c}}) {The Star
  Formation Law in Atomic and Molecular Gas}. \apj 699:850--856

\bibitem[{{Krumholz} et~al.(2011){Krumholz}, {Leroy}, and {McKee}}]{klm11}
{Krumholz} MR, {Leroy} AK, and {McKee} CF (2011) {Which Phase of the
  Interstellar Medium Correlates with the Star Formation Rate?} \apj 731(1):25

\bibitem[{{Krumholz} et~al.(2012){Krumholz}, {Dekel}, and {McKee}}]{krumholz12}
{Krumholz} MR, {Dekel} A, and {McKee} CF (2012) {A Universal, Local Star
  Formation Law in Galactic Clouds, nearby Galaxies, High-redshift Disks, and
  Starbursts}. \apj 745:69

\bibitem[{{Krumholz} et~al.(2018){Krumholz}, {Burkhart}, {Forbes}, and
  {Crocker}}]{krumholz18}
{Krumholz} MR, {Burkhart} B, {Forbes} JC, et~al. (2018) {A unified model for
  galactic discs: star formation, turbulence driving, and mass transport}.
  \mnras 477(2):2716--2740

\bibitem[{{Krumholz} et~al.(2019){Krumholz}, {McKee}, and {Bland
  -Hawthorn}}]{krumholz2019}
{Krumholz} MR, {McKee} CF, and {Bland -Hawthorn} J (2019) {Star Clusters Across
  Cosmic Time}. \araa 57:227--303

\bibitem[{{Krymskii}(1977)}]{Krymskii1977}
{Krymskii} GF (1977) {A regular mechanism for the acceleration of charged
  particles on the front of a shock wave}. Akademiia Nauk SSSR Doklady
  234:1306--1308

\bibitem[{{Kulsrud} and {Pearce}(1969)}]{KulsrudPearce1969}
{Kulsrud} R and {Pearce} WP (1969) {The Effect of Wave-Particle Interactions on
  the Propagation of Cosmic Rays}. \apj 156:445

\bibitem[{{Lada} and {Lada}(2003)}]{LadaLada2003}
{Lada} CJ and {Lada} EA (2003) {Embedded Clusters in Molecular Clouds}. \araa
  41:57--115

\bibitem[{{Lada} et~al.(2013){Lada}, {Lombardi}, {Roman-Zuniga}, {Forbrich},
  and {Alves}}]{lada13}
{Lada} CJ, {Lombardi} M, {Roman-Zuniga} C, et~al. (2013) {Schmidt's Conjecture
  and Star Formation in Molecular Clouds}. \apj 778(2):133

\bibitem[{{Lamers} and {Cassinelli}(1999)}]{LamersCassinelli1999}
{Lamers} HJGLM and {Cassinelli} JP (1999) {Introduction to Stellar Winds}

\bibitem[{{Landau} and {Lifshitz}(1987)}]{landau.87}
{Landau} LD and {Lifshitz} EM (1987) {Fluid Mechanics}

\bibitem[{{Langer}(1976)}]{l76}
{Langer} W (1976) {The carbon monoxide abundance in interstellar clouds.} \apj
  206:699--712

\bibitem[{{Larson}(1969)}]{Larson1969}
{Larson} RB (1969) {Numerical calculations of the dynamics of collapsing
  proto-star}. \mnras 145:271

\bibitem[{{Larson}(1981)}]{larson81}
{Larson} RB (1981) {Turbulence and star formation in molecular clouds}. \mnras
  194:809--826

\bibitem[{{Larson}(1985)}]{Larson1985}
{Larson} RB (1985) {Cloud fragmentation and stellar masses.} \mnras
  214:379--398

\bibitem[{{Latter} and {Black}(1991)}]{latter91}
{Latter} WB and {Black} JH (1991) {Molecular Hydrogen Formation by Excited Atom
  Radiative Association}. \apj 372:161

\bibitem[{Lazarian et~al.(2020)Lazarian, Eyink, Jafari, Kowal, Li, Xu, and
  Vishniac}]{lazarian......20}
Lazarian A, Eyink GL, Jafari A, et~al. (2020) 3d turbulent reconnection:
  Theory, tests, and astrophysical implications. Physics of Plasmas
  27(1):012305, \urlprefix\url{https://doi.org/10.1063/1.5110603}

\bibitem[{{Le Bourlot} et~al.(2012){Le Bourlot}, {Le Petit}, {Pinto}, {Roueff},
  and {Roy}}]{leb12}
{Le Bourlot} J, {Le Petit} F, {Pinto} C, et~al. (2012) {Surface chemistry in
  the interstellar medium. I. H$_{2}$ formation by Langmuir-Hinshelwood and
  Eley-Rideal mechanisms}. \aap 541:A76

\bibitem[{{Lee} et~al.(2009){Lee}, {Hirano}, {Palau}, {Ho}, {Bourke}, {Zhang},
  and {Shang}}]{Lee+2009}
{Lee} CF, {Hirano} N, {Palau} A, et~al. (2009) {Rotation and Outflow Motions in
  the Very Low-Mass Class 0 Protostellar System HH 211 at Subarcsecond
  Resolution}. The Astrophysical Journal 699(2):1584--1594

\bibitem[{Lees and Aluie(2019)}]{lees.19}
Lees A and Aluie H (2019) Baropycnal work: A mechanism for energy transfer
  across scales. Fluids 4(2):92,
  \urlprefix\url{http://dx.doi.org/10.3390/fluids4020092}

\bibitem[{{Lele}(1994)}]{lele94}
{Lele} SK (1994) {Compressibility effects on turbulence}. Annual Review of
  Fluid Mechanics 26:211--254

\bibitem[{{Lequeux}(2005)}]{lequeux2005}
{Lequeux} J (2005) {The Interstellar Medium}

\bibitem[{{Leroy} et~al.(2013){Leroy}, {Walter}, {Sandstrom}, {Schruba},
  {Munoz-Mateos}, {Bigiel}, {Bolatto}, {Brinks}, {de Blok}, {Meidt}, {Rix},
  {Rosolowsky}, {Schinnerer}, {Schuster}, and {Usero}}]{LeroyEtAl2013}
{Leroy} AK, {Walter} F, {Sandstrom} K, et~al. (2013) {Molecular Gas and Star
  Formation in nearby Disk Galaxies}. \aj 146:19

\bibitem[{{Leung}(1975)}]{l75}
{Leung} CM (1975) {Radiation transport in dense interstellar dust clouds.} \apj
  199:340--360

\bibitem[{{Li} et~al.(2012){Li}, {Myers}, and {McKee}}]{lmm12}
{Li} PS, {Myers} A, and {McKee} CF (2012) {Ambipolar Diffusion Heating in
  Turbulent Systems}. \apj 760(1):33

\bibitem[{{Lighthill}(1955)}]{lighthill55}
{Lighthill} MJ (1955) {The Effect of Compressibility on Turbulence}. In: Gas
  Dynamics of Cosmic Clouds, IAU Symposium, vol~2, p 121

\bibitem[{{Liu} et~al.(2011){Liu}, {Koda}, {Calzetti}, {Fukuhara}, and
  {Momose}}]{LiuEtAl2011}
{Liu} G, {Koda} J, {Calzetti} D, et~al. (2011) {The Super-linear Slope of the
  Spatially Resolved Star Formation Law in NGC 3521 and NGC 5194 (M51a)}. \apj
  735:63

\bibitem[{{Longmore} et~al.(2014){Longmore}, {Kruijssen}, {Bastian}, {Bally},
  {Rathborne}, {Testi}, {Stolte}, {Dale}, {Bressert}, and
  {Alves}}]{longmore2014}
{Longmore} SN, {Kruijssen} JMD, {Bastian} N, et~al. (2014) {The Formation and
  Early Evolution of Young Massive Clusters}. Protostars and Planets VI pp
  291--314

\bibitem[{{Lu} et~al.(2018){Lu}, {Zhang}, {Liu}, {Sanhueza}, {Tatematsu},
  {Feng}, {Smith}, {Myers}, {Sridharan}, and {Gu}}]{LuEtAl2018}
{Lu} X, {Zhang} Q, {Liu} HB, et~al. (2018) {Filamentary Fragmentation and
  Accretion in High-mass Star-forming Molecular Clouds}. \apj 855(1):9

\bibitem[{{Lucy} and {Solomon}(1970)}]{LucySolomon1970}
{Lucy} LB and {Solomon} PM (1970) {Mass Loss by Hot Stars}. \apj 159:879

\bibitem[{{Mac Low}(1999)}]{maclow99}
{Mac Low} MM (1999) {The Energy Dissipation Rate of Supersonic,
  Magnetohydrodynamic Turbulence in Molecular Clouds}. \apj 524(1):169--178

\bibitem[{{Mac Low} and {Klessen}(2004)}]{MacLowKlessen2004}
{Mac Low} MM and {Klessen} RS (2004) {Control of star formation by supersonic
  turbulence}. Reviews of Modern Physics 76:125--194

\bibitem[{{Mac Low} et~al.(1998){Mac Low}, {Klessen}, {Burkert}, and
  {Smith}}]{maclow98}
{Mac Low} MM, {Klessen} RS, {Burkert} A, et~al. (1998) {Kinetic Energy Decay
  Rates of Supersonic and Super-Alfv{\'e}nic Turbulence in Star-Forming
  Clouds}. \prl 80(13):2754--2757

\bibitem[{{Machida} and {Hosokawa}(2013)}]{MachidaHosokawa2013}
{Machida} MN and {Hosokawa} T (2013) {Evolution of protostellar outflow around
  low-mass protostar}. \mnras 431(2):1719--1744

\bibitem[{{Machida} et~al.(2008){Machida}, {Inutsuka}, and
  {Matsumoto}}]{Machida+2008}
{Machida} MN, {Inutsuka} Si, and {Matsumoto} T (2008) {High- and Low-Velocity
  Magnetized Outflows in the Star Formation Process in a Gravitationally
  Collapsing Cloud}. \apj 676(2):1088--1108

\bibitem[{{Madore}(1977)}]{madore77}
{Madore} BF (1977) {Numerical simulations of the rate of star formation in
  external galaxies.} \mnras 178:1--9

\bibitem[{{Maio} et~al.(2007){Maio}, {Dolag}, {Ciardi}, and
  {Tornatore}}]{maio07}
{Maio} U, {Dolag} K, {Ciardi} B, et~al. (2007) {Metal and molecule cooling in
  simulations of structure formation}. \mnras 379(3):963--973

\bibitem[{{Marcowith} et~al.(2016){Marcowith}, {Bret}, {Bykov}, {Dieckman},
  {O'C Drury}, {Lemb{\`e}ge}, {Lemoine}, {Morlino}, {Murphy}, {Pelletier},
  {Plotnikov}, {Reville}, {Riquelme}, {Sironi}, and {Stockem
  Novo}}]{MarcowithEtAl2016}
{Marcowith} A, {Bret} A, {Bykov} A, et~al. (2016) {The microphysics of
  collisionless shock waves}. Reports on Progress in Physics 79(4):046901

\bibitem[{{Masunaga} and {Inutsuka}(2000)}]{MasunagaInutsuka2000}
{Masunaga} H and {Inutsuka} Si (2000) {A Radiation Hydrodynamic Model for
  Protostellar Collapse. II. The Second Collapse and the Birth of a Protostar}.
  \apj 531(1):350--365

\bibitem[{{Mathis} et~al.(1977){Mathis}, {Rumpl}, and {Nordsieck}}]{mrn77}
{Mathis} JS, {Rumpl} W, and {Nordsieck} KH (1977) {The size distribution of
  interstellar grains.} \apj 217:425--433

\bibitem[{{Mazzali} et~al.(2014){Mazzali}, {McFadyen}, {Woosley}, {Pian}, and
  {Tanaka}}]{Mazea14}
{Mazzali} PA, {McFadyen} AI, {Woosley} SE, et~al. (2014) {An upper limit to the
  energy of gamma-ray bursts indicates that GRBs/SNe are powered by magnetars}.
  \mnras 443:67--71

\bibitem[{McKee and Ostriker(2007)}]{mckee.07}
McKee CF and Ostriker EC (2007) Theory of star formation. Annual Review of
  Astronomy and Astrophysics 45(1):565--687,
  \urlprefix\url{https://doi.org/10.1146/annurev.astro.45.051806.110602}

\bibitem[{{McKee} and {Ostriker}(1977)}]{McKeeOstriker1977}
{McKee} CF and {Ostriker} JP (1977) {A theory of the interstellar medium -
  Three components regulated by supernova explosions in an inhomogeneous
  substrate}. \apj 218:148--169

\bibitem[{{McKee} and {Zweibel}(1992)}]{mckee.92}
{McKee} CF and {Zweibel} EG (1992) {On the Virial Theorem for Turbulent
  Molecular Clouds}. \apj 399:551

\bibitem[{{Micic} et~al.(2012){Micic}, {Glover}, {Federrath}, and
  {Klessen}}]{micic12}
{Micic} M, {Glover} SCO, {Federrath} C, et~al. (2012) {Modelling H$_{2}$
  formation in the turbulent interstellar medium: solenoidal versus compressive
  turbulent forcing}. \mnras 421(3):2531--2542

\bibitem[{{Mierkiewicz} et~al.(2006){Mierkiewicz}, {Reynolds}, {Roesler},
  {Harlander}, and {Jaehnig}}]{mie06}
{Mierkiewicz} EJ, {Reynolds} RJ, {Roesler} FL, et~al. (2006) {Detection of
  Diffuse Interstellar [O II] Emission from the Milky Way Using Spatial
  Heterodyne Spectroscopy}. \apjl 650(1):L63--L66

\bibitem[{{Mihalas} and {Mihalas}(1984)}]{MihalasMihalas1984}
{Mihalas} D and {Mihalas} BW (1984) {Foundations of radiation hydrodynamics}

\bibitem[{{Minkowski}(1941)}]{Mink41}
{Minkowski} R (1941) {Spectra of Supernovae}. \pasp 53(314):224

\bibitem[{Mittal and Girimaji(2019)}]{mittal.19}
Mittal A and Girimaji SS (2019) Mathematical framework for analysis of internal
  energy dynamics and spectral distribution in compressible turbulent flows.
  Phys Rev Fluids 4:042601,
  \urlprefix\url{https://link.aps.org/doi/10.1103/PhysRevFluids.4.042601}

\bibitem[{{Miura} and {Kida}(1995)}]{miura.95}
{Miura} H and {Kida} S (1995) {Acoustic energy exchange in compressible
  turbulence}. Physics of Fluids 7(7):1732--1742

\bibitem[{{Miville-Desch{\^e}nes} et~al.(2017){Miville-Desch{\^e}nes},
  {Murray}, and {Lee}}]{MivilleDeschenesMurrayLee2017}
{Miville-Desch{\^e}nes} MA, {Murray} N, and {Lee} EJ (2017) {Physical
  Properties of Molecular Clouds for the Entire Milky Way Disk}. \apj 834:57

\bibitem[{{Molina} et~al.(2012){Molina}, {Glover}, {Federrath}, and
  {Klessen}}]{molina2012}
{Molina} FZ, {Glover} SCO, {Federrath} C, et~al. (2012) {The density
  variance-Mach number relation in supersonic turbulence - I. Isothermal,
  magnetized gas}. \mnras 423:2680--2689

\bibitem[{{Momferratos} et~al.(2014){Momferratos}, {Lesaffre}, {Falgarone}, and
  {Pineau des For{\^e}ts}}]{momferratos14}
{Momferratos} G, {Lesaffre} P, {Falgarone} E, et~al. (2014) {Turbulent energy
  dissipation and intermittency in ambipolar diffusion magnetohydrodynamics}.
  \mnras 443(1):86--101

\bibitem[{{Morris} and {Serabyn}(1996)}]{ms96}
{Morris} M and {Serabyn} E (1996) {The Galactic Center Environment}. \araa
  34:645--702

\bibitem[{{Mouschovias} and {Paleologou}(1981)}]{MouschoviasPaleologou1981}
{Mouschovias} TC and {Paleologou} EV (1981) {Ambipolar diffusion in
  interstellar clouds - Time-dependent solutions in one spatial dimension}.
  \apj 246:48--64

\bibitem[{{Mouschovias} and {Spitzer}(1976)}]{MouschoviasSpitzer1976}
{Mouschovias} TC and {Spitzer} L Jr (1976) {Note on the collapse of magnetic
  interstellar clouds}. \apj 210:326--+

\bibitem[{Moyal(1952)}]{moyal52}
Moyal JE (1952) The spectra of turbulence in a compressible fluid; eddy
  turbulence and random noise. Mathematical Proceedings of the Cambridge
  Philosophical Society 48(2):329–344

\bibitem[{{M{\"u}ller} et~al.(2016){M{\"u}ller}, {Heger}, {Liptai}, and
  {Cameron}}]{Muellea16}
{M{\"u}ller} B, {Heger} A, {Liptai} D, et~al. (2016) {A simple approach to the
  supernova progenitor-explosion connection}. \mnras 460(1):742--764

\bibitem[{{Naab} and {Ostriker}(2017)}]{NaabOstriker2017}
{Naab} T and {Ostriker} JP (2017) {Theoretical Challenges in Galaxy Formation}.
  \araa 55:59--109

\bibitem[{{Nagai} et~al.(1998){Nagai}, {Inutsuka}, and
  {Miyama}}]{NagaiEtAl1998}
{Nagai} T, {Inutsuka} Si, and {Miyama} SM (1998) {An Origin of Filamentary
  Structure in Molecular Clouds}. \apj 506(1):306--322

\bibitem[{{Nakamura} and {Li}(2007)}]{Nakamura+2007}
{Nakamura} F and {Li} ZY (2007) {Protostellar Turbulence Driven by Collimated
  Outflows}. The Astrophysical Journal 662(1):395--412

\bibitem[{{Nakano} and {Nakamura}(1978)}]{NakanoNakamura1978}
{Nakano} T and {Nakamura} T (1978) {Gravitational Instability of Magnetized
  Gaseous Disks 6}. \pasj 30:671--680

\bibitem[{{Obukhov}(1941)}]{obukhov41}
{Obukhov} AM (1941) {On the distribution of energy in the spectrum of turbulent
  flow}. Akademiia Nauk SSSR Doklady 32:22--24

\bibitem[{{Offner} and {Arce}(2014)}]{OffnerArce2014}
{Offner} SSR and {Arce} HG (2014) {Investigations of Protostellar Outflow
  Launching and Gas Entrainment: Hydrodynamic Simulations and Molecular
  Emission}. The Astrophysical Journal 784(1):61

\bibitem[{{Offner} and {Chaban}(2017)}]{OffnerChaban2017}
{Offner} SSR and {Chaban} J (2017) {Impact of Protostellar Outflows on
  Turbulence and Star Formation Efficiency in Magnetized Dense Cores}. The
  Astrophysical Journal 847(2):104

\bibitem[{{Offner} and {McKee}(2011)}]{OffnerMcKee2011}
{Offner} SSR and {McKee} CF (2011) {The Protostellar Luminosity Function}. \apj
  736(1):53

\bibitem[{{Offner} et~al.(2014){Offner}, {Clark}, {Hennebelle}, {Bastian},
  {Bate}, {Hopkins}, {Moraux}, and {Whitworth}}]{offner2014}
{Offner} SSR, {Clark} PC, {Hennebelle} P, et~al. (2014) {The Origin and
  Universality of the Stellar Initial Mass Function}. Protostars and Planets VI
  pp 53--75

\bibitem[{{Onsager}(1949)}]{onsager49}
{Onsager} L (1949) {Statistical hydrodynamics}. Il Nuovo Cimento 6(2):279--287

\bibitem[{{Oppenheimer} and {Schaye}(2013)}]{os13}
{Oppenheimer} BD and {Schaye} J (2013) {Non-equilibrium ionization and cooling
  of metal-enriched gas in the presence of a photoionization background}.
  \mnras 434(2):1043--1062

\bibitem[{{Orkisz} et~al.(2017){Orkisz}, {Pety}, {Gerin}, {Bron}, {Guzm{\'a}n},
  {Bardeau}, {Goicoechea}, {Gratier}, {Le Petit}, {Levrier}, {Liszt},
  {{\"O}berg}, {Peretto}, {Roueff}, {Sievers}, and {Tremblin}}]{okrizs...17}
{Orkisz} JH, {Pety} J, {Gerin} M, et~al. (2017) {Turbulence and star formation
  efficiency in molecular clouds: solenoidal versus compressive motions in
  Orion B}. \aap 599:A99

\bibitem[{{Orr} et~al.(2018){Orr}, {Hayward}, {Hopkins}, {Chan},
  {Faucher-Gigu{\`e}re}, {Feldmann}, {Kere{\v{s}}}, {Murray}, and
  {Quataert}}]{orr18}
{Orr} ME, {Hayward} CC, {Hopkins} PF, et~al. (2018) {What FIREs up star
  formation: the emergence of the Kennicutt-Schmidt law from feedback}. \mnras
  478(3):3653--3673

\bibitem[{{Osterbrock}(1989)}]{osterbrock89}
{Osterbrock} DE (1989) {Astrophysics of gaseous nebulae and active galactic
  nuclei}

\bibitem[{{Ostriker} and {Shetty}(2011)}]{OstrikerShetty2011}
{Ostriker} EC and {Shetty} R (2011) {Maximally Star-forming Galactic Disks. I.
  Starburst Regulation Via Feedback-driven Turbulence}. \apj 731:41

\bibitem[{{Ostriker} and {Shu}(1995)}]{OstrikerShu1995}
{Ostriker} EC and {Shu} FH (1995) {Magnetocentrifugally Driven Flows from Young
  Stars and Disks. IV. The Accretion Funnel and Dead Zone}. \apj 447:813

\bibitem[{{Ostriker} et~al.(2010){Ostriker}, {McKee}, and
  {Leroy}}]{OstrikerMcKeeLeroy2010}
{Ostriker} EC, {McKee} CF, and {Leroy} AK (2010) {Regulation of Star Formation
  Rates in Multiphase Galactic Disks: A Thermal/Dynamical Equilibrium Model}.
  \apj 721:975--994

\bibitem[{{Ostriker}(1964)}]{Ostriker1964}
{Ostriker} J (1964) {The Equilibrium of Polytropic and Isothermal Cylinders.}
  \apj 140:1056

\bibitem[{{Padoan} et~al.(2000){Padoan}, {Zweibel}, and {Nordlund}}]{pzn00}
{Padoan} P, {Zweibel} E, and {Nordlund} {\r{A}} (2000) {Ambipolar Drift Heating
  in Turbulent Molecular Clouds}. \apj 540(1):332--341

\bibitem[{{Padoan} et~al.(2004){Padoan}, {Jimenez}, {Nordlund}, and
  {Boldyrev}}]{padoan...04}
{Padoan} P, {Jimenez} R, {Nordlund} {\r{A}}, et~al. (2004) {Structure Function
  Scaling in Compressible Super-Alfv{\'e}nic MHD Turbulence}. \prl
  92(19):191102

\bibitem[{{Padoan} et~al.(2014){Padoan}, {Federrath}, {Chabrier}, {Evans},
  {Johnstone}, {J{\o}rgensen}, {McKee}, and {Nordlund}}]{padoan14}
{Padoan} P, {Federrath} C, {Chabrier} G, et~al. (2014) {The Star Formation Rate
  of Molecular Clouds}. In: {Beuther} H, {Klessen} RS, {Dullemond} CP, et~al.
  (eds) Protostars and Planets VI, p~77

\bibitem[{{Padovani} and {Galli}(2011)}]{PadovaniGalli2011}
{Padovani} M and {Galli} D (2011) {Effects of magnetic fields on the cosmic-ray
  ionization of molecular cloud cores}. \aap 530:A109

\bibitem[{{Padovani} et~al.(2009){Padovani}, {Galli}, and
  {Glassgold}}]{Padovani2009}
{Padovani} M, {Galli} D, and {Glassgold} AE (2009) {Cosmic-ray ionization of
  molecular clouds}. \aap 501:619--631

\bibitem[{{Padovani} et~al.(2013){Padovani}, {Hennebelle}, and
  {Galli}}]{Padovani2013}
{Padovani} M, {Hennebelle} P, and {Galli} D (2013) {Cosmic-ray ionisation in
  collapsing clouds}. \aap 560:A114

\bibitem[{{Padovani} et~al.(2018){Padovani}, {Ivlev}, {Galli}, and
  {Caselli}}]{Padovani2018}
{Padovani} M, {Ivlev} AV, {Galli} D, et~al. (2018) {Cosmic-ray ionisation in
  circumstellar discs}. \aap 614:A111

\bibitem[{{Palau} et~al.(2018){Palau}, {Zapata}, {Rom{\'a}n-Z{\'u}{\~n}iga},
  {S{\'a}nchez-Monge}, {Estalella}, {Busquet}, {Girart}, {Fuente}, and
  {Commer{\c{c}}on}}]{PalauEtAl2018}
{Palau} A, {Zapata} LA, {Rom{\'a}n-Z{\'u}{\~n}iga} CG, et~al. (2018) {Thermal
  Jeans Fragmentation within 1000 au in OMC-1S}. \apj 855(1):24

\bibitem[{{Pan} and {Padoan}(2009)}]{pp09}
{Pan} L and {Padoan} P (2009) {The Temperature of Interstellar Clouds from
  Turbulent Heating}. \apj 692(1):594--607

\bibitem[{{Pan} et~al.(2016){Pan}, {Padoan}, {Haugb{\o}lle}, and
  {Nordlund}}]{pan...16}
{Pan} L, {Padoan} P, {Haugb{\o}lle} T, et~al. (2016) {Supernova Driving. II.
  Compressive Ratio in Molecular-cloud Turbulence}. \apj 825(1):30

\bibitem[{{Pardi} et~al.(2017){Pardi}, {Girichidis}, {Naab}, {Walch}, {Peters},
  {Heitsch}, {Glover}, {Klessen}, {W{\"u}nsch}, and {Gatto}}]{PardiEtAl2017}
{Pardi} A, {Girichidis} P, {Naab} T, et~al. (2017) {The impact of magnetic
  fields on the chemical evolution of the supernova-driven ISM}. \mnras
  465:4611--4633

\bibitem[{{Parker}(1958)}]{Parker1958}
{Parker} EN (1958) {Dynamics of the Interplanetary Gas and Magnetic Fields.}
  \apj 128:664

\bibitem[{{Pelletier} and {Pudritz}(1992)}]{PelletierPudrtiz1992}
{Pelletier} G and {Pudritz} RE (1992) {Hydromagnetic Disk Winds in Young
  Stellar Objects and Active Galactic Nuclei}. \apj 394:117

\bibitem[{{Penston}(1969)}]{Penston1969}
{Penston} MV (1969) {Dynamics of self-gravitating gaseous spheres-III.
  Analytical results in the free-fall of isothermal cases}. \mnras 144:425

\bibitem[{{Peters} et~al.(2010{\natexlab{a}}){Peters}, {Klessen}, {Mac Low},
  and {Banerjee}}]{PetersEtAl2010c}
{Peters} T, {Klessen} RS, {Mac Low} M, et~al. (2010{\natexlab{a}}) {Limiting
  Accretion onto Massive Stars by Fragmentation-induced Starvation}. \apj
  725:134--145

\bibitem[{{Peters} et~al.(2010{\natexlab{b}}){Peters}, {Mac Low}, {Banerjee},
  {Klessen}, and {Dullemond}}]{PetersEtAl2010b}
{Peters} T, {Mac Low} M, {Banerjee} R, et~al. (2010{\natexlab{b}})
  {Understanding Spatial and Spectral Morphologies of Ultracompact H II
  Regions}. \apj 719:831--843

\bibitem[{{Peters} et~al.(2014){Peters}, {Klaassen}, {Mac Low}, {Schr{\"o}n},
  {Federrath}, {Smith}, and {Klessen}}]{PetersEtAl2014}
{Peters} T, {Klaassen} PD, {Mac Low} MM, et~al. (2014) {Collective Outflow from
  a Small Multiple Stellar System}. \apj 788:14

\bibitem[{{Pineda} et~al.(2013){Pineda}, {Langer}, {Velusamy}, and
  {Goldsmith}}]{pineda13}
{Pineda} JL, {Langer} WD, {Velusamy} T, et~al. (2013) {A Herschel [C ii]
  Galactic plane survey. I. The global distribution of ISM gas components}.
  \aap 554:A103

\bibitem[{{Piontek} and {Ostriker}(2004)}]{piontek2004}
{Piontek} RA and {Ostriker} EC (2004) {Thermal and Magnetorotational
  Instability in the Interstellar Medium: Two-dimensional Numerical
  Simulations}. \apj 601:905--920

\bibitem[{{Planck Collaboration} et~al.(2018){Planck Collaboration}, {Aghanim},
  {Akrami}, {Alves}, {Ashdown}, {Aumont}, {Baccigalupi}, {Ballardini},
  {Banday}, {Barreiro}, {Bartolo}, {Basak}, {Benabed}, {Bernard}, {Bersanelli},
  {Bielewicz}, {Bock}, {Bond}, {Borrill}, {Bouchet}, {Boulanger}, {Bracco},
  {Bucher}, {Burigana}, {Calabrese}, {Cardoso}, {Carron}, {Chary}, {Chiang},
  {Colombo}, {Combet}, {Crill}, {Cuttaia}, {de Bernardis}, {de Zotti},
  {Delabrouille}, {Delouis}, {Di Valentino}, {Dickinson}, {Diego}, {Dor{\'e}},
  {Douspis}, {Ducout}, {Dupac}, {Efstathiou}, {Elsner}, {En{\ss}lin},
  {Eriksen}, {Falgarone}, {Fantaye}, {Fernandez-Cobos}, {Ferri{\`e}re},
  {Finelli}, {Forastieri}, {Frailis}, {Fraisse}, {Franceschi}, {Frolov},
  {Galeotta}, {Galli}, {Ganga}, {G{\'e}nova-Santos}, {Gerbino}, {Ghosh},
  {Gonz{\'a}lez-Nuevo}, {G{\'o}rski}, {Gratton}, {Green}, {Gruppuso},
  {Gudmundsson}, {Guillet}, {Handley}, {Hansen}, {Helou}, {Herranz}, {Hivon},
  {Huang}, {Jaffe}, {Jones}, {Keih{\"a}nen}, {Keskitalo}, {Kiiveri}, {Kim},
  {Krachmalnicoff}, {Kunz}, {Kurki-Suonio}, {Lagache}, {Lamarre}, {Lasenby},
  {Lattanzi}, {Lawrence}, {Le Jeune}, {Levrier}, {Liguori}, {Lilje},
  {Lindholm}, {L{\'o}pez-Caniego}, {Lubin}, {Ma}, {Mac{\'\i}as-P{\'e}rez},
  {Maggio}, {Maino}, {Mandolesi}, {Mangilli}, {Marcos-Caballero}, {Maris},
  {Martin}, {Mart{\'\i}nez-Gonz{\'a}lez}, {Matarrese}, {Mauri}, {McEwen},
  {Melchiorri}, {Mennella}, {Migliaccio}, {Miville-Desch{\^e}nes}, {Molinari},
  {Moneti}, {Montier}, {Morgante}, {Moss}, {Natoli}, {Pagano}, {Paoletti},
  {Patanchon}, {Perrotta}, {Pettorino}, {Piacentini}, {Polastri}, {Polenta},
  {Puget}, {Rachen}, {Reinecke}, {Remazeilles}, {Renzi}, {Ristorcelli},
  {Rocha}, {Rosset}, {Roudier}, {Rubi{\~n}o-Mart{\'\i}n}, {Ruiz-Granados},
  {Salvati}, {Sandri}, {Savelainen}, {Scott}, {Sirignano}, {Sunyaev},
  {Suur-Uski}, {Tauber}, {Tavagnacco}, {Tenti}, {Toffolatti}, {Tomasi},
  {Trombetti}, {Valiviita}, {Vansyngel}, {Van Tent}, {Vielva}, {Villa},
  {Vittorio}, {Wandelt}, {Wehus}, {Zacchei}, and {Zonca}}]{PlanckXII2018}
{Planck Collaboration}, {Aghanim} N, {Akrami} Y, et~al. (2018) {Planck 2018
  results. XII. Galactic astrophysics using polarized dust emission}. arXiv
  e-prints arXiv:1807.06212

\bibitem[{{Porter} et~al.(2002){Porter}, {Pouquet}, and
  {Woodward}}]{porter..02}
{Porter} D, {Pouquet} A, and {Woodward} P (2002) {Measures of intermittency in
  driven supersonic flows}. \pre 66(2):026301

\bibitem[{{Porter} et~al.(1998){Porter}, {Woodward}, and
  {Pouquet}}]{porter..98}
{Porter} DH, {Woodward} PR, and {Pouquet} A (1998) {Inertial range structures
  in decaying compressible turbulent flows}. Physics of Fluids 10(1):237--245

\bibitem[{{Prasad} and {Tarafdar}(1983)}]{pt83}
{Prasad} SS and {Tarafdar} SP (1983) {UV radiation field inside dense clouds -
  Its possible existence and chemical implications}. \apj 267:603--609

\bibitem[{{Price} et~al.(2012){Price}, {Tricco}, and {Bate}}]{Price+2012}
{Price} DJ, {Tricco} TS, and {Bate} MR (2012) {Collimated jets from the first
  core}. \mnras 423(1):L45--L49

\bibitem[{{Rees}(1976)}]{Rees1976}
{Rees} MJ (1976) {Opacity-limited hierarchical fragmentation and the masses of
  protostars}. \mnras 176:483--486

\bibitem[{{Reipurth} and {Bally}(2001)}]{ReipurthBally2001}
{Reipurth} B and {Bally} J (2001) {Herbig-Haro Flows: Probes of Early Stellar
  Evolution}. Annual Review of Astronomy and Astrophysics 39:403--455

\bibitem[{{Reynolds}(1989)}]{reynolds89}
{Reynolds} RJ (1989) {The Column Density and Scale Height of Free Electrons in
  the Galactic Disk}. \apjl 339:L29

\bibitem[{{Reynolds} et~al.(1973){Reynolds}, {Scherb}, and
  {Roesler}}]{reynolds73}
{Reynolds} RJ, {Scherb} F, and {Roesler} FL (1973) {Observations of Diffuse
  Galactic HA and [n II] Emission}. \apj 185:869--876

\bibitem[{Richardson(1965)}]{richardson22}
Richardson L (1965) Weather Prediction by Numerical Process. Dover books
  explaining science, Dover Publications,
  \urlprefix\url{https://books.google.com/books?id=I4u\_AAAAIAAJ}

\bibitem[{{Richings} et~al.(2014){Richings}, {Schaye}, and
  {Oppenheimer}}]{rso14}
{Richings} AJ, {Schaye} J, and {Oppenheimer} BD (2014) {Non-equilibrium
  chemistry and cooling in the diffuse interstellar medium - I. Optically thin
  regime}. \mnras 440(4):3349--3369

\bibitem[{{Robitaille}(2011)}]{Robitaille2011}
{Robitaille} TP (2011) {HYPERION: an open-source parallelized three-dimensional
  dust continuum radiative transfer code}. \aap 536:A79

\bibitem[{{Rodriguez} et~al.(1980){Rodriguez}, {Moran}, {Ho}, and
  {Gottlieb}}]{Rodriguez+1980}
{Rodriguez} LF, {Moran} JM, {Ho} PTP, et~al. (1980) {Radio observations of
  water vapor, hydroxyl, silicon monoxide, ammonia, carbon monoxide, and
  compact H II regions in the vicinities of suspected Herbig-Haro objects.} The
  Astrophysical Journal 235:845--865

\bibitem[{{Rogers} and {Pittard}(2013)}]{RogersPittard2013}
{Rogers} H and {Pittard} JM (2013) {Feedback from winds and supernovae in
  massive stellar clusters - I. Hydrodynamics}. \mnras 431(2):1337--1351

\bibitem[{{Roser} et~al.(2003){Roser}, {Swords}, {Vidali}, {Manic{\`o}}, and
  {Pirronello}}]{roser03}
{Roser} JE, {Swords} S, {Vidali} G, et~al. (2003) {Measurement of the Kinetic
  Energy of Hydrogen Molecules Desorbing from Amorphous Water Ice}. \apjl
  596(1):L55--L58

\bibitem[{{Rybicki} and {Lightman}(1986)}]{rybicki86}
{Rybicki} GB and {Lightman} AP (1986) {Radiative Processes in Astrophysics}

\bibitem[{Sagaut and Cambon(2018)}]{sagaut.18}
Sagaut P and Cambon C (2018) Homogeneous Turbulence Dynamics, 2nd edn. Springer
  International Publishing AG, Gewerbestrasse 11, 6330 Cham, Switzerland

\bibitem[{{Salpeter}(1955)}]{Salpeter1955}
{Salpeter} EE (1955) {The Luminosity Function and Stellar Evolution.} \apj
  121:161

\bibitem[{{Sarkar} et~al.(1991){Sarkar}, {Erlebacher}, {Hussaini}, and
  {Kreiss}}]{sarkar...91}
{Sarkar} S, {Erlebacher} G, {Hussaini} MY, et~al. (1991) {The analysis and
  modelling of dilatational terms in compressible turbulence}. Journal of Fluid
  Mechanics 227:473--493

\bibitem[{{Schaye}(2004)}]{schaye04}
{Schaye} J (2004) {Star Formation Thresholds and Galaxy Edges: Why and Where}.
  \apj 609:667--682

\bibitem[{{Schinnerer} et~al.(2019){Schinnerer}, {Hughes}, {Leroy}, {Groves},
  {Blanc}, {Kreckel}, {Bigiel}, {Chevance}, {Dale}, {Emsellem}, {Faesi},
  {Glover}, {Grasha}, {Henshaw}, {Hygate}, {Kruijssen}, {Meidt}, {Pety},
  {Querejeta}, {Rosolowsky}, {Saito}, {Schruba}, {Sun}, and
  {Utomo}}]{schinnerer19}
{Schinnerer} E, {Hughes} A, {Leroy} A, et~al. (2019) {The Gas-Star Formation
  Cycle in Nearby Star-forming Galaxies. I. Assessment of Multi-scale
  Variations}. \apj 887(1):49

\bibitem[{{Schmidt}(1959)}]{Schmidt1959}
{Schmidt} M (1959) {The Rate of Star Formation.} \apj 129:243

\bibitem[{{Schmidt} and {Grete}(2019)}]{schmidt.19}
{Schmidt} W and {Grete} P (2019) {Kinetic and internal energy transfer in
  implicit large-eddy simulations of forced compressible turbulence}. \pre
  100(4):043116

\bibitem[{Schmidt et~al.(2008)Schmidt, Federrath, and Klessen}]{schmidt..08}
Schmidt W, Federrath C, and Klessen R (2008) Is the scaling of supersonic
  turbulence universal? Phys Rev Lett 101:194505,
  \urlprefix\url{https://link.aps.org/doi/10.1103/PhysRevLett.101.194505}

\bibitem[{{Schneider} et~al.(2015){Schneider}, {Bontemps}, {Girichidis},
  {Rayner}, {Motte}, {Andr{\'e}}, {Russeil}, {Abergel}, {Anderson},
  {Arzoumanian}, {Benedettini}, {Csengeri}, {Didelon}, {Di Francesco},
  {Griffin}, {Hill}, {Klessen}, {Ossenkopf}, {Pezzuto}, {Rivera-Ingraham},
  {Spinoglio}, {Tremblin}, and {Zavagno}}]{SchneiderEtAl2015b}
{Schneider} N, {Bontemps} S, {Girichidis} P, et~al. (2015) {Detection of two
  power-law tails in the probability distribution functions of massive GMCs}.
  \mnras 453:L41--L45

\bibitem[{{Sch{\"o}ier} et~al.(2005){Sch{\"o}ier}, {van der Tak}, {van
  Dishoeck}, and {Black}}]{sch05}
{Sch{\"o}ier} FL, {van der Tak} FFS, {van Dishoeck} EF, et~al. (2005) {An
  atomic and molecular database for analysis of submillimetre line
  observations}. \aap 432(1):369--379

\bibitem[{{Schruba} et~al.(2010){Schruba}, {Leroy}, {Walter}, {Sandstrom}, and
  {Rosolowsky}}]{SchrubaEtAl2010}
{Schruba} A, {Leroy} AK, {Walter} F, et~al. (2010) {The Scale Dependence of the
  Molecular Gas Depletion Time in M33}. \apj 722:1699--1706

\bibitem[{{Schwarz} et~al.(2010){Schwarz}, {Beetz}, {Dreher}, and
  {Grauer}}]{schwarz...10}
{Schwarz} C, {Beetz} C, {Dreher} J, et~al. (2010) {Lyapunov exponents and
  information dimension of the mass distribution in turbulent compressible
  flows}. Physics Letters A 374(8):1039--1042

\bibitem[{{Seifried} et~al.(2011){Seifried}, {Schmidt}, and {Niemeyer}}]{ssn11}
{Seifried} D, {Schmidt} W, and {Niemeyer} JC (2011) {Forced turbulence in
  thermally bistable gas: a parameter study}. \aap 526:A14

\bibitem[{{Seifried} et~al.(2012){Seifried}, {Pudritz}, {Banerjee}, {Duffin},
  and {Klessen}}]{Seifried+2012}
{Seifried} D, {Pudritz} RE, {Banerjee} R, et~al. (2012) {Magnetic fields during
  the early stages of massive star formation - II. A generalized outflow
  criterion}. \mnras 422(1):347--366

\bibitem[{{Shaviv} and {Regev}(1994)}]{ShavivRegev1994}
{Shaviv} NJ and {Regev} O (1994) {Interface dynamics and domain growth in
  thermally bistable fluids}. \pre 50(3):2048--2056

\bibitem[{{Shepherd} and {Churchwell}(1996)}]{ShepherdChurchwell1996}
{Shepherd} DS and {Churchwell} E (1996) {Bipolar Molecular Outflows in Massive
  Star Formation Regions}. The Astrophysical Journal 472:225

\bibitem[{{Shu}(1977)}]{Shu1977}
{Shu} FH (1977) {Self-similar collapse of isothermal spheres and star
  formation.} \apj 214:488--497

\bibitem[{{Shu}(1992)}]{ShuAstroGas1992}
{Shu} FH (1992) {The physics of astrophysics. Volume II: Gas dynamics.}

\bibitem[{{Shu} et~al.(1988){Shu}, {Lizano}, {Ruden}, and {Najita}}]{Shu+1988}
{Shu} FH, {Lizano} S, {Ruden} SP, et~al. (1988) {Mass Loss from Rapidly
  Rotating Magnetic Protostars}. \apjl 328:L19

\bibitem[{{Shu} et~al.(1995){Shu}, {Najita}, {Ostriker}, and
  {Shang}}]{Shu+1995}
{Shu} FH, {Najita} J, {Ostriker} EC, et~al. (1995) {Magnetocentrifugally Driven
  Flows from Young Stars and Disks. V. Asymptotic Collimation into Jets}. The
  Astrophysical Journal 455:L155

\bibitem[{{Skilling} and {Strong}(1976)}]{SkillingStrong1976}
{Skilling} J and {Strong} AW (1976) {Cosmic ray exclusion from dense molecular
  clouds.} \aap 53(2):253--258

\bibitem[{{Slyz} et~al.(2005){Slyz}, {Devriendt}, {Bryan}, and
  {Silk}}]{SlyzEtAl2005}
{Slyz} AD, {Devriendt} JEG, {Bryan} G, et~al. (2005) {Towards simulating star
  formation in the interstellar medium}. \mnras 356:737--752

\bibitem[{{Smartt}(2009)}]{Smartt09}
{Smartt} SJ (2009) {Progenitors of Core-Collapse Supernovae}. \araa
  47(1):63--106

\bibitem[{{Smith}(2014)}]{Smith2014}
{Smith} N (2014) {Mass Loss: Its Effect on the Evolution and Fate of High-Mass
  Stars}. \araa 52:487--528

\bibitem[{{Snell} et~al.(1980){Snell}, {Loren}, and {Plambeck}}]{Snell+1980}
{Snell} RL, {Loren} RB, and {Plambeck} RL (1980) {Observations of CO in L 1551
  : evidence for stellar wind driven shocks.} The Astrophysical Journal
  239:L17--L22

\bibitem[{{Somerville} and {Dav{\'e}}(2015)}]{SomervilleDave2015}
{Somerville} RS and {Dav{\'e}} R (2015) {Physical Models of Galaxy Formation in
  a Cosmological Framework}. \araa 53:51--113

\bibitem[{{Spitzer}(1942)}]{Spitzer1942}
{Spitzer} L Jr (1942) {The Dynamics of the Interstellar Medium. III. Galactic
  Distribution.} \apj 95:329

\bibitem[{{Spruit}(2013)}]{Spruit2013}
{Spruit} HC (2013) {Essential Magnetohydrodynamics for Astrophysics}. arXiv
  e-prints arXiv:1301.5572

\bibitem[{{Stacy} and {Bromm}(2013)}]{StacyBromm2013}
{Stacy} A and {Bromm} V (2013) {Constraining the statistics of Population III
  binaries}. \mnras 433(2):1094--1107

\bibitem[{{Stacy} et~al.(2016){Stacy}, {Bromm}, and {Lee}}]{StacyBrommLee2016}
{Stacy} A, {Bromm} V, and {Lee} AT (2016) {Building up the Population III
  initial mass function from cosmological initial conditions}. \mnras
  462(2):1307--1328

\bibitem[{{Stahler} et~al.(1980){Stahler}, {Shu}, and {Taam}}]{Stahler+1980}
{Stahler} SW, {Shu} FH, and {Taam} RE (1980) {The evolution of protostars. I -
  Global formulation and results}. \apj 241:637--654

\bibitem[{{Stecher} and {Williams}(1967)}]{sw67}
{Stecher} TP and {Williams} DA (1967) {Photodestruction of Hydrogen Molecules
  in H I Regions}. \apjl 149:L29

\bibitem[{{Stephens} and {Dalgarno}(1973)}]{sd73}
{Stephens} TL and {Dalgarno} A (1973) {Kinetic Energy in the Spontaneous
  Radiative Dissociation of Molecular Hydrogen}. \apj 186:165--168

\bibitem[{{Sternberg} and {Dalgarno}(1995)}]{sd95}
{Sternberg} A and {Dalgarno} A (1995) {Chemistry in Dense Photon-dominated
  Regions}. \apjs 99:565

\bibitem[{{Sternberg} et~al.(2014){Sternberg}, {Le Petit}, {Roueff}, and {Le
  Bourlot}}]{sternberg2014}
{Sternberg} A, {Le Petit} F, {Roueff} E, et~al. (2014) {H I-to-H$_{2}$
  Transitions and H I Column Densities in Galaxy Star-forming Regions}. \apj
  790:10

\bibitem[{{Stone} and {Zweibel}(2010)}]{StoneZweibel2010}
{Stone} JM and {Zweibel} EG (2010) {Ambipolar Diffusion-mediated Thermal Fronts
  in the Neutral Interstellar Medium}. \apj 724(1):131--139

\bibitem[{{Stone} et~al.(1998){Stone}, {Ostriker}, and {Gammie}}]{Stone98}
{Stone} JM, {Ostriker} EC, and {Gammie} CF (1998) {Dissipation in Compressible
  Magnetohydrodynamic Turbulence}. \apjl 508(1):L99--L102

\bibitem[{{Strong} et~al.(2007){Strong}, {Moskalenko}, and
  {Ptuskin}}]{StrongMoskalenkoPtuskin2007}
{Strong} AW, {Moskalenko} IV, and {Ptuskin} VS (2007) {Cosmic-Ray Propagation
  and Interactions in the Galaxy}. Annual Review of Nuclear and Particle
  Science 57:285--327

\bibitem[{{Subramanian}(2019)}]{Subramanian2019}
{Subramanian} K (2019) {From Primordial Seed Magnetic Fields to the Galactic
  Dynamo}. Galaxies 7(2):47

\bibitem[{{Sun} et~al.(2018){Sun}, {Leroy}, {Schruba}, {Rosolowsky}, {Hughes},
  {Kruijssen}, {Meidt}, {Schinnerer}, {Blanc}, {Bigiel}, {Bolatto}, {Chevance},
  {Groves}, {Herrera}, {Hygate}, {Pety}, {Querejeta}, {Usero}, and
  {Utomo}}]{sun18}
{Sun} J, {Leroy} AK, {Schruba} A, et~al. (2018) {Cloud-scale Molecular Gas
  Properties in 15 Nearby Galaxies}. \apj 860:172

\bibitem[{{Susa}(2013)}]{Susa2013}
{Susa} H (2013) {The Mass of the First Stars}. \apj 773(2):185

\bibitem[{{Sytine} et~al.(2000){Sytine}, {Porter}, {Woodward}, {Hodson}, and
  {Winkler}}]{sytine....00}
{Sytine} IV, {Porter} DH, {Woodward} PR, et~al. (2000) {Convergence Tests for
  the Piecewise Parabolic Method and Navier-Stokes Solutions for Homogeneous
  Compressible Turbulence}. Journal of Computational Physics 158(2):225--238

\bibitem[{{Takahashi} et~al.(2013){Takahashi}, {Ho}, {Teixeira}, {Zapata}, and
  {Su}}]{TakahashiEtAl2013}
{Takahashi} S, {Ho} PTP, {Teixeira} PS, et~al. (2013) {Hierarchical
  Fragmentation of the Orion Molecular Filaments}. \apj 763(1):57

\bibitem[{{Tanaka} et~al.(2017){Tanaka}, {Tan}, and {Zhang}}]{Tanaka+2017}
{Tanaka} KEI, {Tan} JC, and {Zhang} Y (2017) {The Impact of Feedback During
  Massive Star Formation by Core Accretion}. \apj 835(1):32

\bibitem[{{Tegmark} et~al.(1997){Tegmark}, {Silk}, {Rees}, {Blanchard}, {Abel},
  and {Palla}}]{teg97}
{Tegmark} M, {Silk} J, {Rees} MJ, et~al. (1997) {How Small Were the First
  Cosmological Objects?} \apj 474:1

\bibitem[{{Tielens}(2010)}]{tielens2010}
{Tielens} AGGM (2010) {The Physics and Chemistry of the Interstellar Medium}

\bibitem[{{Tielens} and {Hollenbach}(1985)}]{th85}
{Tielens} AGGM and {Hollenbach} D (1985) {Photodissociation regions. I. Basic
  model.} \apj 291:722--746

\bibitem[{{Tomida} et~al.(2010){Tomida}, {Tomisaka}, {Matsumoto}, {Ohsuga},
  {Machida}, and {Saigo}}]{Tomida+2010}
{Tomida} K, {Tomisaka} K, {Matsumoto} T, et~al. (2010) {Radiation
  Magnetohydrodynamics Simulation of Proto-stellar Collapse: Two-component
  Molecular Outflow}. \apjl 714(1):L58--L63

\bibitem[{{Toomre}(1964)}]{Toomre1964}
{Toomre} A (1964) {On the gravitational stability of a disk of stars}. \apj
  139:1217--1238

\bibitem[{{Troland} and {Crutcher}(2008)}]{TrolandCrutcher2008}
{Troland} TH and {Crutcher} RM (2008) {Magnetic Fields in Dark Cloud Cores:
  Arecibo OH Zeeman Observations}. \apj 680(1):457--465

\bibitem[{{Troland} and {Heiles}(1986)}]{TrolandHeiles1986}
{Troland} TH and {Heiles} C (1986) {Interstellar magnetic field strengths and
  gas densities Observational and theoretical perspectives}. \apj 301:339--345

\bibitem[{{van der Kruit} and {Freeman}(2011)}]{VanDerKruitFeeman2011}
{van der Kruit} PC and {Freeman} KC (2011) {Galaxy Disks}. \araa 49(1):301--371

\bibitem[{{van der Tak} and {van Dishoeck}(2000)}]{vv00}
{van der Tak} FFS and {van Dishoeck} EF (2000) {Limits on the cosmic-ray
  ionization rate toward massive young stars}. \aap 358:L79--L82

\bibitem[{{van Dishoeck}(1987)}]{vd87}
{van Dishoeck} EF (1987) {Photodissociation processes of astrophysical
  molecules.} In: {Vardya} MS and {Tarafdar} SP (eds) Astrochemistry, IAU
  Symposium, vol 120, pp 51--65

\bibitem[{{van Dishoeck} and {Black}(1988)}]{vdb88}
{van Dishoeck} EF and {Black} JH (1988) {The Photodissociation and Chemistry of
  Interstellar CO}. \apj 334:771

\bibitem[{{van Loo} et~al.(2007){van Loo}, {Falle}, {Hartquist}, and
  {Moore}}]{vanloo2007}
{van Loo} S, {Falle} SAEG, {Hartquist} TW, et~al. (2007) {Shock-triggered
  formation of magnetically-dominated clouds}. \aap 471:213--218

\bibitem[{{Veilleux} et~al.(2005){Veilleux}, {Cecil}, and
  {Bland-Hawthorn}}]{VeilleuxCecilBlandHawthorn2005}
{Veilleux} S, {Cecil} G, and {Bland-Hawthorn} J (2005) {Galactic Winds}. \araa
  43:769--826

\bibitem[{{Vink} et~al.(2001){Vink}, {de Koter}, and {Lamers}}]{Vink+2001}
{Vink} JS, {de Koter} A, and {Lamers} HJGLM (2001) {Mass-loss predictions for O
  and B stars as a function of metallicity}. \aap 369:574--588

\bibitem[{{Visser} et~al.(2009){Visser}, {van Dishoeck}, and
  {Black}}]{visser09}
{Visser} R, {van Dishoeck} EF, and {Black} JH (2009) {The photodissociation and
  chemistry of CO isotopologues: applications to interstellar clouds and
  circumstellar disks}. \aap 503(2):323--343

\bibitem[{{von Weizs{\"a}cker}(1951)}]{weizsacker51}
{von Weizs{\"a}cker} CF (1951) {The Evolution of Galaxies and Stars.} \apj
  114:165

\bibitem[{{Wagner} et~al.(2012){Wagner}, {Falkovich}, {Kritsuk}, and
  {Norman}}]{wagner...12}
{Wagner} R, {Falkovich} G, {Kritsuk} AG, et~al. (2012) {Flux correlations in
  supersonic isothermal turbulence}. Journal of Fluid Mechanics 713:482--490

\bibitem[{{Walch} et~al.(2015){Walch}, {Girichidis}, {Naab}, {Gatto}, {Glover},
  {W{\"u}nsch}, {Klessen}, {Clark}, {Peters}, {Derigs}, and
  {Baczynski}}]{WalchEtAl2015}
{Walch} S, {Girichidis} P, {Naab} T, et~al. (2015) {The SILCC (SImulating the
  LifeCycle of molecular Clouds) project - I. Chemical evolution of the
  supernova-driven ISM}. \mnras 454:238--268

\bibitem[{{Walch} et~al.(2012){Walch}, {Whitworth}, {Bisbas}, {W{\"u}nsch}, and
  {Hubber}}]{WalchEtAl2012}
{Walch} SK, {Whitworth} AP, {Bisbas} T, et~al. (2012) {Dispersal of molecular
  clouds by ionizing radiation}. \mnras 427:625--636

\bibitem[{{Wang} et~al.(2010{\natexlab{a}}){Wang}, {Wang}, {Xiao}, {Shi}, and
  {Chen}}]{wang....10}
{Wang} J, {Wang} LP, {Xiao} Z, et~al. (2010{\natexlab{a}}) {A hybrid numerical
  simulation of isotropic compressible turbulence}. Journal of Computational
  Physics 229(13):5257--5279

\bibitem[{{Wang} et~al.(2012){Wang}, {Shi}, {Wang}, {Xiao}, {He}, and
  {Chen}}]{wang.....12}
{Wang} J, {Shi} Y, {Wang} LP, et~al. (2012) {Scaling and Statistics in
  Three-Dimensional Compressible Turbulence}. \prl 108(21):214505

\bibitem[{Wang et~al.(2013)Wang, Yang, Shi, Xiao, He, and Chen}]{wang.....13}
Wang J, Yang Y, Shi Y, et~al. (2013) Cascade of kinetic energy in
  three-dimensional compressible turbulence. Phys Rev Lett 110:214505,
  \urlprefix\url{https://link.aps.org/doi/10.1103/PhysRevLett.110.214505}

\bibitem[{{Wang} et~al.(2017{\natexlab{a}}){Wang}, {Gotoh}, and
  {Watanabe}}]{wang..17c}
{Wang} J, {Gotoh} T, and {Watanabe} T (2017{\natexlab{a}}) {Scaling and
  intermittency in compressible isotropic turbulence}. Physical Review Fluids
  2(5):053401

\bibitem[{{Wang} et~al.(2017{\natexlab{b}}){Wang}, {Gotoh}, and
  {Watanabe}}]{wang..17b}
{Wang} J, {Gotoh} T, and {Watanabe} T (2017{\natexlab{b}}) {Shocklet statistics
  in compressible isotropic turbulence}. Physical Review Fluids 2(2):023401

\bibitem[{{Wang} et~al.(2017{\natexlab{c}}){Wang}, {Gotoh}, and
  {Watanabe}}]{wang..17a}
{Wang} J, {Gotoh} T, and {Watanabe} T (2017{\natexlab{c}}) {Spectra and
  statistics in compressible isotropic turbulence}. Physical Review Fluids
  2(1):013403

\bibitem[{{Wang} et~al.(2019){Wang}, {Wan}, {Chen}, {Xie}, {Wang}, and
  {Chen}}]{wang.....19}
{Wang} J, {Wan} M, {Chen} S, et~al. (2019) {Cascades of temperature and entropy
  fluctuations in compressible turbulence}. Journal of Fluid Mechanics
  867:195--215

\bibitem[{{Wang} et~al.(2010{\natexlab{b}}){Wang}, {Li}, {Abel}, and
  {Nakamura}}]{Wang+2010}
{Wang} P, {Li} ZY, {Abel} T, et~al. (2010{\natexlab{b}}) {Outflow Feedback
  Regulated Massive Star Formation in Parsec-Scale Cluster-Forming Clumps}. The
  Astrophysical Journal 709(1):27--41

\bibitem[{{Ward} et~al.(2019){Ward}, {Kruijssen}, and {Rix}}]{ward2019}
{Ward} JL, {Kruijssen} JMD, and {Rix} HW (2019) {Not all stars form in clusters
  -- $Gaia$-DR2 uncovers the origin of OB associations}. \mnras~submitted
  arXiv:1910.06974

\bibitem[{{Webb} et~al.(1985){Webb}, {Axford}, and
  {Forman}}]{WebbAxfordForman1985}
{Webb} GM, {Axford} WI, and {Forman} MA (1985) {Cosmic-ray acceleration at
  stellar wind terminal shocks}. \apj 298:684--709

\bibitem[{{Weber} and {Davis}(1967)}]{WeberDavis1967}
{Weber} EJ and {Davis} J Leverett (1967) {The Angular Momentum of the Solar
  Wind}. \apj 148:217--227

\bibitem[{{Weingartner} and {Draine}(2001{\natexlab{a}})}]{wd01a}
{Weingartner} JC and {Draine} BT (2001{\natexlab{a}}) {Dust Grain-Size
  Distributions and Extinction in the Milky Way, Large Magellanic Cloud, and
  Small Magellanic Cloud}. \apj 548(1):296--309

\bibitem[{{Weingartner} and {Draine}(2001{\natexlab{b}})}]{wd01b}
{Weingartner} JC and {Draine} BT (2001{\natexlab{b}}) {Photoelectric Emission
  from Interstellar Dust: Grain Charging and Gas Heating}. \apjs
  134(2):263--281

\bibitem[{{Whelan} et~al.(2009){Whelan}, {Ray}, {Podio}, {Bacciotti}, and
  {Randich}}]{Whelan+2009}
{Whelan} ET, {Ray} TP, {Podio} L, et~al. (2009) {Classical T Tauri-like Outflow
  Activity in the Brown Dwarf Mass Regime}. The Astrophysical Journal
  706(2):1054--1068

\bibitem[{{White} et~al.(2019){White}, {Oliver}, {Mabey}, {K{\"u}hn-Kauffeldt},
  {Bott}, {D{\"o}hl}, {Bell}, {Bingham}, {Clarke}, {Foster}, {Giacinti},
  {Graham}, {Heathcote}, {Koenig}, {Kuramitsu}, {Lamb}, {Meinecke}, {Michel},
  {Miniati}, {Notley}, {Reville}, {Ryu}, {Sarkar}, {Sakawa}, {Selwood},
  {Squire}, {Scott}, {Tzeferacos}, {Woolsey}, {Schekochihin}, and
  {Gregori}}]{white-ea-19}
{White} TG, {Oliver} MT, {Mabey} P, et~al. (2019) {Supersonic plasma turbulence
  in the laboratory}. Nature Communications 10:1758

\bibitem[{{Whitworth} and {Summers}(1985)}]{WhitworthSummers1985}
{Whitworth} A and {Summers} D (1985) {Self-similar condensation of spherically
  symmetric self-gravitating isothermal gas clouds}. \mnras 214:1--25

\bibitem[{{Whitworth} et~al.(2007){Whitworth}, {Bate}, {Nordlund}, {Reipurth},
  and {Zinnecker}}]{WhitworthEtAl2007}
{Whitworth} A, {Bate} MR, {Nordlund} {\r{A}}, et~al. (2007) {The Formation of
  Brown Dwarfs: Theory}. In: {Reipurth} B, {Jewitt} D, and {Keil} K (eds)
  Protostars and Planets V, p 459

\bibitem[{{Whitworth} and {Stamatellos}(2006)}]{WhitworthStamatellos2006}
{Whitworth} AP and {Stamatellos} D (2006) {The minimum mass for star formation,
  and the origin of binary brown dwarfs}. \aap 458(3):817--829

\bibitem[{{Wiener} et~al.(2013){Wiener}, {Zweibel}, and
  {Oh}}]{WienerZweibelOh2013}
{Wiener} J, {Zweibel} EG, and {Oh} SP (2013) {Cosmic Ray Heating of the Warm
  Ionized Medium}. \apj 767:87

\bibitem[{Wiener(1930)}]{wiener30}
Wiener N (1930) Generalized harmonic analysis. Acta Math 55:117--258,
  \urlprefix\url{https://doi.org/10.1007/BF02546511}

\bibitem[{{Wiersma} et~al.(2009){Wiersma}, {Schaye}, and {Smith}}]{wiersma09}
{Wiersma} RPC, {Schaye} J, and {Smith} BD (2009) {The effect of photoionization
  on the cooling rates of enriched, astrophysical plasmas}. \mnras
  393(1):99--107

\bibitem[{{Wolfire} et~al.(1995){Wolfire}, {Hollenbach}, {McKee}, {Tielens},
  and {Bakes}}]{wolf95}
{Wolfire} MG, {Hollenbach} D, {McKee} CF, et~al. (1995) {The Neutral Atomic
  Phases of the Interstellar Medium}. \apj 443:152

\bibitem[{{Wolfire} et~al.(2003){Wolfire}, {McKee}, {Hollenbach}, and
  {Tielens}}]{wolfire2003}
{Wolfire} MG, {McKee} CF, {Hollenbach} D, et~al. (2003) {Neutral Atomic Phases
  of the Interstellar Medium in the Galaxy}. \apj 587:278--311

\bibitem[{{Wurster} and {Li}(2018)}]{WursterLi2019}
{Wurster} J and {Li} ZY (2018) {The role of magnetic fields in the formation of
  protostellar discs}. Frontiers in Astronomy and Space Sciences 5:39

\bibitem[{Yaglom(1949)}]{yaglom49}
Yaglom AM (1949) The field of acceleration in turbulent flow. Dokl Akad Nauk
  SSSR 67(5):795--798

\bibitem[{{Yeung}(2019)}]{yeung19}
{Yeung} PK (2019) {Advancing understanding of turbulence through extreme-scale
  computation}. In: APS Division of Fluid Dynamics Meeting Abstracts, APS
  Meeting Abstracts, p E01.001

\bibitem[{Yeung et~al.(2015)Yeung, Zhai, and Sreenivasan}]{yeung..15}
Yeung PK, Zhai XM, and Sreenivasan KR (2015) Extreme events in computational
  turbulence. Proceedings of the National Academy of Sciences of the United
  States of America 112(41):12633--12638,
  \urlprefix\url{https://www.jstor.org/stable/26465472}

\bibitem[{{Yorke} and {Sonnhalter}(2002)}]{YorkeSonnhalter2002}
{Yorke} HW and {Sonnhalter} C (2002) {On the Formation of Massive Stars}. \apj
  569(2):846--862

\bibitem[{{Zhang} et~al.(2005){Zhang}, {Hunter}, {Brand}, {Sridharan},
  {Cesaroni}, {Molinari}, {Wang}, and {Kramer}}]{Zhang+2005}
{Zhang} Q, {Hunter} TR, {Brand} J, et~al. (2005) {Search for CO Outflows toward
  a Sample of 69 High-Mass Protostellar Candidates. II. Outflow Properties}.
  The Astrophysical Journal 625(2):864--882

\bibitem[{{Zhang} et~al.(2016){Zhang}, {Arce}, {Mardones}, {Cabrit}, {Dunham},
  {Garay}, {Noriega-Crespo}, {Offner}, {Raga}, and {Corder}}]{Zhang+2016}
{Zhang} Y, {Arce} HG, {Mardones} D, et~al. (2016) {ALMA Cycle 1 Observations of
  the HH46/47 Molecular Outflow: Structure, Entrainment, and Core Impact}. The
  Astrophysical Journal 832(2):158

\bibitem[{{Zhang} et~al.(2019){Zhang}, {Arce}, {Mardones}, {Cabrit}, {Dunham},
  {Garay}, {Noriega-Crespo}, {Offner}, {Raga}, and {Corder}}]{Zhang+2019}
{Zhang} Y, {Arce} HG, {Mardones} D, et~al. (2019) {An Episodic Wide-angle
  Outflow in HH 46/47}. arXiv e-prints arXiv:1908.00689

\bibitem[{{Zhao} and {Aluie}(2018)}]{zhao.18}
{Zhao} D and {Aluie} H (2018) {Inviscid criterion for decomposing scales}.
  Physical Review Fluids 3(5):054603

\bibitem[{{Zrake} and {MacFadyen}(2012)}]{zrake.12}
{Zrake} J and {MacFadyen} AI (2012) {Numerical Simulations of Driven
  Relativistic Magnetohydrodynamic Turbulence}. \apj 744(1):32

\bibitem[{{Zweibel}(2013)}]{Zweibel2013}
{Zweibel} EG (2013) {The microphysics and macrophysics of cosmic rays}. Physics
  of Plasmas 20(5):055501

\bibitem[{{Zweibel}(2017)}]{Zweibel2017}
{Zweibel} EG (2017) {The basis for cosmic ray feedback: Written on the wind}.
  Physics of Plasmas 24(5):055402

\end{thebibliography}

\end{document}